\documentclass[12pt,preprint]{aastex}

\begin{document}
\shorttitle{X-rays from Cepheus B} \shortauthors{Getman et al.}
\slugcomment{Accepted for publication in the Astrophysical
Journal}

\title{{\it Chandra} Study of the Cepheus B Star Forming Region: Stellar Populations and the Initial Mass Function}

\author{Konstantin V.\ Getman\altaffilmark{1}, Eric D.\
Feigelson\altaffilmark{1}, Leisa Townsley\altaffilmark{1}, Patrick
Broos\altaffilmark{1}, Gordon Garmire\altaffilmark{1}, Masahiro
Tsujimoto\altaffilmark{1,2}}

\altaffiltext{1}{Department of Astronomy \& Astrophysics, 525 Davey
Laboratory, Pennsylvania State University, University Park PA 16802}

\altaffiltext{2}{Department of Physics, Rikkyo University, 3-34-1,
Nishi-Ikebukuro, Toshima, Tokyo, 171-8501, Japan}

\begin{abstract}
Cepheus B (Cep B) molecular cloud and a portion of the nearby Cep
OB3b OB association, one of the most active regions of star
formation within 1 kpc, has been observed with the ACIS detector
on board the {\it Chandra X-ray Observatory}. We detect 431 X-ray
sources, of which $89\%$ are confidently identified as clustered
pre-main sequence (PMS) stars. Two main results are obtained.
First, we provide the best census to date for the stellar
population of the region.  We identify many members of two rich
stellar clusters: the lightly obscured Cep OB3b association, and
the deeply embedded cluster in Cep B whose existence was
previously traced only by a handful of radio sources and T Tauri
stars. Second, we find a discrepancy between the X-ray Luminosity
Functions (XLFs) of the Cep OB3b and the Orion Nebula Cluster
(ONC). This may be due to different Initial Mass Functions (IMFs)
of two regions (excess of $\simeq 0.3$ M$_\odot$ stars), or
different age distributions.

Several other results are obtained. A diffuse X-ray component seen
in the field is attributed to the integrated emission of
unresolved low mass PMS stars. The X-ray emission from HD 217086
(O7n), the principle ionizing source of the region, follows the
standard model involving many small shocks in an unmagnetized
radiatively accelerated wind. The X-ray source \#294 joins a
number of similar superflare PMS stars where long magnetic
structures may connect the protoplanetary disk to the stellar
surface.
\end{abstract}


\section{Introduction}

\begin{figure}
\centering
\includegraphics[angle=0.,width=6.5in]{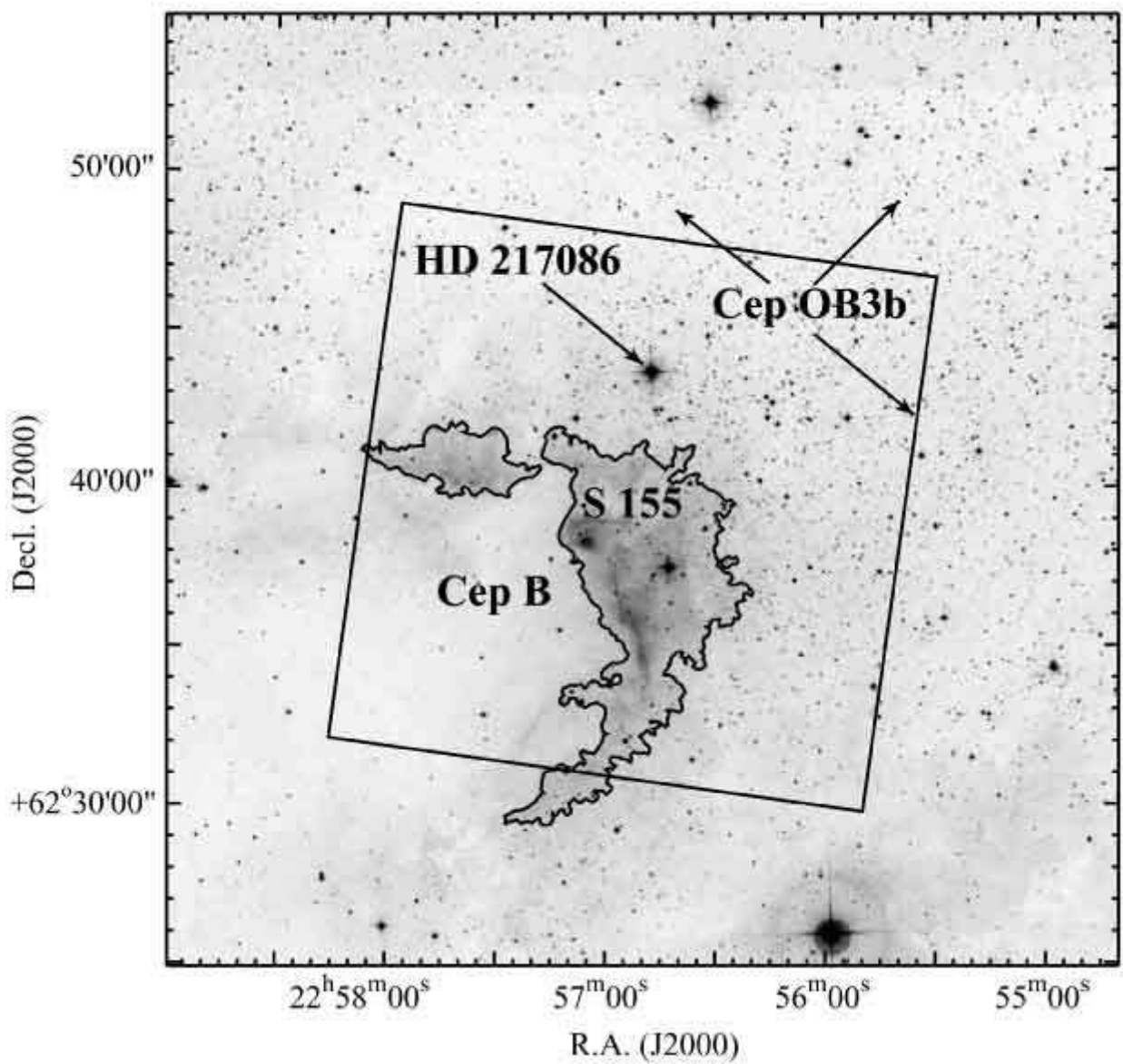}
\caption{$R$-band image covering $0.5\degr \times 0.5\degr$ of the
Cepheus B (Cep B) and Cep OB3b neighborhood from the Digital Sky
Survey. The {\it Chandra} $17\arcmin \times 17\arcmin$ ACIS-I is
outlined by the square. Cep B, the hottest component of the
Cepheus Molecular Cloud, is at the bottom-left corner of the
$Chandra$ field. To the north and west lies Cep OB3b, the youngest
of two subgroups of the stellar Cep OB3 association. The interface
between Cep B and Cep OB3 is delineated by the HII region,
Sharpless 155 (S 155). The most massive and optically bright star
in our field, HD 217086 (O7n), is labelled. North is up, east is
to the left. \label{fig_intro_fig}}
\end{figure}

Cepheus B (Cep B), the hottest CO component of the Cepheus
molecular cloud \citep{Sargent77,Sargent79}, is located near the
northwestern edge of the cloud at a distance of 725 pc from the
Sun \citep{Blaauw59,Crawford70}. Outside Cep B, to the north and
west of the cloud, lies the stellar Cep OB3 association, composed
of two subgroups of stars of different ages with the youngest (Cep
OB3b) lying closer to the molecular cloud \citep{Blaauw64,Jordi96}
(see Figure \ref{fig_intro_fig}).

\citet{Sargent79} considered Cep OB3  to be a good example of
sequential star formation, following the model of
\citet{Elmegreen77}. The interface between the molecular cloud and
the OB star association is clearly delineated by the optically
visible H{\sc II} region Sharpless 155 (S 155), where cloud
material is ionized and heated by the radiation field of the
youngest generation of the Cep OB association \citep{Panagia81}.
The photodissociation region (PDR) at S 155 is favorably oriented
to reveal the progression of star formation. As the surface of the
cloud is being eroded by the early-type stars, the cloud edge
moves eastward across the observer's field of view with new stars
emerging from the obscuring molecular cloud. Radio continuum
\citep{Felli78, Testi95}, far-infrared, CO \citep{Minchin92}, and
near-infrared studies \citep{Moreno-Corral93, Testi95} reveal a
few young stars embedded in the Cep B cloud behind the PDR.
Sources located on the edge of the molecular cloud and exhibiting
high extinction were suggested to represent a third generation of
star formation triggered by the H{\sc II} region shock propagating
into the cloud.

The low mass pre-main sequence (PMS) stellar population of Cep
OB3b can be effectively discriminated using X-ray emission due to
the heavy contamination in optical and near-infrared (ONIR)
surveys by both foreground and background Galactic field stars.
X-ray surveys are complementary to ONIR surveys because they trace
magnetic activity (mainly plasma heated in violent magnetic
reconnection flares) rather than photospheric or circumstellar
disk blackbody emission, and PMS X-ray emission is elevated
$10^{1}-10^{4}$ above main sequence (MS) levels
\citep{Preibisch05}. A {\it ROSAT} X-ray study of Cep OB3b
identified over 50 likely PMS stars with $A_V < 3$ mag and $L_X
\ga 10^{30}$ ergs s$^{-1}$ \citep{Naylor99}.  Ten of these {\it
ROSAT} sources are coincident with optical sources, and four of
these are spectroscopically confirmed classical and weak-lined T
Tauri stars \citep{Pozzo03}.

Much more can be achieved with the current generation of X-ray
telescopes, particularly the {\it Chandra X-ray Observatory} with
its excellent high-resolution mirrors and low-noise detectors.
Studies of the Orion Nebula region demonstrate that {\it Chandra}
images can penetrate up to $A_V \simeq 500$ mag into the cloud
\citep{Getman05, Grosso05}, considerably deeper than near-infrared
surveys in the $JHKL$ bands. $Chandra$ is also effective in
resolving crowded fields down to $\simeq 0.7$\arcsec\/ scales
\citep{Getman05b}.  In X-rays, OB stars are often not much
brighter than PMS stars, so multiple systems associated with OB
stars can be identified \citep{Stelzer05}.  $Chandra$ X-ray
studies are particularly effective in uncovering heavily obscured
low-mass cloud populations and in discriminating PMS populations
from unrelated older stars.

We present here a {\it Chandra} study of the region using its
Advanced CCD Imaging Spectrometer (ACIS) detector.  Over 380 PMS
stars in two rich clusters are found, concentrated in the lightly
obscured Cep OB3b cluster and a previously unknown embedded
cluster, and we infer the existence of an additional $700-800$
low-mass members. Only $\simeq 7-9$\% of the detected X-ray
sources are attributable to contaminating Galactic and
extragalactic populations.  The observation and source lists are
described in \S \ref{data_reduction_section}. Identification with
2MASS counterparts and their NIR properties are considered in \S
\ref{counterparts_and_properties_section}. X-ray fluxes and their
errors are determined in \S \ref{flux_det_section}. X-ray sources
unrelated to the cloud are discussed in \S
\ref{unidentified_main_section}. The discovery of two rich
clusters is described in \S \ref{new_cluster_section}.  Their
stellar populations are discussed in \S \ref{imf_xlf_section}
using measured X-ray luminosity functions (XLFs) and Initial Mass
Functions (IMFs). The origin of the soft diffuse X-ray emission
discovered here is discussed in \S \ref{diffuse_emission_section}.
Finally, we address two individual stars of particular interest:
HD 217086, the most massive star in Cep OB3b (\S
\ref{OB_section}), and {\it Chandra} source \#294 which exhibited
an unusually luminous and hot X-ray flare (\S
\ref{interesting_sources_section}).

\section{The {\it Chandra} observation and source list \label{data_reduction_section}}

\subsection{Observation and data reduction \label{data_reduction_sub1}}

The observation of Cep B and Cep OB3b was obtained on 2003 March
$11.51-11.88$ with the ACIS camera \citep{Garmire03} on-board {\it
Chandra} \citep{Weisskopf02}. We consider here only results
arising from the imaging array (ACIS-I) of four abutted $1024
\times 1024$ pixel front-side illuminated charge-coupled devices
(CCDs) covering about $17\arcmin \times 17 \arcmin$ on the sky.
The S2 and S3 detectors in the spectroscopic array (ACIS-S) were
also operational, but as the telescope point spread function (PSF)
is considerably degraded far off-axis, the S2-S3 data are omitted
from the analysis. The aim point of the array is
$22^{\rm{h}}56^{\rm{m}}49\fs4$, $+62\arcdeg39\arcmin55\farcs6$
(J2000), and the satellite roll angle (i.e., orientation of the
CCD array relative to the north-south direction) is $7\fdg9$. The
total net exposure time of our observation is 30 ks with no
background flaring or data losses.

Data reduction, source detection, and extraction have been done
using procedures similar to those described in detail by
\citet{Getman05}. Briefly, the data were partially corrected for
CCD charge transfer inefficiency \citep{Townsley02}, cleaned with
``grade", ``status", and ``good-time interval" filters, trimmed of
background events outside of the $0.5-8.0$ keV band, and cleaned
of bad pixel columns with energies of $< 0.7$ keV, left by the
standard processing (see Appendix B of \citet{Townsley03} for more
details). Based on several dozen matches between bright $Chandra$
and 2MASS point sources a correction to the absolute astrometry
was applied and the slight PSF broadening from the $Chandra$ X-ray
Center's (CXC's) software randomization of positions was removed.
Source searching was performed with data image and exposure maps
constructed at three spatial resolutions (0.5, 1.0, and
$1.4\arcsec$ per pixel) using the $CIAO$ {\it wavdetect} tool.
Those procedures were performed with $CIAO$ software package 3.1,
$LHEASOFT$ 5.3, the Penn State CTI corrector version 1.42, and the
$ACIS Extract$ ($AE$) package version 3.67. The latter two tools
were developed at Penn State\footnote{Descriptions and codes for
CTI correction and {\rm{\it ACIS Extract}} can be found at
\url{http://www.astro.psu.edu/users/townsley/cti/} and
\url{http://www.astro.psu.edu/xray/docs/TARA/ae\_users\_guide.html},
respectively.}.

\subsection{Source detection and characterization}

\begin{figure}
\centering
\includegraphics[angle=0.,width=6.5in]{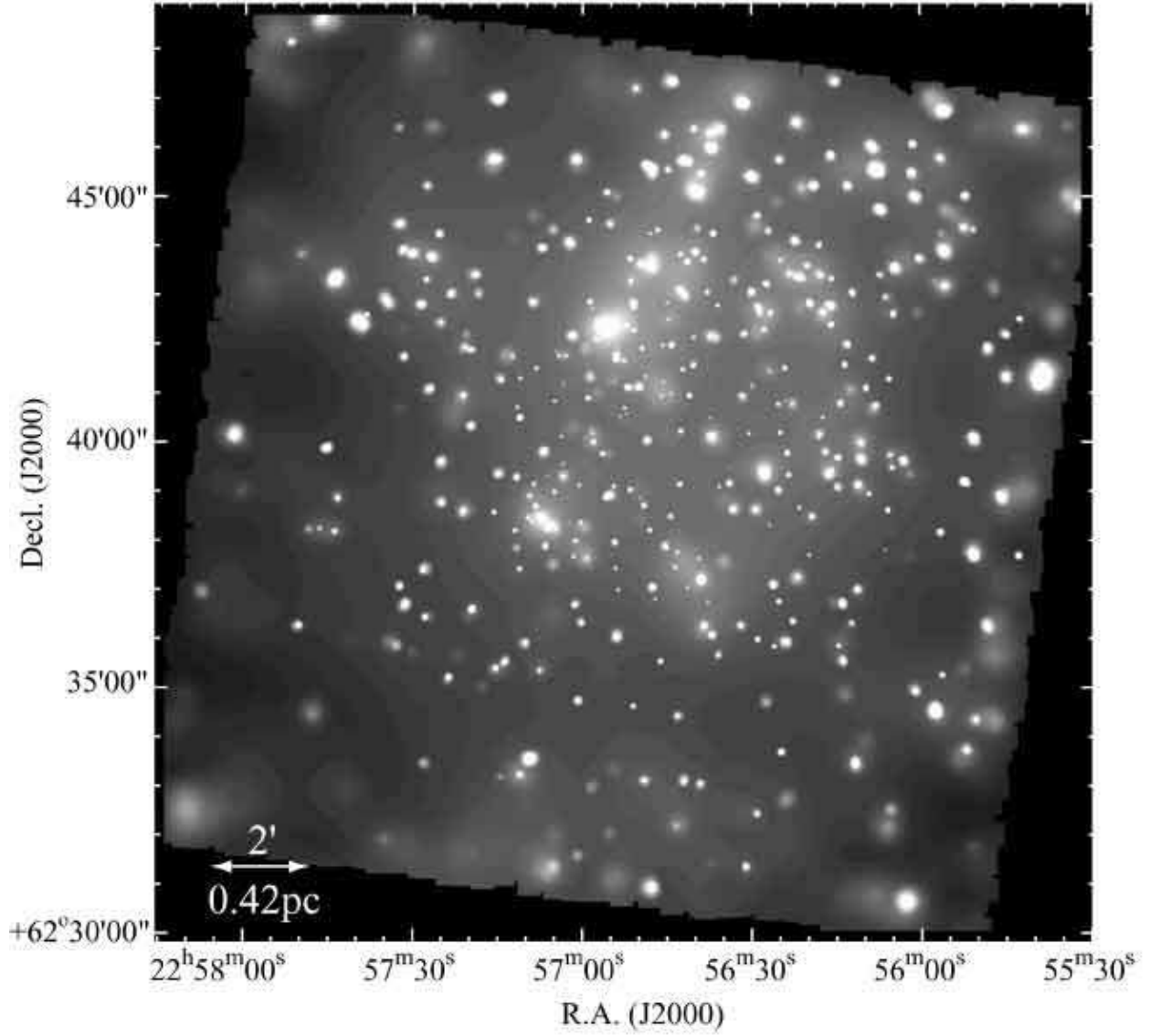}
\caption{The adaptively smoothed Cepheus B ACIS-I image in the
full $0.5-8.0$ keV band. Smoothing has been performed at the
$2.5\sigma$ level and gray scales are
logarithmic.\label{full_smoothed_fig}}
\end{figure}

Figure \ref{full_smoothed_fig} shows the resulting smoothed image
of the ACIS-I field. More than 400 point sources are
seen\footnote{We do not show here the raw image of the field as it
gives a misleading impression of high background and low
resolution.}. A wavelet-based source detection procedure ({\it
wavdetect}) is applied. It is followed by visual examination to
locate other candidate sources, mainly close doubles and sources
near the detection threshold. Using {\it AE}, photons are
extracted from the polygonal contours of $\sim 90\%$ encircled
energy using detailed position-dependent models of the PSF.
Background is measured locally in source-free regions.  The list
of candidate sources is trimmed to omit sources with fewer than
three background-subtracted counts in the PSF (i.e. $NetFull/PSF <
3$; these and other quantities are defined in the notes of Table
\ref{tbl_bsp}).  This resulted in a catalog of 431 sources, which
are listed in Table \ref{tbl_bsp}. Twenty-seven of the 31 weakest
sources with $3<NetFull/PSF<5$ counts (median source significance
$Signif = 1.1$, which corresponds \citep{Getman05} to
$2.2\,\sigma$ detection threshold of {\it PWDetect}) have
near-infrared (NIR) counterparts (\S
\ref{counterparts_and_properties_section}); we thus believe most
of these are real despite their marginal statistical significance.
One hundred and eight sources with $5 < NetFull/PSF < 10$ have
median $Signif = 1.8$ ($3.6\,\sigma$ detection threshold of {\it
PWDetect}).

The {\it AE} package provides a variety of source characteristics
including celestial position, off-axis angle, net and background
counts within the PSF-based extraction area, source significance
assuming Poisson statistics, effective exposure (corrected for
telescope vignetting, satellite dithering, and bad CCD pixels),
median energy after background subtraction, a variability
indicator extracted from the one-sided Kolmogorov-Smirnov
statistics, and occasional anomalies related to chip gap or field
edge positions. These are all reported in Table \ref{tbl_bsp}; see
\citet{Getman05} for a fuller description of these quantities.

\section{2MASS counterparts and their properties \label{counterparts_and_properties_section}}

\subsection{Identification with catalogued 2MASS sources \label{2MASS counterparts_section}}

An automated cross-correlation between the {\it Chandra} source
positions and 2MASS NIR catalogs was made using a search radius of
$1\arcsec$ within $\simeq 3.5\arcmin$ of the ACIS field center,
and a search radius of $2\arcsec$ in the outer regions of the
field where the {\it Chandra} PSF deteriorates. To evaluate the
merits of this automated procedure, we performed a careful visual
examination of each source. In 99\% of the cases, the
identifications are clear and unambiguous. Seventeen matches near
the edges of the field with {\it Chandra}-2MASS separations
greater than $2\arcsec$ were added to the catalog. The result is
that 385 out of 431 (89\%) {\it Chandra} sources have NIR
counterparts in the 2MASS catalog (Table \ref{tbl_nir}).

After a correction to the {\it Chandra} boresight of $\Delta {\rm
R.A.} = +0.13\arcsec$ and $\Delta {\rm Decl.}=-0.14\arcsec$,
mentioned in section \ref{data_reduction_sub1}, to bring it into
the 2MASS/Hipparcos reference frame, the median {\it
Chandra}-2MASS separations are better than $0.21\arcsec$ at the
central part of the field (off-axis angle $< 3.5\arcmin$), and
$0.47\arcsec$ at larger off-axis angles.  This excellent
astrometric agreement between 2MASS and {\it Chandra} positions is
similar to that achieved in the Orion Nebula Cluster field
\citep{Getman05}.

\subsection{X-ray sources with un-catalogued 2MASS counterparts \label{uncataloged_2MASS}}

\begin{figure}
\centering
\includegraphics[angle=0.,width=6.3in]{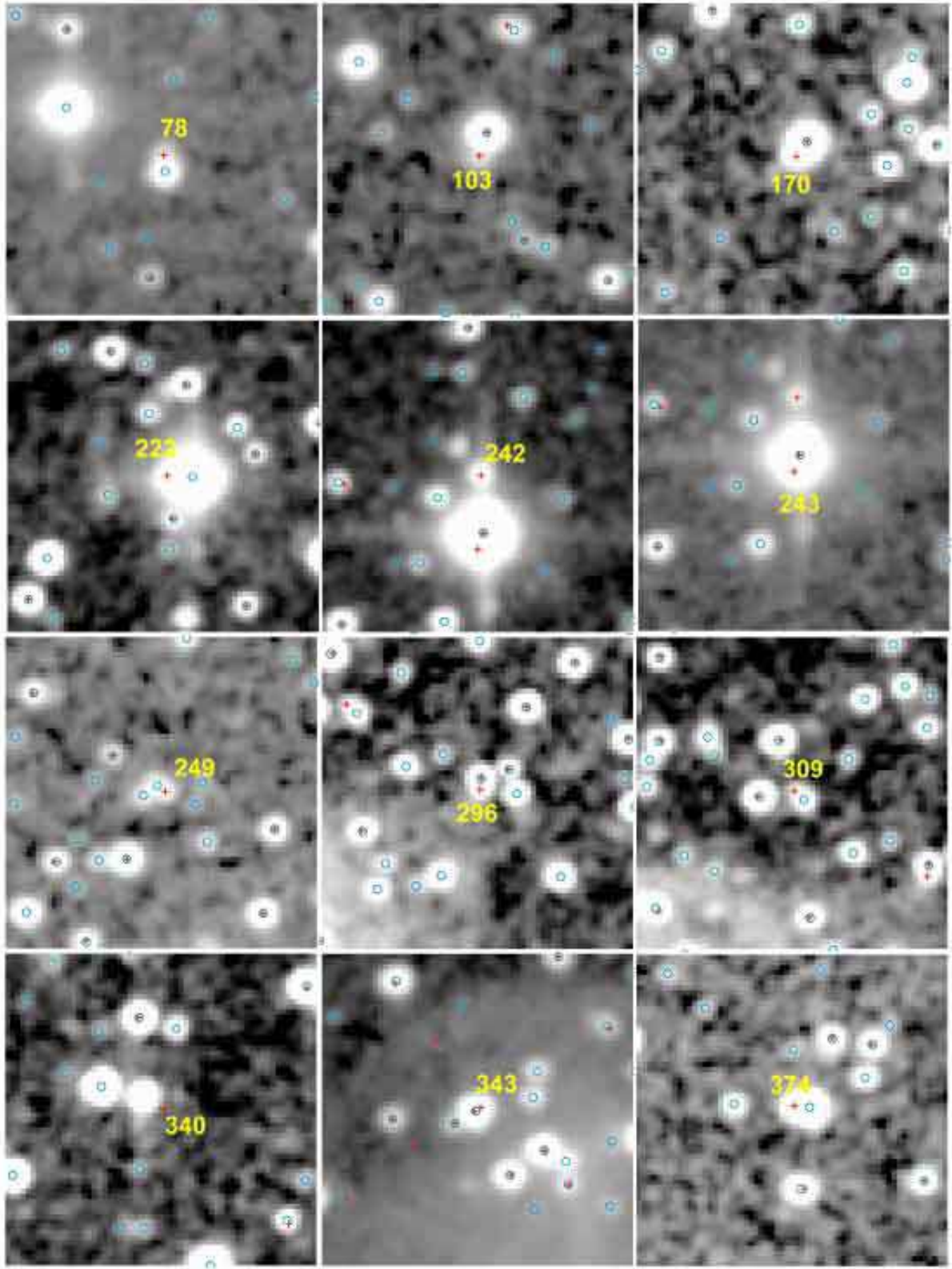}
\caption{Lightly absorbed $Chandra$ stars which are not resolved
in the 2MASS catalog. Each panel shows a $1\arcmin \times
1\arcmin$ neighborhood (1 pixel $= 1\arcsec$) around $Chandra$
sources (marked by a red $+$ and labelled with their source number
in yellow) lying in close proximity to bright 2MASS stars. Stars
registered in 2MASS catalog are marked by cyan
$\circ$.\label{soft_unident_fig}}
\end{figure}

Of the 46 X-ray sources not associated with 2MASS catalogued stars
14 are soft ($MedE < 2$ keV) and 32 are hard ($MedE > 2$ keV).
Visual examination of the 2MASS atlases shows that 12 soft (\#\#
78, 103, 170, 222, 242, 243, 249, 296, 309, 340, 343, 374) and 2
hard (\#\# 395 and 403) X-ray sources without identification lie
very close to bright NIR stars. As an example, expanded views of
the 2MASS atlas around 12 soft X-ray sources are shown in Figure
\ref{soft_unident_fig}. The nearby catalogued 2MASS stars are
often flagged as an extended NIR source. In the case of \#\# 242
and 243, {\it Chandra} found two un-catalogued companions to a
bright 2MASS star. X-ray source \# 340 is suspiciously close to an
unregistered bright 2MASS source, but does not coincide with it;
this may be emission from an HH shock, but the lack of optical
data does not allow us to speculate on it. The only two soft X-ray
sources that are truly not seen in NIR 2MASS are \#\# 26 and 214.
These two are the hardest among the soft sample and perhaps are
too absorbed to be detected in NIR. Later in the membership study
(\S \ref{unidentified_main_section}) we consider these 14 soft
sources, including \#\# 26, 214, 340, and two hard sources, \#\#
395 and 403, as likely members of the Cepheus populations.

\subsection{NIR properties of X-ray detected sources \label{nir_properties_section}}

\begin{figure}
\centering
\includegraphics[angle=0.,width=6.5in]{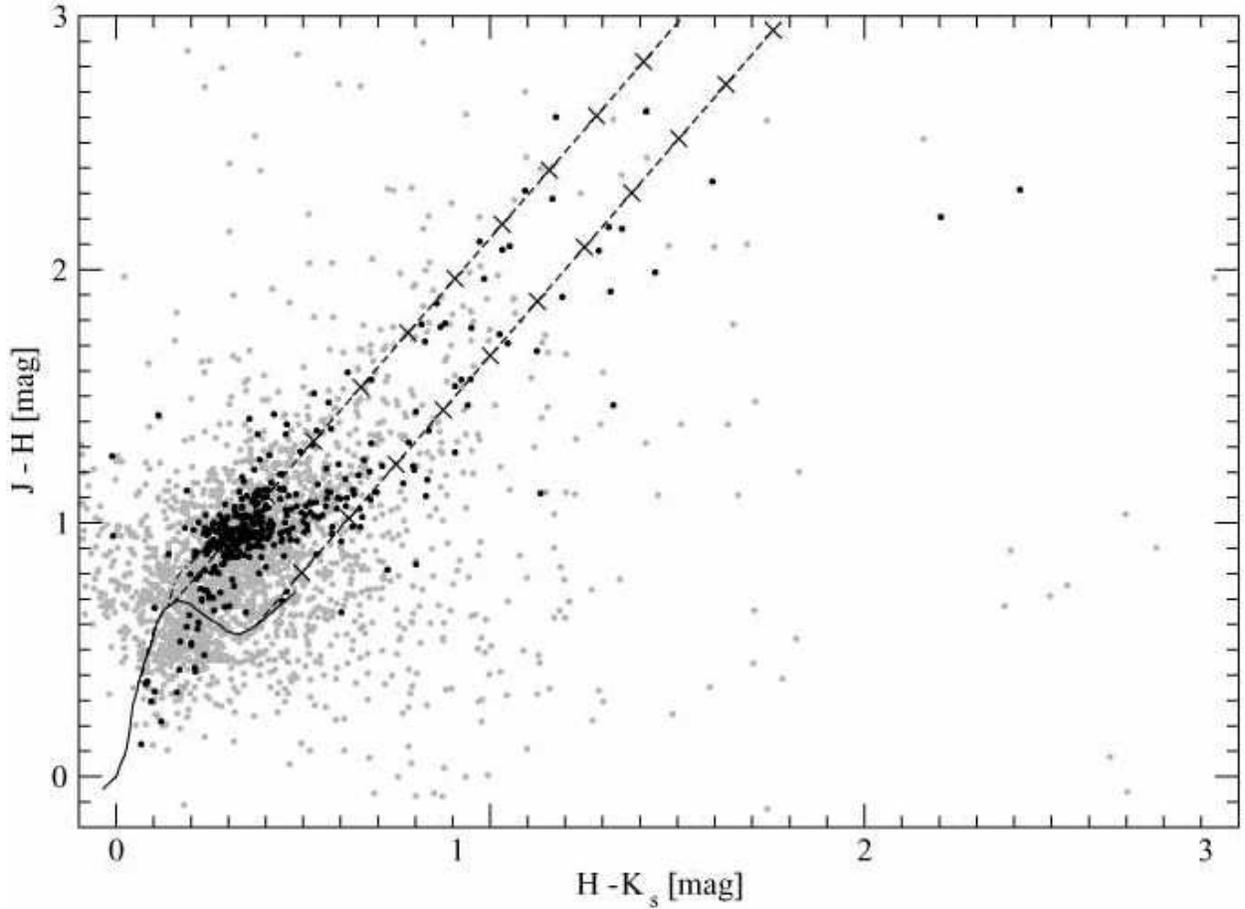}
\caption{NIR color-color diagram shows $JHK$ colors of all ($\sim
2700$) 2MASS sources (grey dots) inside the {\it Chandra} ACIS-I
field, and 385 2MASS sources with possible {\it Chandra}
counterparts (black dots). The solid and dot-dashed curves show
sites of intrinsic $JHK_s$ colors of main sequence and giant stars
respectively. The dashed lines are reddening vectors originating
at M0 V (left line) and M6.5 V (right line), and marked at
intervals of $A_V = 2$ mag.\label{cc_cmd_fig}}
\end{figure}

\begin{figure}
\centering
\includegraphics[angle=0.,width=6.5in]{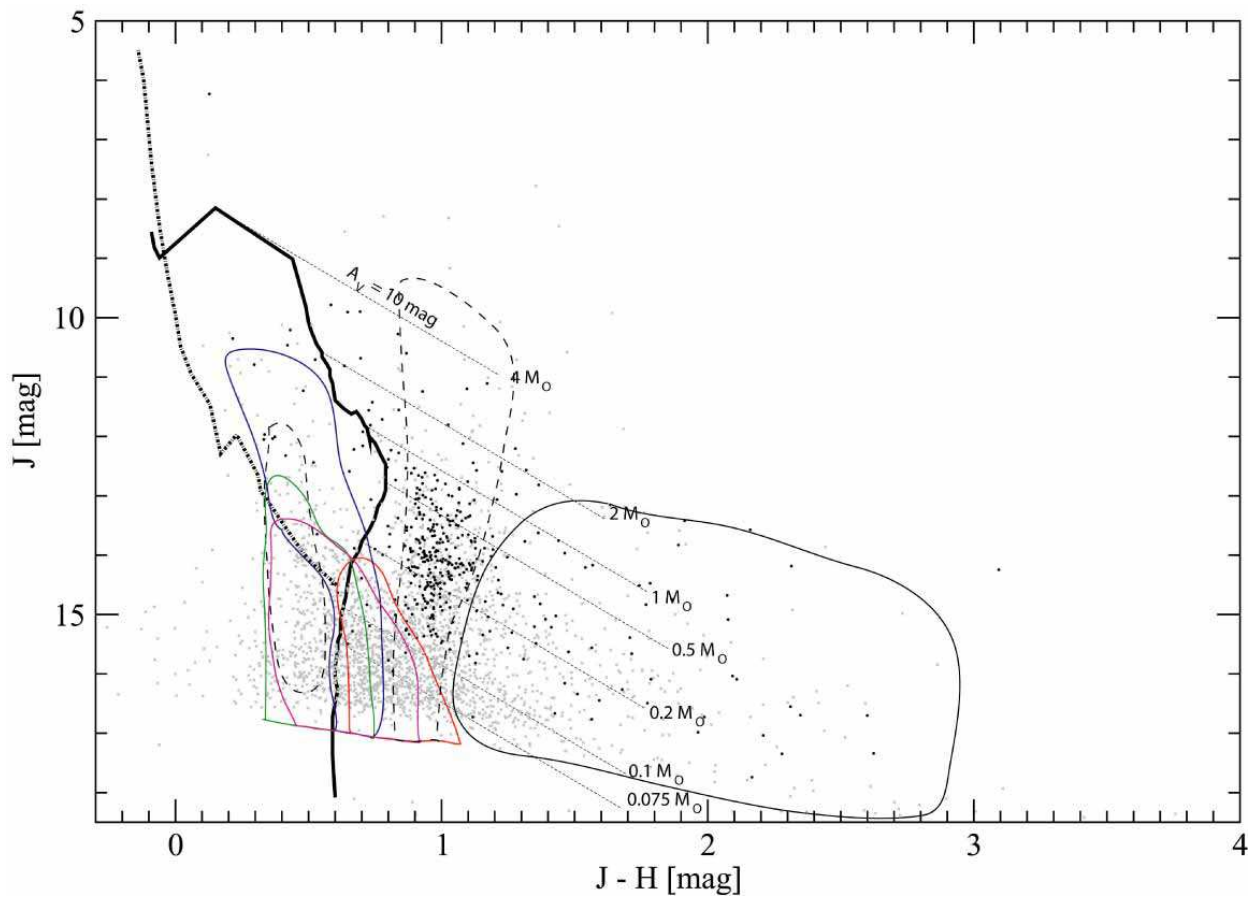}
\caption{NIR color-magnitude diagram for the same set of sources
as in Figure \ref{cc_cmd_fig}.  Black dots show $Chandra$ sources
with 2MASS counterparts, and grey dots show all 2MASS sources in
the field.  Thick black lines give the unabsorbed ZAMS (dotted
line) and 1\,Myr isochrones (solid line), from which $A_{V} \sim
10$ mag reddening vectors are shown for various star masses. The
remaining curves show various expected subpopulations of
foreground and background stellar populations obtained from the
Besan\c{c}on simulations. Details are given in footnote
\ref{cmd_besancon_footnote}.\label{cmd_besancon_fig}}
\end{figure}

Figures \ref{cc_cmd_fig} and \ref{cmd_besancon_fig} compare NIR
properties of all cataloged 2MASS sources in the ACIS-I field with
the 385 detected in X-rays (not including un-cataloged 2MASS
sources from \S \ref{uncataloged_2MASS}). The latter sources are
clustered in NIR color space at $A_V \sim 2-4$ mag with only $\sim
10$ having NIR colors consistent with foreground main sequence
stars. NIR colors for $\sim 30$, 8\% of the X-ray selected
population, show $K$-band excesses that indicate the presence of
dusty protoplanetary disks.

\citet{Pozzo03} found the age of $\sim 1$ Myr for four out of six
PMS candidates within their 1a and 2b fields (see their Figure
15), which partially overlap our ACIS-I field. In Figure
\ref{cmd_besancon_fig} we show the 1~Myr isochrone in the
$J$~vs.~$(J-H)$ color-magnitude diagram  with reddening loci for
PMS stars of different masses (see footnote
\ref{cmd_besancon_footnote} for references). We compare locations
of our X-ray sources on the color-magnitude diagram with the 1~Myr
PMS isochrone and estimate their masses, bolometric luminosities,
and local absorptions. Most of the X-ray sources are consistent
with PMS stars with masses of $0.1-1$ M$_\odot$ at a distance of
725 pc \citep{Testi95} subject to $1<A_V<3$ mag absorption.
Several dozen appear to be more obscured PMS stars with
$5<A_V<15$.

Table \ref{tbl_nir} provides information on the 385 2MASS
counterparts, their offsets from {\it Chandra} positions,
photometric magnitudes, photometry quality and confusion flags.
With the assumption that most of the detected X-ray sources are
$\sim 1$ Myr old PMS stars we also give our estimates of stellar
masses, absorptions, and bolometric luminosities based on the NIR
color-magnitude diagram in Figure \ref{cmd_besancon_fig}.

\subsection{Understanding the background stellar population \label{besancon}}

We examined Monte Carlo simulations of the Galactic stellar
population expected in the direction of our ACIS field
$(l,b)=(110.2, +2.7)$ based on the stellar population synthesis
model of \citet{Robin03}\footnote{This was computed by the Web
service provided by the \citet{Robin03} group at Besan\c{c}on at
\url{http://bison.obs-besancon.fr/modele}.  In addition to the
smooth absorption component provided in the model, we added a
cloud at $d=725$ pc with $A_V \sim 2.5$ mag, as inferred from the
average NIR colors of {\it Chandra} sources shown in Figures
\ref{cc_cmd_fig}-\ref{cmd_besancon_fig}.  As we have not
considered the patchy denser molecular cloud which covers $\sim
1/4$ of ACIS-I FOV, the inferred background population may be
somewhat overestimated.  We refer to this hereafter as the
Besan\c{c}on model.}.  Within the ACIS field-of-view, the
Besan\c{c}on model predicts $\sim 100-150$ MS foreground ($d <
725$ pc) and $\sim 1400-3100$ background MS, giants, and
sub-giants within the 2MASS sensitivity limit around $J < 16-17$
mag.

The simulations help us to understand the bimodal distribution of
2MASS stars in the color-magnitude diagram (this was previously
noticed from $V$ versus $V-I$ optical diagrams by
\citet{Pozzo03}). The model indicates that, due to its sensitivity
limits, 2MASS will preferentially detect the M, K, and G MS stars
at distances of $\sim 0.8 - 3.6$ kpc, but will detect large
populations of A and F MS stars, and sub-giants and  giants of all
spectral types, located at distances of $\sim 5.4 - 5.7$ kpc. Loci
for these sub-populations are shown in the NIR color-magnitude
diagram of Figure \ref{cmd_besancon_fig}\footnote{We give here
details concerning the $J$ vs. $(J-H)$ color magnitude diagram in
Figure \ref{cmd_besancon_fig}.  Black dots show $Chandra$ sources
with 2MASS counterparts, and grey dots show all 2MASS sources in
the field. The ZAMS (thick black dotted line on left) is obtained
from Tables 7.6 and 15.7 of \citet{Cox00}. The 1\,Myr isochrone
(thick black line on right) is from \citet{Baraffe98} for $0.02\,
M_{\odot} \leqslant M \leqslant 1.4\, M_{\odot}$ and
\citet{Siess00} for $1.4\, M_{\odot} \leqslant M \leqslant 7.0\,
M_{\odot}$. Reddening vectors of $A_{V} \sim 10$ mag (dashed
lines) are given for $0.075\,M_{\odot}, 0.1\,M_{\odot},
0.2\,M_{\odot}, 0.5\,M_{\odot}, 1.0\,M_{\odot}, 2.0\,M_{\odot}$,
and $4.0\,M_{\odot}$.  Overlayed on this color-magnitude diagram
are outlines of the distributions of expected foreground and
background populations obtained from Besan\c{c}on simulations of
the Galactic stellar populations in the direction of Cepheus.
They clearly explain the two-horn shape of the color distribution
of 2MASS stars. The horn around $J-H \simeq 0.5-0.8$ consists of
G-K-M main sequence background stars (red locus), A-F main
sequence background stars (green locus), background sub-giants
(luminosity class IV, purple locus), younger ($< 1$ Gyr)
background giants (luminosity class III, black dashed locus around
$J-H=0.5$), and foreground main sequence stars (blue locus).  The
horn around $J-H \simeq 0.8-1.2$, which includes most of the
$Chandra$ sources, has older background giants ($>1$ Gyr, black
dashed locus around $J-H=1.0$). The solid black locus covering
heavily obscured stars with $A_V \ga 5$ mag contains a mixture of
embedded PMS stars and highly reddened background main sequence
stars, sub-giants and giants. \label{cmd_besancon_footnote}}. The
region occupied by most of our X-ray sources, PMS stars with $A_V
\sim 2-4$ mag, is also the characteristic location for $\sim
400-500$ older ($> 1$ Gyr) background giants.  None of those
giants is expected to emit X-rays detectable in our observation
(\S \ref{galactic_contamination_section}).

\section{X-ray spectroscopy and fluxes \label{flux_det_section}}

\begin{figure}
\centering
\includegraphics[angle=0.,width=6.0in]{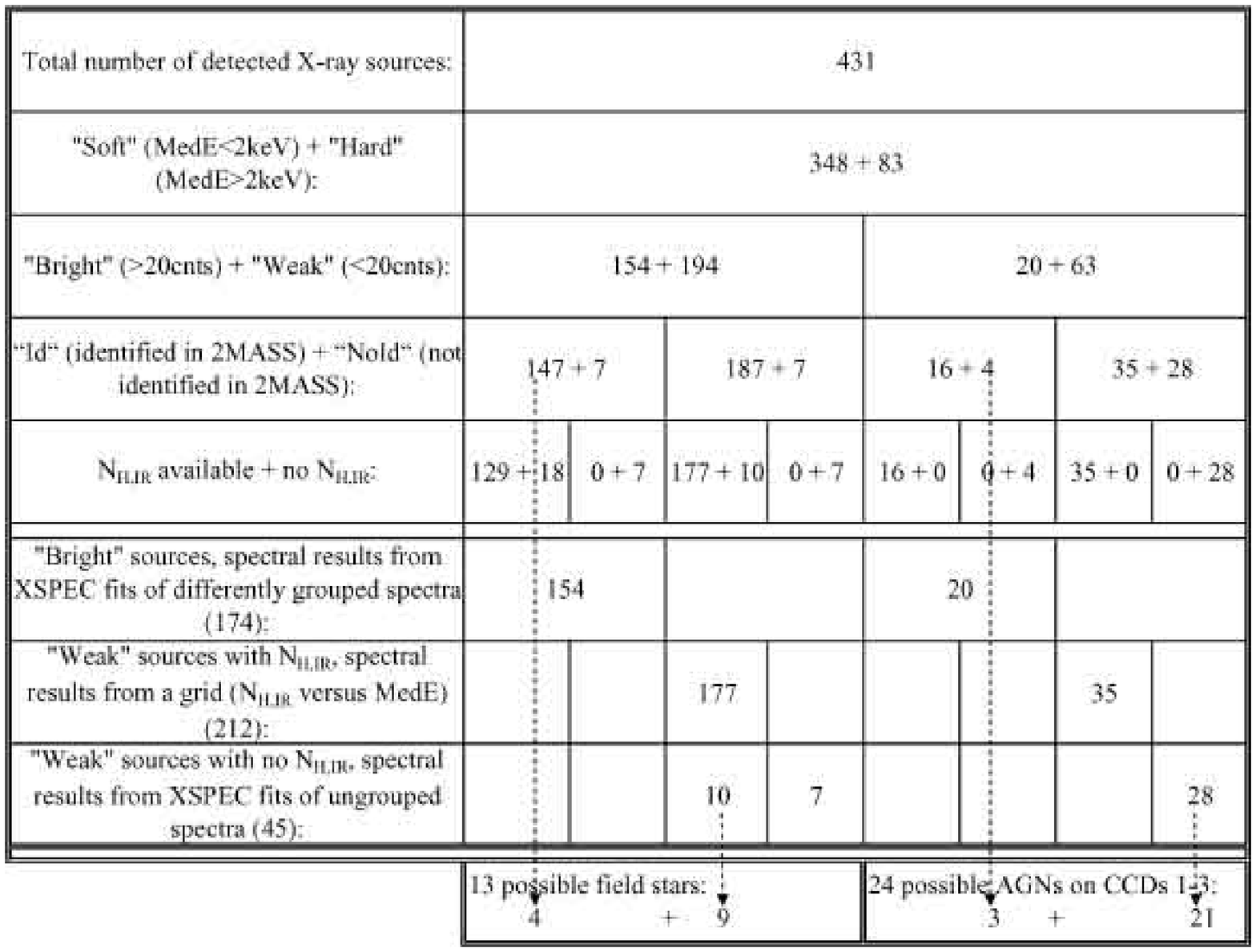}
\caption{X-ray source classification.  The first five rows
classify sources based on X-ray photometry and NIR
counterpart-photometry information.  The next three rows indicate
classes of sources with different spectral analysis methods
applied. The last row shows classes of sources likely unrelated to
the Cepheus region derived in \S\S \ref{hard_unidentified_section}
and \ref{foreground_possible_section}.\label{source_class_fig}}
\end{figure}

The XLF of stars in young stellar clusters exhibits a range of
$10^5$ in X-ray luminosity \citep{Feigelson05}.  In relatively
short exposures of distant clusters, the majority of sources have
very few counts while a few may have thousands of counts.
Consequently, no single method is well-adapted to characterization
of source spectra and fluxes.   We describe here three different
approaches applied to three flux levels, which are summarized in
Figure \ref{source_class_fig}: for 174 sources with $>20$ counts
(sixth row of the Figure), we perform regressions to spectral
models giving direct estimations of plasma temperatures,
line-of-sight absorption, and broad-band fluxes; for 212 weak
sources with absorptions estimated from NIR
colors\footnote{$N_{H,IR}$ was derived from $A_{V,IR}$ reported in
Table \ref{tbl_nir} using the relationship $N_H = 2.0 \times
10^{21} A_V$ \citep{Ryter96}.} (seventh row of the Figure), we
perform regressions with plasma temperature as a parameter; and
for 45 weak X-ray sources without $A_{V,IR}$ estimates (eighth row
of the Figure), we obtain non-unique spectral fits for the sole
purpose of estimating observed broad-band fluxes.  This last
category of sources mostly consists of sources unrelated to the
Cepheus star formation region (\S\S
\ref{hard_unidentified_section} and
\ref{foreground_possible_section}).  Resulting spectral fits and
broad-band luminosities appear in Table \ref{tbl_spe}. Following
\citet{Getman05}, we report observed hard-band ($2-8$ keV, $\log
L_h$) and full-band ($0.5-8$ keV, $\log L_t$) luminosities, and
intrinsic luminosities after correction for absorption ($\log
L_{h,c}$, $\log L_{t,c}$).  We omit soft band luminosities due to
the absorption seen in virtually all sources.

\subsection{Spectral fits for sources with $>20$ counts
\label{bright_fits_section}}

For 174 sources with greater than 20 net counts (154 lightly
absorbed and 20 heavily absorbed), automated spectral fitting was
performed using {\it AE} and {\it XSPEC} \citep{Arnaud96}. We
applied a one-temperature optically thin thermal plasma emission
model \citep[APEC]{Smith01} assuming a uniform density and 0.3
times solar elemental abundances which is typical of young stellar
X-ray sources \citep[e.g.][]{Imanishi01,Feigelson02a}. To explore
the whole range of various alternative models that successfully
fit the data we performed spectral fits on grouped (with different
minimum number of grouped counts, starting from 1, using the
$\chi^2$-statistic\footnote{We tested $\chi^2$-statistics with
different error weighting methods, such as standard, gehrels, and
churazov, and found that the best-fit values of the model
parameters remain similar for a spectrum of the same grouping.})
as well as on un-grouped (using the C-statistic) spectra. We then
performed a visual inspection of all spectral fits to choose the
``most appropriate" spectral fit based on subjective judgment as
outlined below.

\begin{figure}
\centering
\includegraphics[angle=0.,height=8.3in]{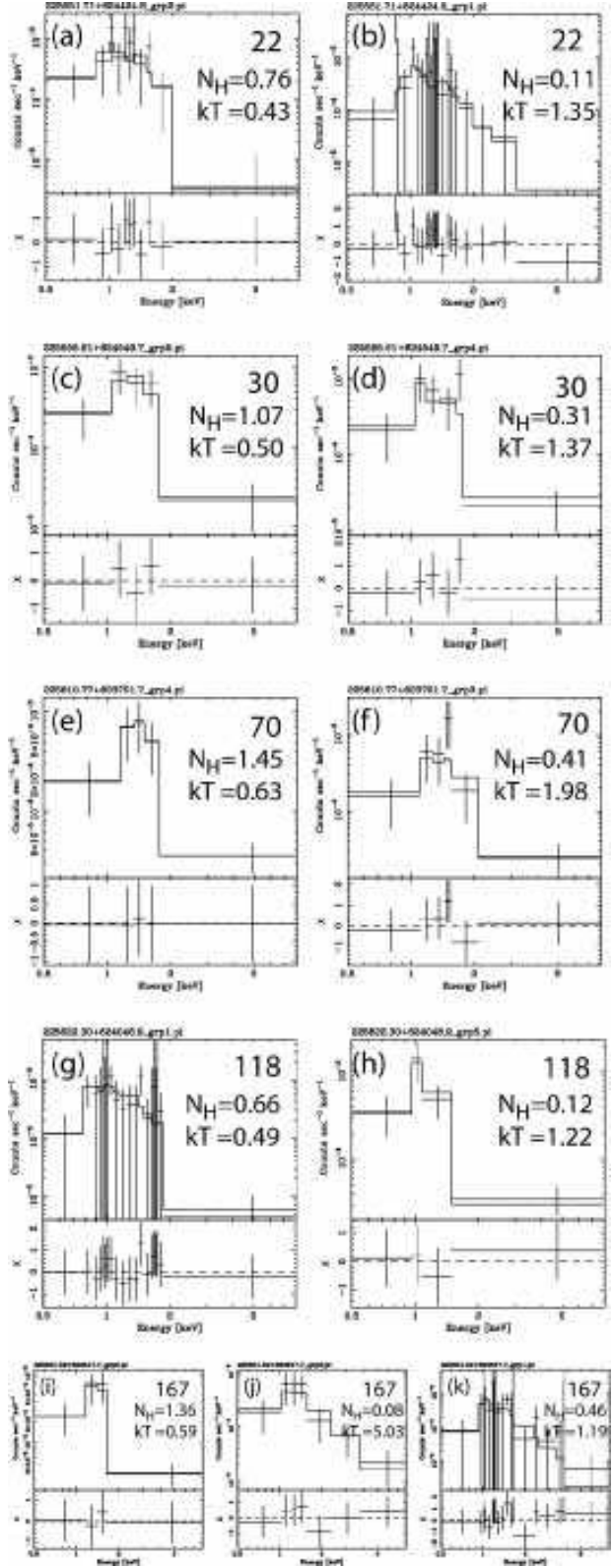}
\caption{Spectra of five lightly absorbed X-ray sources with $20 <
NetFull < 23$ counts exemplifying a common ambiguity in spectral
fitting. Each row shows alternative models which successfully fit
the data; the right-hand panel shows the fit we use with $kT$
values typical for COUP PMS stars. \label{spectra_soft_fig}}
\end{figure}

\begin{figure}
\centering
\includegraphics[angle=0.,height=8.3in]{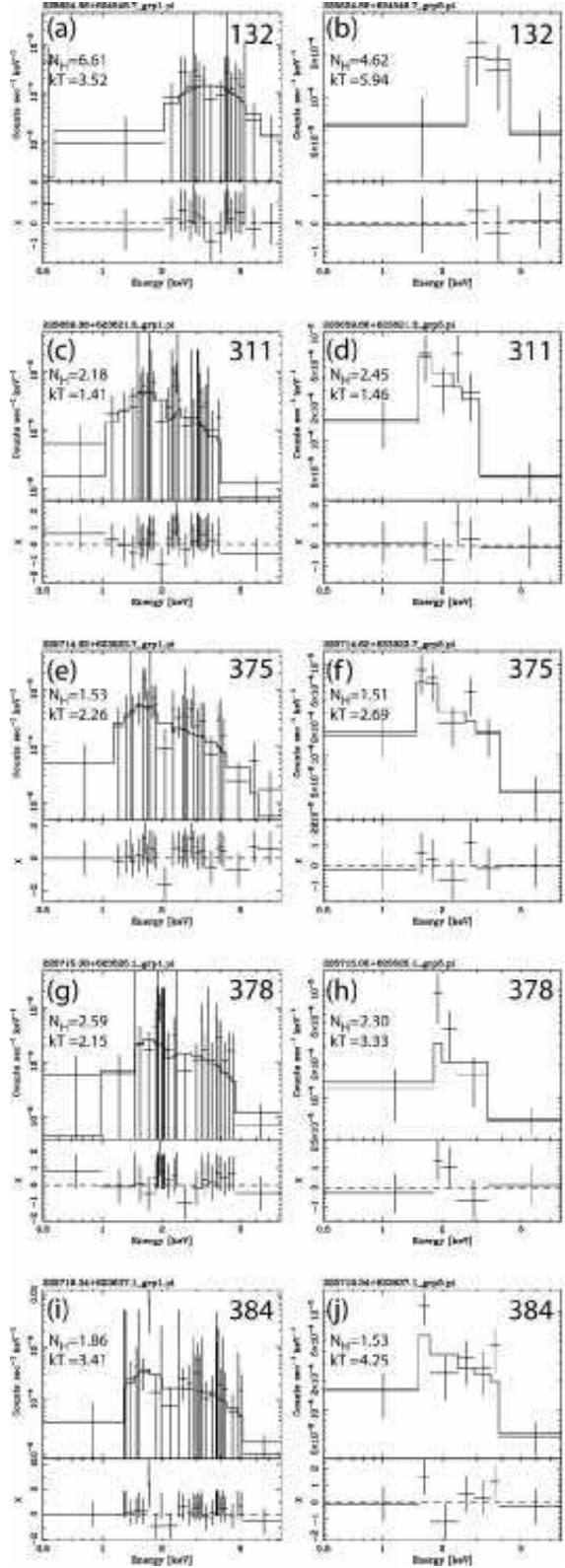}
\caption{Spectra of five heavily absorbed X-ray sources with
$NetFull > 20$ counts exemplifying stable spectral fitting over
different grouping schemes. The finest grouping is to the left and
the coarsest available grouping is to the right. As a final fit
for highly absorbed sources we thus arbitrarily report a fit from
one of the available groupings. \label{spectra_hard_fig}}
\end{figure}

For lightly absorbed sources (roughly, sources with median energy
$MedE < 2$ keV), data can often be successfully fit with two or
more qualitatively different spectral models. Spectral fits of
five sources with $20-23$ net counts are shown in Figure
\ref{spectra_soft_fig} to exemplify this common ambiguity in
spectral fitting. Sources can be often successfully fit by two
different types of models: high $N_H$ and low $kT$, or lower $N_H$
with higher $kT$. Source \# 167 can be fit by three different
classes of models. This problem was also noted by \citet{Getman05}
for the Orion Nebula Cluster (COUP project) where many more
photons are available. In these situations, we adopted a model
where the plasma energy most closely matches the typical energy of
PMS stars in Orion: for unabsorbed stars with NIR counterparts,
the median plasma temperature (from one-temperature fits) is $kT
\sim 1.6$ keV; for absorbed stars with NIR counterparts, $kT \sim$
2.6 keV; and for absorbed stars without NIR counterparts, $kT \sim
4.4$ keV \citep{Getman05b}.  In Figure \ref{spectra_soft_fig}, we
thus preferred the right-hand panel fits.  For highly absorbed
sources (all except 4 with $>30$ net counts), data with different
groupings are successfully fit by similar class of models (Figure
\ref{spectra_hard_fig}), and we arbitrarily report a fit from one
of the available groupings.

\subsection{Spectral fits for sources with $<20$ counts
\label{fits_less_20_counts}}

\begin{figure}
\centering
\includegraphics[angle=0.,width=6.5in]{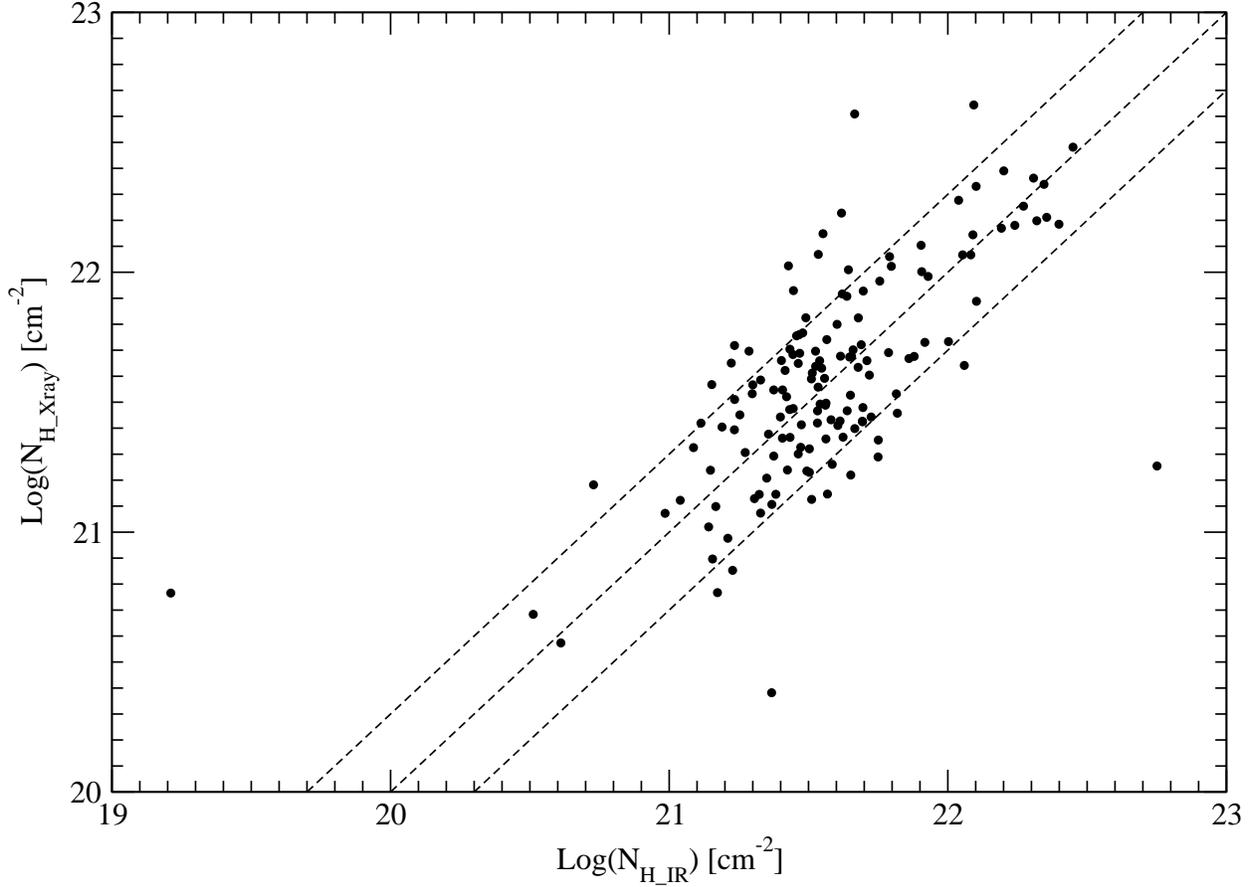}
\caption{Plot of absorbing column densities obtained from X-ray
spectral fits ($N_{H\_Xray}$) against column densities obtained
from NIR photometry ($N_{H\_IR}$). The sample consists of 129
bright soft and 16 bright hard sources (see row 5 of Figure
\ref{source_class_fig}). $N_{H,IR}$ is derived from the visual
absorption $A_{V}$  using $N_{H} = 2.0 \times 10^{21} A_V$
\citep{Ryter96} where $A_V$ is obtained by dereddening source
colors to the 1 Myr PMS isochrone in the 2MASS color-magnitude
diagram. The dashed lines show $\log (N_{H\_Xray}/N_{H\_IR})$ of
-0.3, 0, and 0.3. \label{compare_nhs_fig}}
\end{figure}

Figure \ref{compare_nhs_fig} compares absorption column densities
derived from our X-ray spectra of bright ($NetFull > 20$ cts)
sources with those derived from the 2MASS $J$~vs.$(J-H)$ color
magnitude diagram.  NIR sources were  dereddened to the 1 Myr PMS
isochrone. We see that the X-ray and NIR absorption columns are
statistically equivalent; no bias is seen and scatter is generally
within $\pm 0.3$ in $\log N_{H}$\footnote{A similar correlation
exists for COUP data, but it has not been reported.}. On this
basis, we can reliably substitute $\log N_{H,IR}$ for $\log
N_{H,Xray}$ in spectral fitting of 212 fainter sources (177 soft
and 35 hard) which have 2MASS NIR photometry.

\begin{figure}
\centering
\includegraphics[angle=0.,width=5.5in]{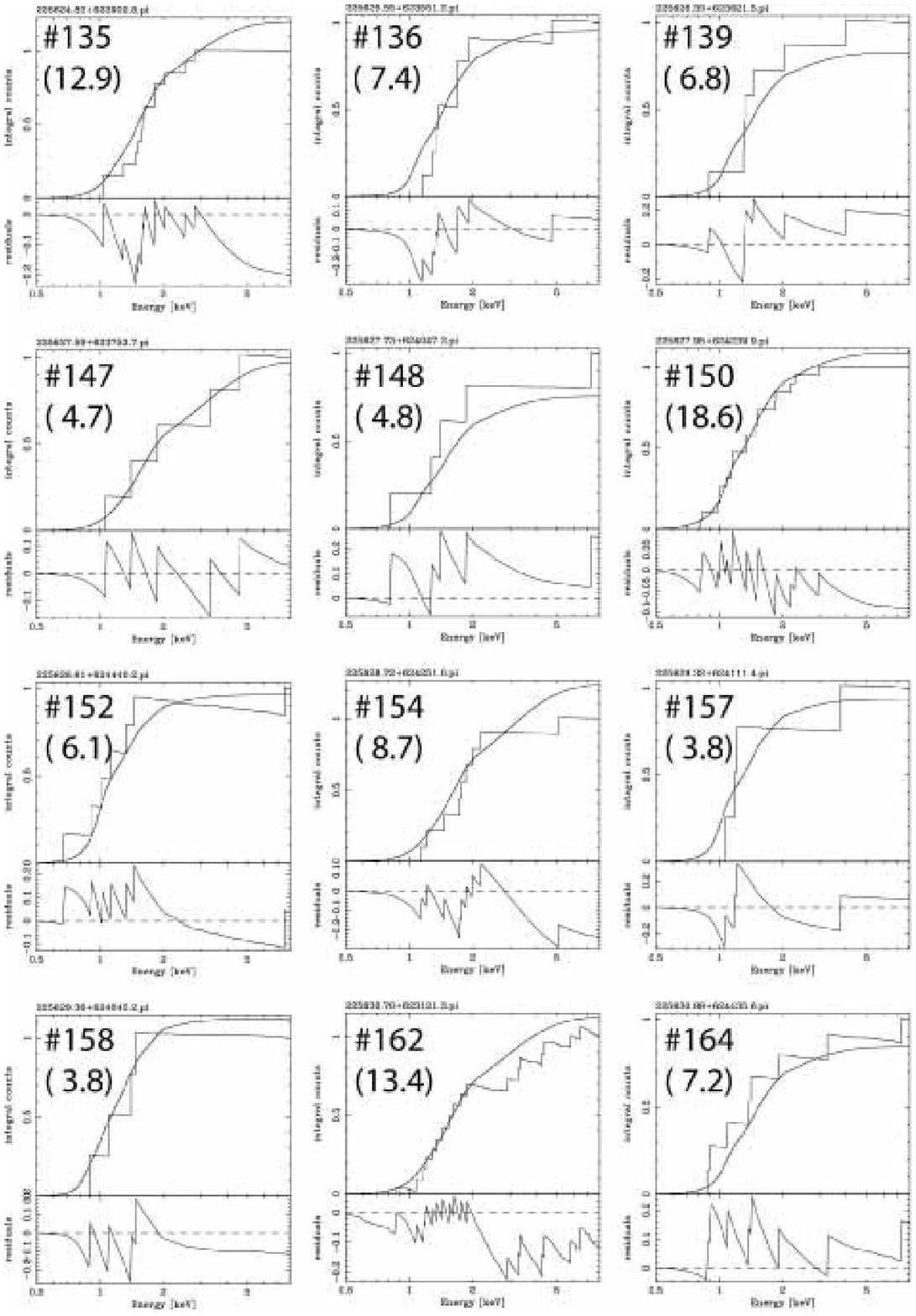}
\caption{Cumulative spectra for a sample of 12 consecutive sources
of 177 weak soft sources with available $N_{H,IR}$ information
(row 7 in Figure \ref{source_class_fig}). Upper panels show
observed counts (step function) and model spectrum (smooth curve);
lower panels show model residuals. $N_{H,Xray}$ was set equal to
$N_{H,IR}$. The source number and its net counts appear at the top
left.\label{weak_soft_spectra fig}}
\end{figure}

\begin{figure}
\centering
\includegraphics[angle=0.,width=5.5in]{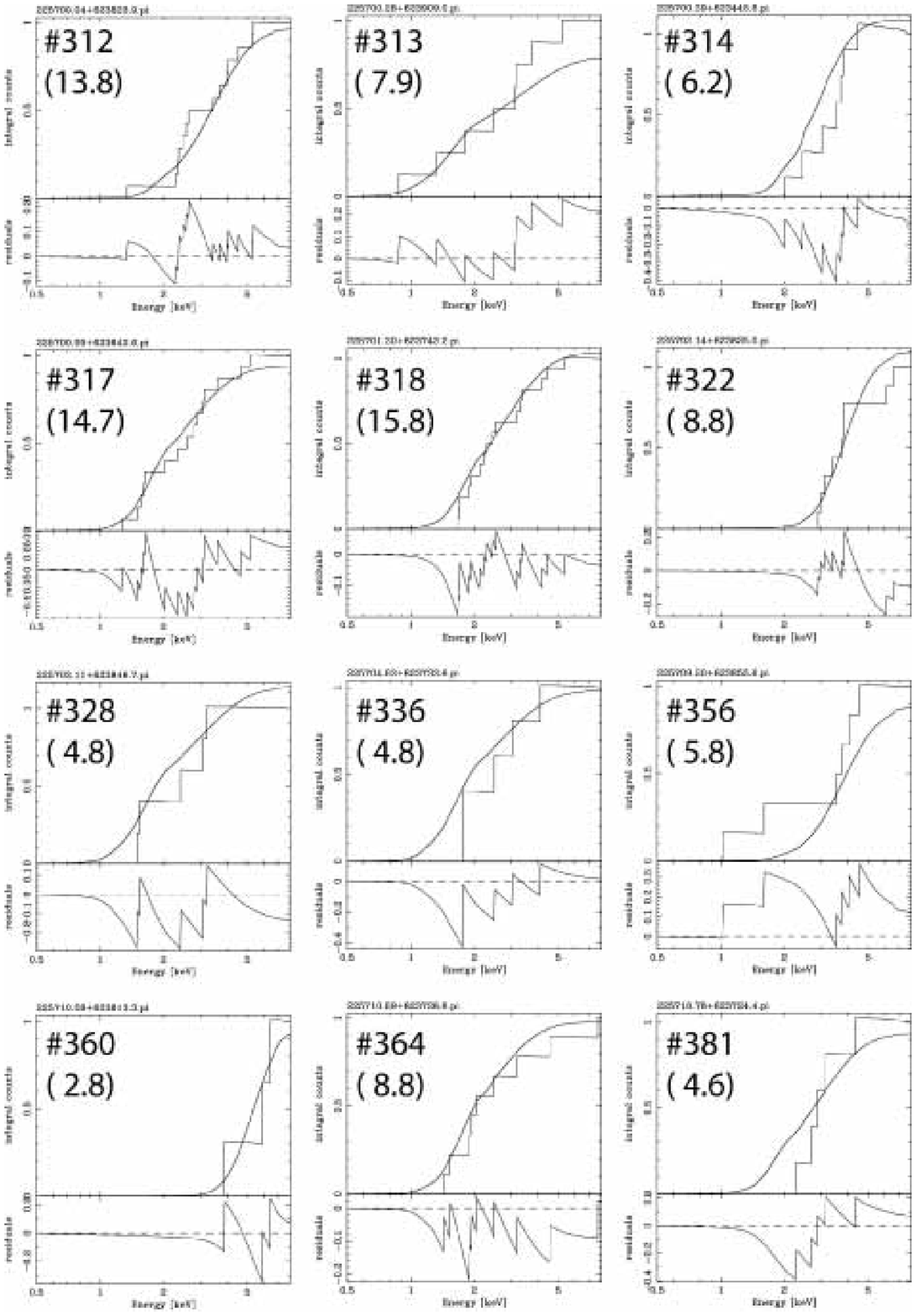}
\caption{Cumulative spectra for a sample of 12 consecutive sources
of 35 weak hard sources with available $N_{H,IR}$ information (row
7 in Figure \ref{source_class_fig}). Upper panels show observed
counts (step function) and model spectrum (smooth curve); lower
panels show model residuals. $N_{H,Xray}$ was set equal to
$N_{H,IR}$. The source number and its net counts appear at the top
left.\label{weak_hard_spectra fig}}
\end{figure}

To estimate plasma temperatures in these faint sources, we use the
median energy of the extracted photons.  The COUP study showed
that $MedE$ is a predictor of $\log N_H$ in strong PMS stars
superposed by a weaker dependence on $kT$ \citep{Feigelson05}. To
test this method, we simulated a grid of absorbed plasma models
(wabs*APEC) using the {\it fakeit} command in {\it XSPEC}.  We
then passed simulated spectra through {\it AE} to perform
photometric analysis including calculation of $MedE$. For the 212
faint sources with NIR photometry, we fixed column densities to
$N_{H,IR}$ and placed them on the simulated ($N_{H}$ versus
$MedE$) plane to roughly estimate their plasma energies. With the
$N_H$ and $kT$ parameters frozen we then performed $XSPEC$ fits to
derive the spectrum normalization and to calculate broad-band
X-ray fluxes. Examples of these spectral fits are shown in Figures
\ref{weak_soft_spectra fig} and \ref{weak_hard_spectra fig}. The
resulting spectral fits appear satisfactory, even for very weak
sources of as few as 3-5 net counts.

Finally, we consider the 45 weak sources (17 soft and 28 hard)
that lack sufficient 2MASS photometry to estimate $N_{H,IR}$. Most
of those sources are probably unrelated to the Cepheus region (see
\S \ref{unidentified_main_section}). Here we report one of the
possible spectral fits from a wide range of models in order to
integrate under the model and to estimate observed broad-band
fluxes. The spectral models for these 45 weak sources are used
here simply as spline fits with no astrophysical meaning, mainly
to provide future researchers with the ability to derive much more
reliable observed X-ray fluxes. Table \ref{tbl_spe} reports
thermal models for all 45 sources; Table \ref{tbl_pow24} reports
power law models for 21 of these sources (see last row in Figure
\ref{source_class_fig}) that are AGN candidates (see \S
\ref{hard_unidentified_section}).

\subsection{Flux uncertainties
\label{flux_uncertainties_section}}

Formal errors on X-ray luminosities for young stellar members of
the Cepheus region have been estimated through Monte Carlo
simulations. For each source: (1) we randomly draw a distribution
of $\log N_H$, comprised of 1000 points, with the mean equal to
that reported from spectral analysis and $\sigma \sim 0.3$,
obtained in \S \ref{fits_less_20_counts}; (2) we randomly draw a
count rate ($CR$) Poisson distribution of 1000 points with the
mean equal to the measured value; (3) using the MARX ray-trace
simulator, we simulate a very bright source with spectral source's
characteristics, $N_H$, $kT$, we slice the simulated source on
thousand of chunks, comprising the number of counts of the real
source, and derive the characteristic distribution of source's
median energy ($MedE$); (4) we pass the obtained distributions of
$\log N_H$, $CR$, and $MedE$ through the simulated database of
optically thin plasma model (see \S \ref{fits_less_20_counts}) to
report the standard deviations of X-ray fluxes in Table
\ref{tbl_spe}.

\begin{figure}
\centering
\includegraphics[angle=0.,width=5.5in]{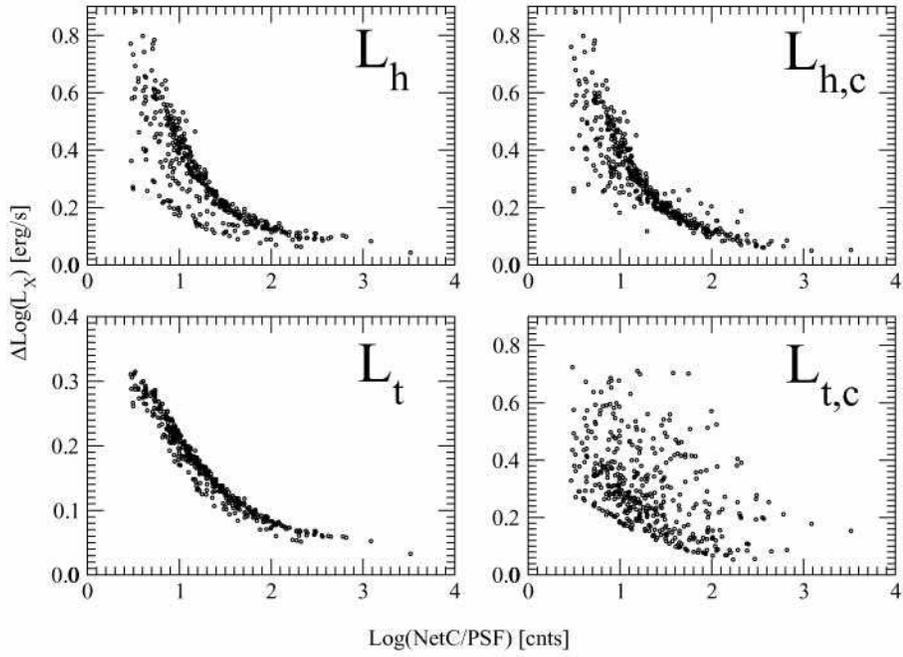}
\caption{Plots of formal errors in X-ray luminosities $\Delta \log
(L_X)$, derived from our Monte Carlo simulations (see \S
\ref{flux_uncertainties_section}), as a function of source
strength $\log (NetFull/PSF)$. The panels show results for four
measures of source luminosity: observed hard band luminosities
($L_h$; $2.0-8.0$ keV), hard band luminosities corrected for
absorption ($L_{h,c}$), observed total band luminosities ($L_t$,
$0.5-8.0$ keV), and total band luminosities corrected for
absorption ($L_{t,c}$).\label{dlx_vs_nc fig}}
\end{figure}

The simulations indicate that $1\, \sigma$ errors on $MedE$ are in
general better than $0.2$ keV for brighter ($NetFull> 20$ counts)
sources, but become as large as $\sim 1$ keV for weaker sources.
X-ray observed luminosities $L_t$ ($L_h$, $L_{h,c}$) are known
better than 0.2 (0.3,0.3) dex for brighter sources, but become as
much uncertain as 0.3 (0.6,0.6) dex for weaker sources (Figure
\ref{dlx_vs_nc fig}). The uncertainty on intrinsic total band
luminosity $L_{t,c}$ in addition to count rate is affected by
absorption, but in general for most of the sources it is better
than 0.4 dex.

The scientific reliability of these luminosities, however, is
usually lower than the formal statistical uncertainty for several
reasons.  First, all PMS stars exhibit strong temporal
variability. From Orion studies, we estimate that $\sim$10\% of
the stars will be caught in a flare state during the 8 hours of
Cepheus B exposure \citep{Wolk05}. Second, the $L_{t,c}$ will
systematically underestimate the true emission for obscured or
embedded populations, due to the absence of soft band spectral
components. Third, if the distance of 725 pc adopted in this paper
were revised to 800 pc, quoted in \citet{Pozzo03}, then the PMS
stellar X-ray luminosities should systematically be increased by
$\simeq 0.1$ dex.

\section{X-ray sources unrelated to the cloud \label{unidentified_main_section}}

\subsection{Unidentified X-ray sources and extragalactic contamination \label{hard_unidentified_section}}

\begin{figure}
\centering
\includegraphics[angle=0.,width=6.3in]{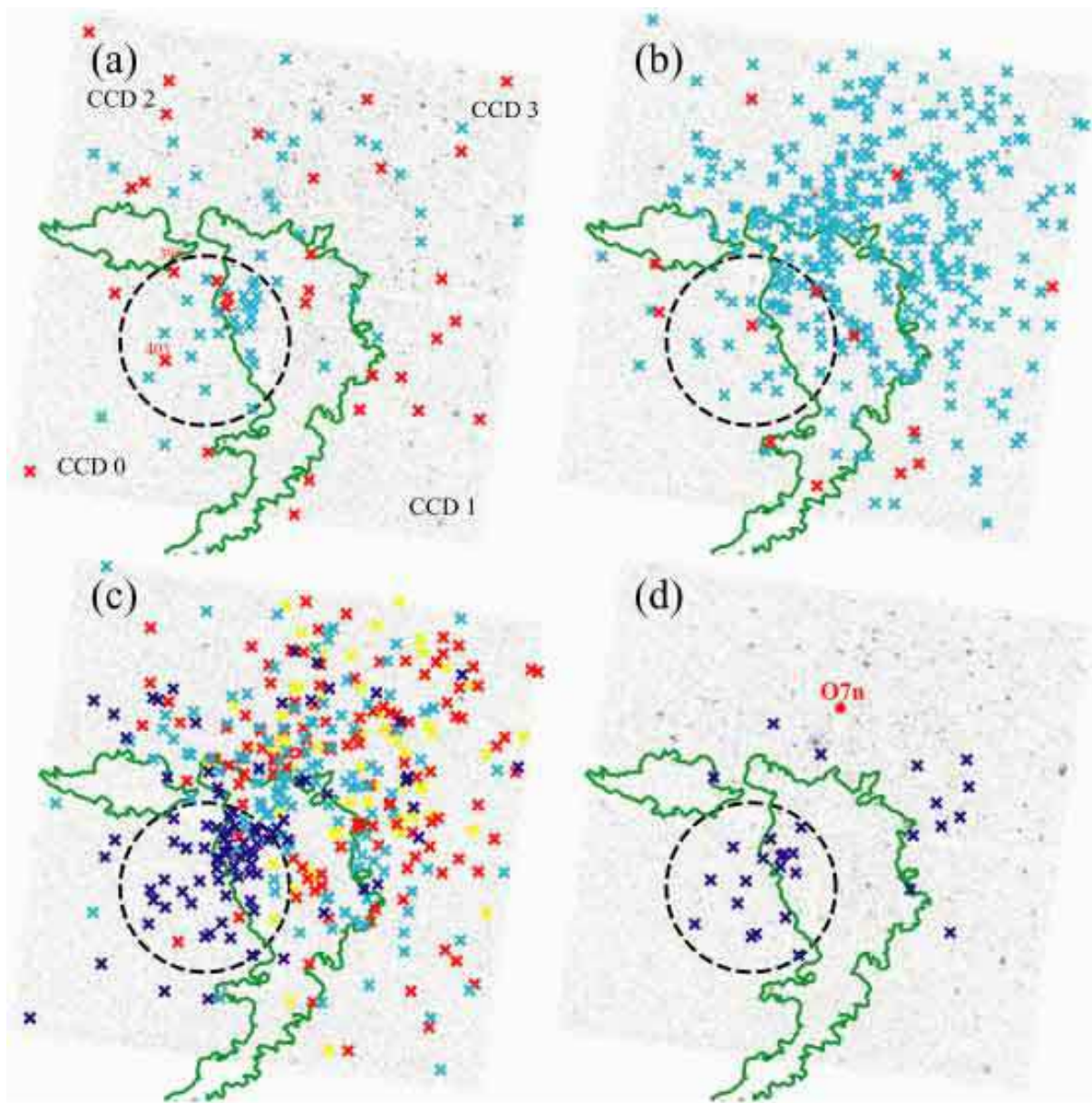}
\caption{Spatial distribution of different classes of X-ray
sources superposed on a grey-scale low-resolution image of the
$17\arcmin \times 17\arcmin$ {\it Chandra} field. In each panel,
the embedded cluster is outlined by a black dashed circle of
$3\arcmin$ radius. An outline of the HII region and ionization
front, seen in the optical DSS image is marked by the green
contour. (a) Distribution of all 83 hard ($MedE
> 2$ keV) X-ray sources including NIR-identified (cyan $\times$) and
un-identified (red $\times$) sources. The four ACIS-I CCD chips
are labelled; the molecular cloud lies mostly on CCD 0. (b)
Distribution of all 348 soft ($MedE < 2$ keV) X-ray sources
including probable members of the region (cyan $\times$) and
possible foreground stars (red $\times$). (c) $N_H$-stratified
distribution of all probable cloud members; i.e., excluding the 24
(red $\times$) probable AGNs in CCDs 1, 2, 3 from panel (a) and
the 13 (red $\times$) probable field stars from panel (b).  Color
code: yellow $\times$ have $\log N_H < 21.3$, red $\times$ have
$21.3 < \log N_H < 21.6$, cyan $\times$ have $21.6 < \log N_H <
21.9$, and blue $\times$ have $\log N_H > 21.9$. (d) Distribution
of all 29 $K$-band excess sources (blue $\times$).  HD 217086 O7n
star is marked by the red dot.\label{spat_distrib_fig}}
\end{figure}

Out of 83 ``hard" ($MedE > 2$ keV, Figure \ref{source_class_fig})
sources 30 have not been identified with 2MASS NIR counterparts
(this does not include \#\# 395, 403 from \S
\ref{uncataloged_2MASS}). Twenty four of these sources without
2MASS counterparts lie outside of the Cep B molecular cloud, which
covers the southeastern ACIS chip (CCD0) (Figure
\ref{spat_distrib_fig}(a)). We argue here these are most likely
extragalactic sources.  No more than 3 extragalactic sources are
likely present behind the dark cloud itself; most of the hard
sources in the cloud are concentrated around the PDF ionization
front (see Figure \ref{spat_distrib_fig}(a)) or in the new heavily
absorbed cluster found in the ACIS image (\S
\ref{new_cluster_section}). We therefore omit consideration of
extragalactic contamination on the southeastern ACIS chip.

We evaluate the expected contamination by extragalactic X-ray
sources using the methods of \citet{Getman05b}. Confining analysis
to the hard $(2-8)$ keV band, we construct Monte Carlo simulations
of the extragalactic population by placing artificial sources
randomly across the field.  Incident fluxes are drawn from the
X-ray background $\log N - \log S$ distribution described by
\citet{Moretti03} assuming a power law photon index distribution
consistent with flux-dependencies described by \citet{Brandt01}.
The spectrum of each simulated source was passed through a uniform
absorbing column density $\log N_{H} \sim 22.2$ cm$^{-2}$. This is
the estimated sum of two components: absorption of $\log N_{H}
\sim 21.7$ cm$^{-2}$ from Earth to the Cepheus region taken from
the median of soft X-ray sources away from the dark cloud and a
H{\sc I} column density of $\log N_{H} \sim 22.1$ cm$^{-2}$
through the entire Galactic disk in the direction of Cepheus
\citep{Dickey90}\footnote{This was obtained from NASA's HEASARC
tool located at
\url{http://heasarc.gsfc.nasa.gov/cgi-bin/Tools/w3nh/w3nh.pl}}.
After applying local background levels found in our ACIS image, we
then remove very weak extragalactic sources that would have fallen
below our source detection threshold.

\begin{figure}
\centering
\includegraphics[angle=0.,width=6.5in]{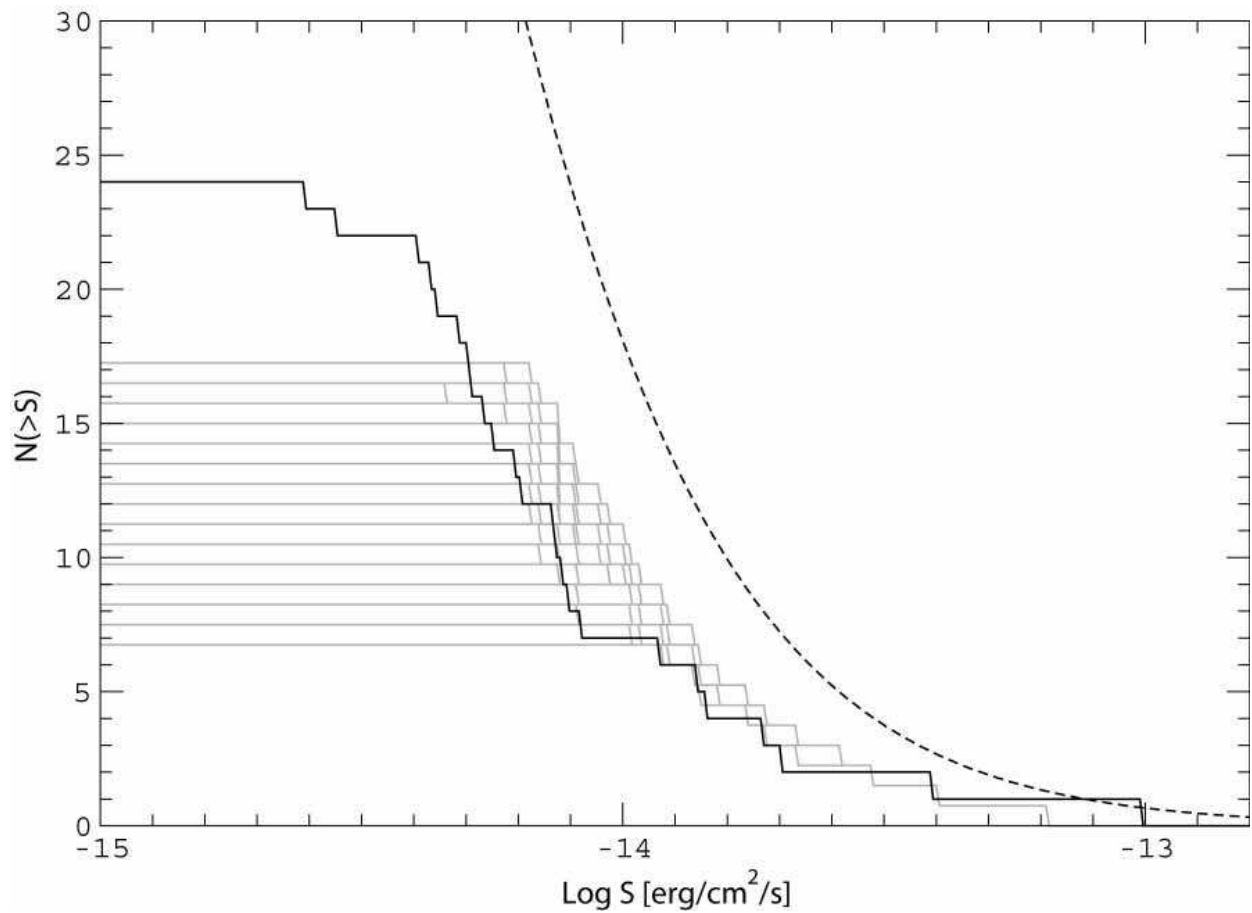}
\caption{Evaluation of extragalactic source contamination using
the X-ray source $\log N - \log S$ diagram. $N(>S)$ gives the
number of sources in the three CCDs outside of the molecular
cloud. $S$ is the source flux in the hard $2-8$ keV band. The
black solid line shows the observed distribution for the 24
heavily absorbed sources without 2MASS counterparts.  The dashed
line shows the expected distribution for an unobscured
extragalactic population \citep{Moretti03}. The grey lines show
examples of extragalactic source simulations, which include source
spectral variations, large scale Galactic absorption, absorption
local to Cepheus, and variations in the {\it Chandra} field X-ray
background. \label{n_logs_fig}}
\end{figure}

The grey lines in Figure \ref{n_logs_fig} show the predicted $N -
\log S$ distribution of extragalactic sources from several typical
simulations with different detection thresholds. Here N is the
total number of extragalactic sources expected to be detected and
$S = F_h$ is the observed source flux in the hard band. This
prediction is compared with the observed hard fluxes of 24 Cepheus
hard unidentified sources (black solid line in Figure
\ref{n_logs_fig}; values given in Table \ref{tbl_pow24}). The
simulations predict that a range of $7-17$ extragalactic sources
would have been detected in these three CCD chips by our data
processing, which is roughly half of the 24 observed unidentified
hard Cepheus sources.  The others may be physically associated
with the Cepheus region --- for example new embedded sources in
the molecular cloud, which may extend towards the southwestern
(part of CCD1) and northeastern (part of CCD2) parts of the ACIS-I
field (Figure \ref{spat_distrib_fig}(a)). Since the observed
source counts lie somewhat below the simulations for $S < 1 \times
10^{-14}$ erg s$^{-1}$ cm$^{-2}$, it is plausible that $3-5$ of
the hard sources with 2MASS counterparts are bright extragalactic
sources rather than PMS stars in the cloud.

As we know that roughly half of the 24 hard unidentified sources
are extragalactic contaminants, but we can not make individual
determinations, we flag all 24 sources as possible AGN
contaminants (flag `a' in Table \ref{tbl_spe}).  They are omitted
from further consideration of the Cepheus PMS population.

\subsection{Foreground stellar contamination
\label{foreground_possible_section}}

\begin{figure} \centering
\includegraphics[angle=0.,width=6.5in]{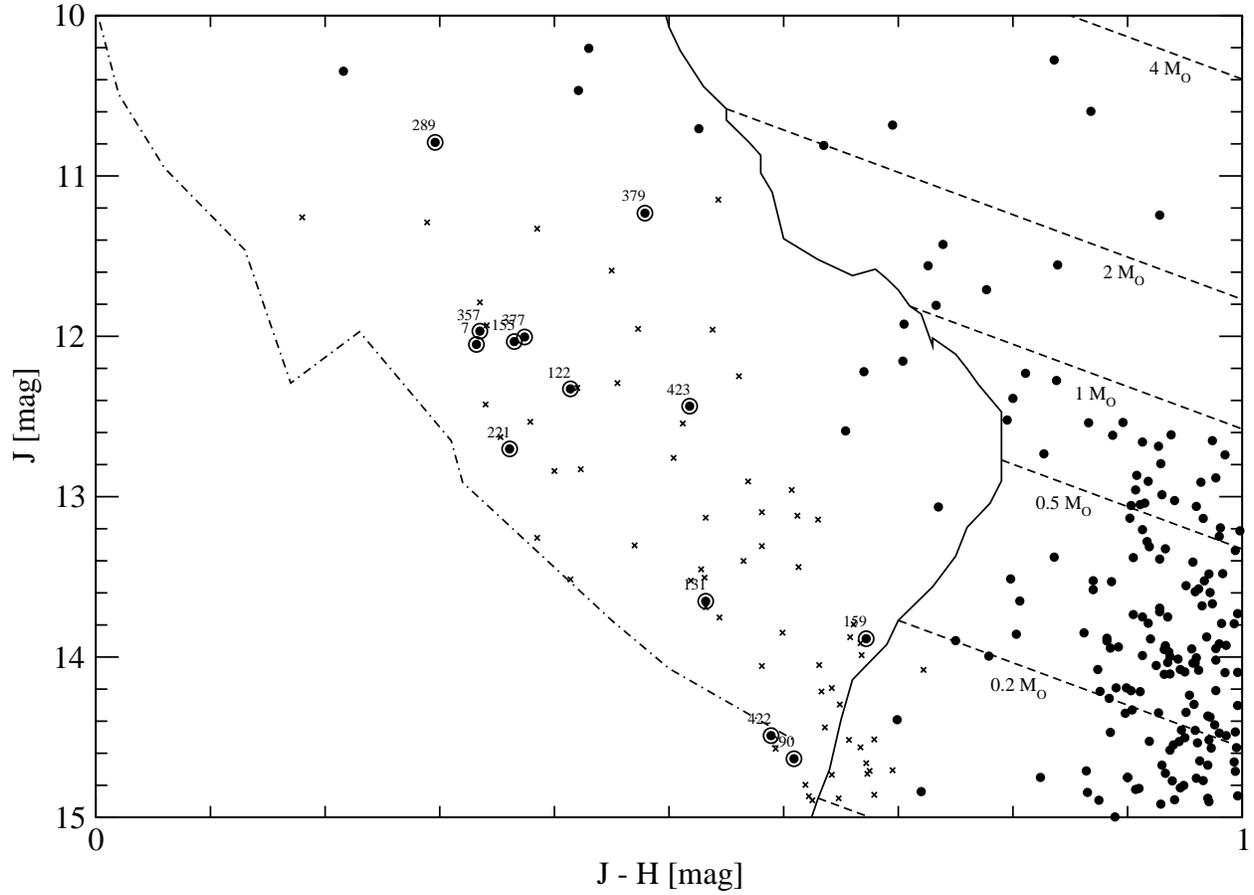}
\caption{Expanded view of the foreground star locus area on the
NIR color-magnitude diagram shown in Figure
\ref{cmd_besancon_fig}. {\it Chandra} sources are marked by black
dots. The $\times$ symbols show simulated star locations of
foreground stars based on the Besan\c{c}on Galactic star model.
Thirteen likely X-ray foreground stars are labelled and outlined
by circles. \label{fg_cmd_fig}}
\end{figure}

Figure \ref{fg_cmd_fig} shows an expanded view of a portion of the
NIR color-magnitude diagram in Figure \ref{cmd_besancon_fig}.
Thirteen lightly-absorbed {\it Chandra} stars lie between the 1
Myr isochrone and the ZAMS within the locus of Besan\c{c}on
simulated foreground stars and thus are unlikely to be Cep B cloud
members. These are also indicated schematically at the bottom left
of the X-ray source classification diagram shown in Figure
\ref{source_class_fig}. Although X-ray surveys of star forming
regions are very effective in preferentially detecting PMS stars
over MS stars, even a very low detection rate of older stars will
lead to interlopers in the X-ray image due to the huge number of
field stars.  Field stars detected in X-ray surveys will
preferentially be younger members of the disk population, as
stellar X-ray emission decays rapidly after $\sim 1$ Gyr
\citep{Feigelson04,Preibisch05b}.

Half of the field stars in the Besan\c{c}on simulation have ages
greater than 2 Gyr while half have ages ranging from 0.15 to 2
Gyr. Most reside at distances from 400 to 725 pc; 65\% are M
stars, 24\% are K stars, 9\% are G stars, and 2\% are more
massive. We convolved the Beson\c{c}on model populations with the
X-ray luminosity functions of stars in the solar neighborhood
measured from {\it ROSAT} surveys \citep{Schmitt95, Schmitt97,
Hunsch99}. Luminosities were adjusted to account for the different
{\it ROSAT} and {\it Chandra} spectral bands following the stellar
hardness-luminosity relation \citep{Gudel98}.  The resulting
predicted X-ray flux of the foreground Besan\c{c}on sample in the
$(0.5-8.0)$ keV band is $F_X \sim 2 \pm 1 \times 10^{-13}$ erg
s$^{-1}$ cm$^{-2}$.  Following the same procedure we used for
extragalactic sources (\S \ref{hard_unidentified_section}), we
applied our ACIS detection process to these simulated field stars.
The results suggest that $5-10$ field stars may be detected in the
ACIS-I field. A typical Monte Carlo run predicts 2 F stars, 3 G-K
stars, and 4 M stars.  Half are typically between 450 and 725 pc.
This prediction may rise to 10-15 stars if one includes an
age-dependence to the X-ray luminosities \citep{Preibisch05b} into
the simulations, since $\sim 1/4$ of the stars have ages $< 1$
Gyr.

The 13 {\it Chandra} sources highlighted in Figure
\ref{fg_cmd_fig} have NIR magnitudes and colors lying within the
locus of expected foreground stars.  The spatial distribution of
the 13 soft-spectrum stars is relatively uniform in the ACIS field
(see the red $\times$ symbols in Figure \ref{spat_distrib_fig}b)
and they have soft X-ray spectra expected from MS coronal emission
(sources with flag `f' for `foreground' in Table \ref{tbl_spe}).
We omit these X-ray sources from further consideration of the
Cepheus population.

\subsection{Galactic background contamination}
\label{galactic_contamination_section}

The Besan\c{c}on model is again convolved with X-ray luminosity
functions to simulate the number of stars behind the cloud that
may enter our X-ray sample.  We consider only plausible classes of
X-ray emitters (e.g., omitting MS A stars and subgiants) and adopt
a limiting magnitude $V < 22$.  The model then predicts $\sim
1700$ F dwarfs, $\sim 550$ G dwarfs, $\sim 1950$ K dwarfs, $\sim
25$ M dwarfs, and $\sim 632$ giants. We again use the dwarf XLFs
established for the solar neighborhood
\citep{Schmitt95,Schmitt97}. For giants we assume the lognormal
XLF of $\sim$ $<$28.9$>$ $\pm 0.9$, obtained from Table 2 of
\citet{Pizzolato02} for a sample of 120 post-main-sequence
late-type stars within 100 pc. Rarely, simulations result in up to
5(7) dwarfs (giants) detected in our $Chandra$ observation, while
typical runs suggest 0(3) dwarfs (giants). As we can not make
individual determinations for 3 possible background giants among
our PMS stars, we ignore Galactic background contamination in our
further consideration of the Cepheus PMS population. The
simulations yield the total typical observed X-ray flux of the
Galactic background population of up to $F_X \sim 2 \times
10^{-14}$ erg cm$^{-2}$ s$^{-1}$ that may be contributing to the
X-ray diffuse emission seen in the ACIS-I field.

\section{Discovery of two rich stellar clusters}\label{new_cluster_section}

Cepheus B, the hottest CO component of the Cepheus molecular
cloud, is located at the northwestern edge of the cloud near the
Cepheus OB3 association. The association is thought to be composed
of two subgroups of different ages, with the youngest lying closer
to the molecular cloud \citep{Sargent79}.  The interface between
the molecular cloud and the OB association is delineated by the
optically bright H{\sc II} region S 155 with ionization fronts
bounding the dust/molecular cloud \citep{Felli78}. \citet{Testi95}
detected four radio structures --- one extended source tracing an
ionization front and three compact sources inside the molecular
cloud --- accompanied by three very red near-infrared objects.
They suggested these are manifestations of a previously
unrecognized young embedded stellar cluster at the interface of
the H{\sc II} region and the molecular cloud.

\citet{Moreno-Corral93} claimed the discovery of more than 50 NIR
sources, with 60\% of them located close to or within the Cep B
molecular cloud. Unfortunately, it seems that positions for those
sources reported in their Table 5 are incorrect, as none of them
matches known 2MASS or $Chandra$ objects considering any possible
systematic shift. Thus we omit the results of this work from
further consideration.

\begin{figure}
\centering
\includegraphics[angle=0.,width=5.5in]{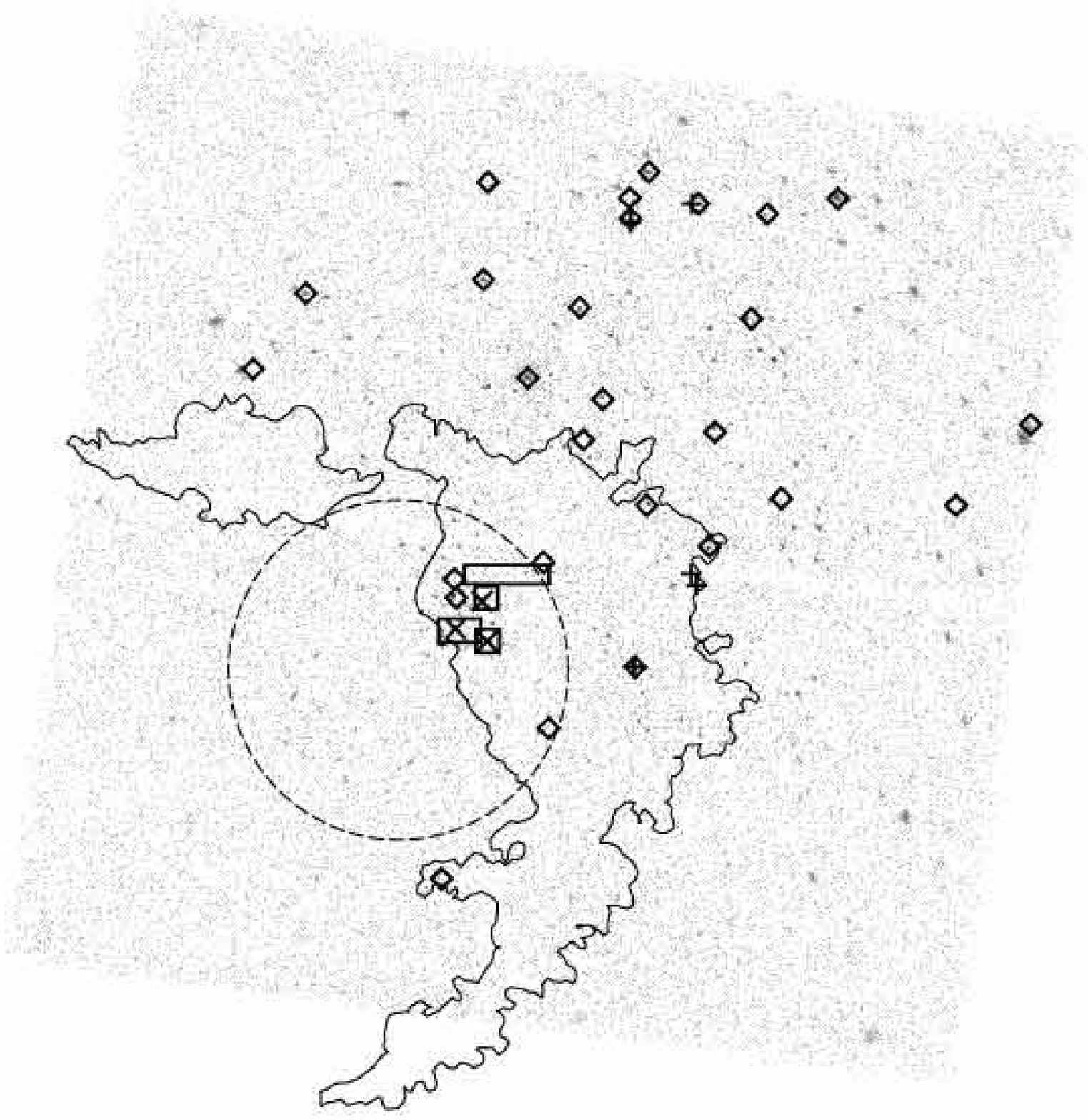}
\caption{Positions of previously reported young members of the
region, which fall within the ACIS-I field. Four VLA compact radio
sources ($\sq$) and 3 red NIR objects ($\times$) from
\citep{Testi95}, 27 $ROSAT$ sources from \citep{Naylor99}
($\diamond$), and 5 PMS with optical spectra  from \citep{Pozzo03}
($+$). Lines are the same as in Figure \ref{spat_distrib_fig}: the
dashed circle outlines the embedded cluster of X-ray stars, and
the solid contour outlines the HII region and ionization front
seen in optical DSS image.\label{prev_catalogs_fig}}
\end{figure}

\citet{Naylor99} found more than 50 $ROSAT$ X-ray sources in Cep
OB3b; these revealed a low-mass stellar population accompanying
the high-mass stars of the OB association. Optical CCD photometry
and spectroscopy identified 10 $ROSAT$ sources in their catalog,
of which 4 were confirmed as T Tauri and 1 was suggested as a
possible PMS star \citep{Pozzo03}. Six (five) more optical stars
were associated with T-Tauri (possible PMS stars) totalling 10 (6)
confirmed T-Tauri (possible PMS). The results of \citet{Testi95},
\citet{Naylor99} and \citet{Pozzo03} are summarized in Figure
\ref{prev_catalogs_fig} which shows the previously known PMS stars
lying in the ACIS-I field: 4 radio (boxes); 3 red NIR sources
($\times$); 27 out of 50 $ROSAT$ (marked as diamonds)\footnote{All
except one of these $ROSAT$ sources has one or multiple $Chandra$
counterparts. $ROSAT$ source \# 41 is missed, probably because it
lies on a gap between CCD chips in the ACIS array. One of the
candidate optical PMS stars similarly lies on a chip gap.}; and 5
out of 16 optical PMS stars ($+$). The four stars from
\citet{Pozzo03} with Chandra counterparts are 7a (Chandra \#193),
8 (\# 156), 12 (\# 197), and 106 (\# 160). \citet{Pozzo03}
recognized the incompleteness and low spatial resolution of the
$ROSAT$ data and anticipated our $Chandra$ X-ray observation which
would be capable of reaching a much larger PMS population.

The popular scenario for star birth in the S 155/Cepheus B region
\citep{Minchin92} is that the ultraviolet (UV) radiation from the
O7n star HD 217086, the brightest member of Cep OB3b cluster,
ionizes ambient material and induces a burst of star formation. We
investigate this model by searching for the next (third)
generation of young stars directly behind the ionization front.
Figure \ref{spat_distrib_fig}c shows the $N_H$-stratified
distribution of the $Chandra$ source population, excluding field
and extragalactic candidates, overlayed on an outline of the
optical H{\sc II} region.

A clear congregation of $\sim 64$ highly-absorbed YSOs is seen in
a $\sim 3\arcmin$ ($\sim 0.6$ pc) radius area, delineated by the
black circle, at the H{\sc II}-molecular cloud interface. The
remaining $\sim 321$ $Chandra$ stars with moderate or low
absorption are located outside the molecular cloud, although $\sim
9$ highly absorbed stars are dispersed throughout the field.
Figure \ref{spat_distrib_fig}d shows the distribution of $Chandra$
stars with a $K$-band excess inferred from 2MASS photometry
(Figure \ref{cc_cmd_fig}). These stars are likely to be younger
than the stars without bright inner dusty disks.  These are
preferentially associated with the cluster of highly-absorbed
$Chandra$ stars: 26\% of stars within the black circle have
$K$-band excesses compared to 4\% elsewhere in the field.  We
recognize that the UV radiation from the O7n star could evaporate
nearby circumstellar disks, but this likely affects only the outer
cooler disk regions.

Our findings support the triggered star formation scenario of
\citet{Minchin92} with an unambiguous discovery of the
third-generation embedded star cluster forming close to the
ionization front. This is based on the spatial concentration of
highly-absorbed $Chandra$ sources and their high proportion of
$K$-band excesses just behind the S 155 ionization front.   It
seems unlikely that this cluster was formed by spontaneous
self-gravitational collapse of molecular cloud material.  We
detect $\sim 64$ members of this embedded cluster (henceforth
denoted as ``Cep B embedded cluster''), which are marked ``e'' (=
``embedded'') in column 13 of Table \ref{tbl_spe}. The 321 members
of the older second generation cluster (hence denoted as ``Cep
OB3b cluster within ACIS-I field'' = ``Cep OB3b unobscured
cluster'') are marked ``l'' (=``lightly absorbed'').

\section{XLFs, KLFs and IMFs} \label{imf_xlf_section}

A major goal of studies of young stellar clusters is to quantify
and understand the distribution of masses arising from star
formation processes \citep{Corbelli05}.  This effort is subject to
a variety of difficulties, but as X-ray surveys present
$different$ problems than those arising at other wavebands they
can be valuable for IMF studies \citep{Feigelson05a}. In regions
like Cepheus B, NIR surveys are overwhelmed by unrelated field
stars (e.g. Figure \ref{cc_cmd_fig}).  A common procedure is to
subtract statistically a NIR source counts from a nearby control
field and convert the resulting $K$-band luminosity function (KLF)
into masses using theoretical stellar models to estimate a cluster
IMF \citep{Lada95}. We use a similar procedure on the Cepheus B
region below (\S \ref{KLF_sub.sec}). X-ray surveys, on the other
hand, suffer relatively little contamination; we found that no
more than 5\% of $Chandra$ sources in the Cepheus B field are
likely extragalactic contaminants and no more than $4$\% are
likely stellar contaminants (\S \ref{unidentified_main_section}).
Infrared surveys also are biased towards selection of young stars
with heavy protoplanetary disks, while X-ray surveys are mostly
unbiased with respect to disks \citep[although a small bias
against accretion is present,][]{Preibisch05}.  X-ray surveys
penetrate deeper into obscuring material than optical or NIR
surveys, but are less complete in detecting the lowest mass
objects.

In this section, we first examine the empirical stellar XLF of the
Cepheus population and compare it to the well-studied Orion Nebula
Cluster (ONC) XLF (\S \ref{XLF_sub.sec}).  We then use
independently derived masses for Cepheus members to estimate the
IMF (\S \ref{IMF_sub.sec}) and compare this with an IMF created
from the KLF (\S \ref{KLF_sub.sec}). The result is surprising: the
Cep OB3 cluster --- at least the part of it seen in the ACIS-I
field
--- appears to have a deficiency in higher mass stars compared to
the ONC and standard IMFs.

\subsection{Cep OB3b and ONC XLFs \label{XLF_sub.sec}}

\begin{figure}
\centering
\includegraphics[angle=0.,width=6.5in]{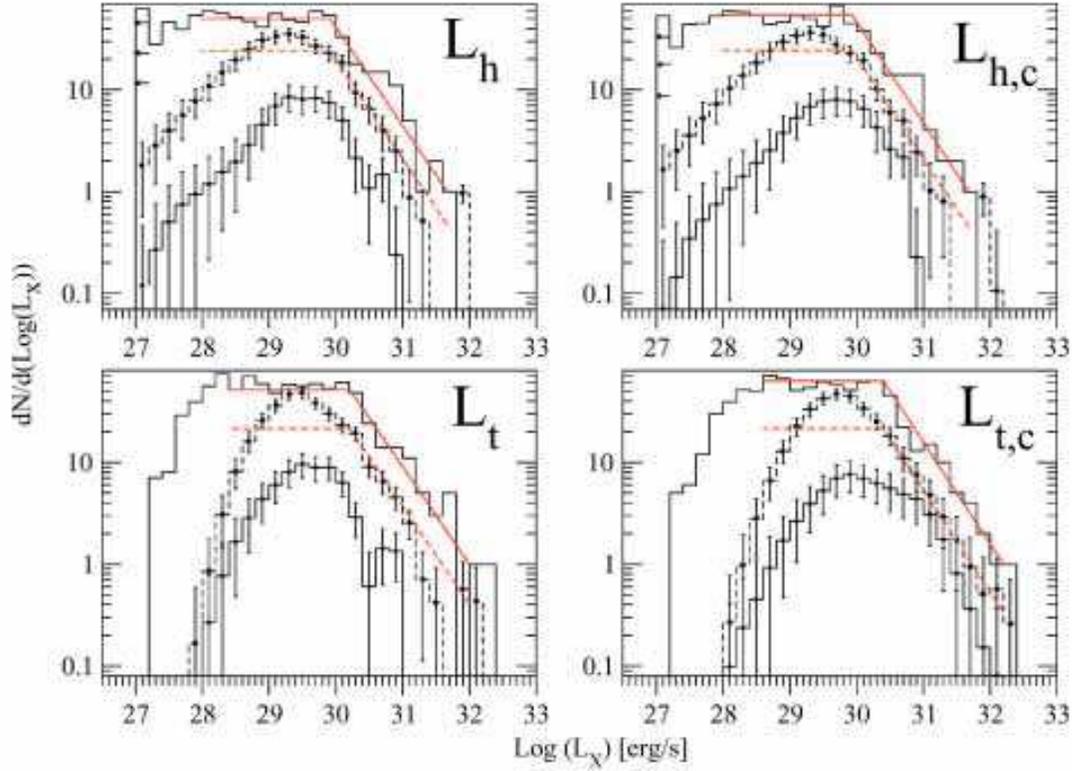}
\caption{Comparison between XLFs of the ONC and Cepheus
populations (\S \ref{XLF_sub.sec}). The upper solid histograms
show the COUP unobscured cool sample of 839 stars
\citep{Feigelson05}; the middle dashed histogram shows the Cep
OB3b unobscured sample of 321 stars; the lower solid histogram
shows the Cep B obscured sample of 64 stars. X-ray luminosities
are the same as in Figure \ref{dlx_vs_nc fig}. Error bars indicate
68\% confidence intervals ($1 \sigma$) from Monte Carlo simulated
distributions when X-ray luminosities are randomly drawn from
Gaussian distributions with source's measured $\log L_X$ and
source's estimated errors $\Delta \log L_X $. The broken red lines
are added to aid the eye; they are based on the shape of the ONC
XLFs and scaled downward to match the Cep OB3b unobscured
population.\label{xlf_fig}}
\end{figure}

Comparison of XLFs between poorly characterized young stellar
clusters like Cep OB3b and a well-characterized cluster like the
ONC can give insight into any missing population of stars.  This
was done by \citet{Grosso05} to estimate the total population of
the embedded Becklin-Neugebauer-Kleinman-Low OMC-1 population.
Figure \ref{xlf_fig} shows such a comparison among the unobscured
COUP population (upper solid line), the Cep OB3b population
(dashed line with error bars), and the Cep B embedded population
(bottom solid line with error bars). The ONC sample (= 839 COUP
lightly obscured cool stars) consists of COUP cloud members
excluding OB stars and sources with absorbing column density $\log
N_H > 22.0$ cm$^{-2}$ ($A_V \gtrsim 5$ mag) \citep{Feigelson05}.
XLF error bars for the Cep OB3b and Cep B embedded populations
show the range containing 68\% ($1\, \sigma$ equivalent) from
Monte Carlo simulated XLF distributions where individual X-ray
luminosities are randomly drawn from Gaussian distributions with
mean equal to the measured source's $\log L_X$ and variances based
on observed $\Delta \log L_X $ values. The error bars thus include
both the $\sqrt{N}$ counting error and the $L_X$-dependent
measurement error.  Regression lines to COUP's XLFs were added to
figures by eye (red solid lines) and scaled vertically to obtain a
predicted XLF for the Cep OB3b population (red dashed lines).

\begin{figure}
\centering
\includegraphics[angle=0.,width=6.5in]{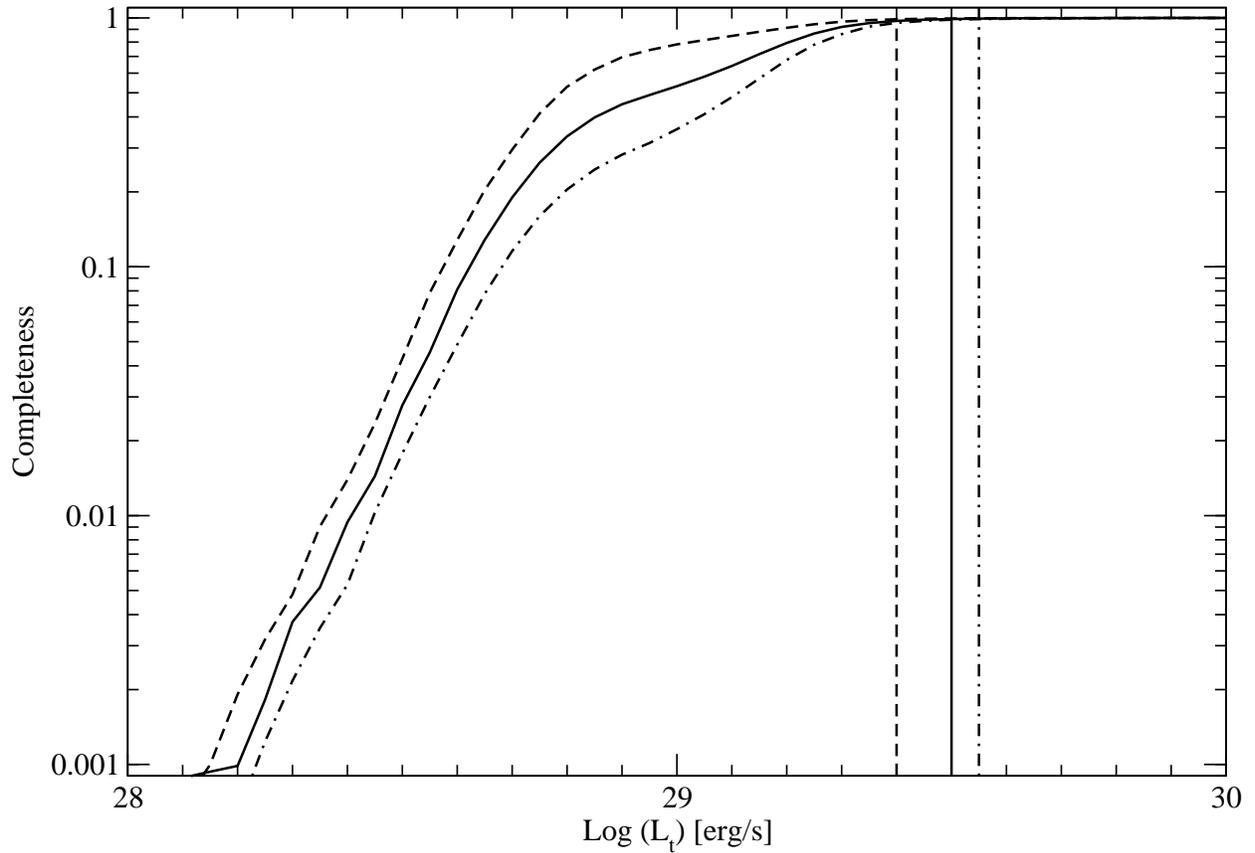}
\caption{Detection probability for a point source at a given
observed total-band luminosity, as derived from our Monte-Carlo
simulations of the Cep OB3 ACIS observation. Three completeness
curves correspond to three different detection thresholds: most
representative to the current observation (solid), more
conservative (dot-dashed), and less conservative (dashed). Three
vertical lines indicate corresponding completeness limits: 29.5
(solid), 29.55 (dot-dashed), and 29.4 (dashed).
\label{completeness_fig}}
\end{figure}

It is clear from Figure \ref{xlf_fig} that the Cep OB3b cluster
XLF shows a significant excess of stars around $\log L_{t,c} \sim
29.7$, $\log L_t \sim 29.5$, and $\log L_{h,c} \sim \log L_h \sim
29.3$ compared to the scaled ONC XLF. This excess is not related
to incompleteness of the current observation. Figure
\ref{completeness_fig}\footnote{Figure \ref{completeness_fig} is
derived from Monte-Carlo simulations performed by randomly drawing
sources of a given luminosity scattered throughout the ACIS image
and applying local background levels found in the image, similar
to the AGN simulations in \S \ref{hard_unidentified_section}.
Observed total band luminosity $L_t$ was chosen as it is more
readily converted to the observed source count rate assuming a
typical distribution of source median energy, and the source
significance then is easily calculated. Source significance
thresholds are then evaluated to mimic the behavior of our source
detection procedure. \label{completeness_note_label}} shows that
the completeness limit for the observed total band luminosity is
around $\log L_t = 29.5$ erg s$^{-1}$, and is still above 50\% at
$\log L_t = 29.0$ erg s$^{-1}$. This is far below the range $\log
L_t \ga 30.2$ erg s$^{-1}$ over which the ONC and Cep OB3b XLFs
were scaled.  We are thus confident that the relative amplitudes
of the XLFs in Figure \ref{xlf_fig} are correct. In fact, the
excess of Cep OB3b X-ray stars over the ONC XLF becomes even
stronger if the completeness curve from Figure
\ref{completeness_fig} is applied to the incomplete Cep OB3b
$L_t$-XLF of Figure \ref{xlf_fig}.

Table \ref{tbl_xlf_detected} summarizes total X-ray luminosities
for all X-ray detected stars in the Cep OB3b unobscured and Cep B
embedded populations within the ACIS-I field. It thus appears that
the Cep OB3b population within the ACIS-I field has an XLF with a
different shape than seen in the ONC.  Instead of a two-powerlaw
model with break around $L_{t,c} \sim 2.5 \times 10^{30}$ erg
s$^{-1}$, the Cep OB3b XLF with the sensitivity limit of $\log
L_{t,c,lim} \sim 10^{29}$ erg s$^{-1}$ can be roughly modelled as
a lognormal with mean around $<\log L_{t,c}> \sim 29.8$ erg
s$^{-1}$ and standard deviation of $\sigma(\log L_{t,c}) \sim
0.6$. The mean and standard deviation of cluster XLFs are perhaps
observation biased - for example for more sensitive observations
of NGC 1333 and IC 348 with $\log L_{t,c,lim} \sim 10^{28}$ erg
s$^{-1}$ their XLF shapes are wider with the standard deviation of
$\sigma(\log L_{t,c}) \sim 0.9$, peaking at lower luminosities of
$<\log L_{t,c}> \sim 29.5$ erg s$^{-1}$ \citep{Feigelson05a}.

\subsection{Cep OB3b and ONC IMFs \label{IMF_sub.sec}}

To compare the Cepheus IMF with that of the ONC, we need to adopt
compatible procedures for estimating masses. To be consistent with
the current work and to derive masses for the whole sample, we
re-derive\footnote{Table 9 in \citet{Getman05} provides masses for
half of the COUP sample based on optical spectroscopy.} COUP
stellar masses here using NIR photometry (Table 10 in
\citet{Getman05}). Similar to Cepheus masses, ONC mass estimates
are based on star locations in the $J$ vs. $J-H$ color-magnitude
diagram and theoretical stellar model tracks derived by
\citet{Baraffe98} (for $0.02\, M_{\odot} \leqslant M \leqslant
1.4\, M_{\odot}$) and \citet{Siess00} (for $1.4\, M_{\odot}
\leqslant M \leqslant 7.0\, M_{\odot}$).  We are aware that these
mass estimates are subject to significant uncertainties and may be
incompatible with masses obtained by other methods such as optical
spectroscopy.  However, here we are not so much interested in
exact masses for individual objects as in the general trends of
mass distributions and consistent treatment for comparison between
clusters.

\begin{figure}
\centering
\includegraphics[angle=0.,width=5.5in]{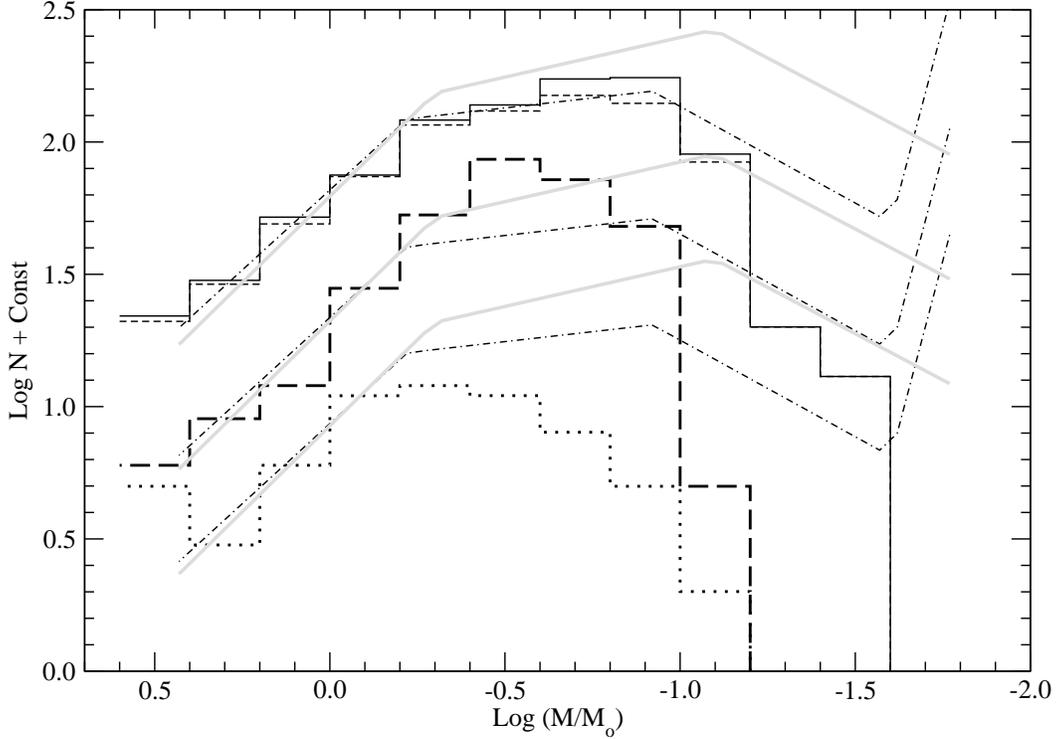}
\caption{Comparison between IMFs of the Orion Nebula Cluster and
the Cepheus X-ray sources where stellar masses for both clusters
are derived from NIR photometry (\S \ref{IMF_sub.sec}). The upper
two histograms show the ONC derived from the X-ray COUP sample:
the dashed is the 839-star COUP unobscured, and the solid shows
the addition of 82 $K_s$-bright members un-detected in X-rays. The
middle dashed histogram is for the 321-star X-ray unobscured Cep
OB3b population obtained here. The lower dotted histogram is for
the 64-star obscured Cep B population. IMF models are shown as
broken lines: dashed-dotted lines show three scaled versions of
the ONC IMF \citep{Muench02}, and grey lines show three scaled
versions of Galactic field IMF \citep{Kroupa02}.\label{imf_fig}}
\end{figure}

The IMF for the COUP 839 sample is shown as an upper dashed
histogram in Figure \ref{imf_fig}. The COUP sample can be extended
to account for incompleteness from instrumental causes: low local
ACIS exposure time, high local background due to readout streaks
from bright ONC sources, and contamination from the point spread
function wings of close bright X-ray sources. There are 82 such
ONC stars with $A_V \lesssim 5$ mag and $K_s < 13$ mag,
corresponding to $M > 0.1\, M_{\odot}$ in Orion \citep{Getman05b}.
The resulting IMF of the 921 ($839+82$) extended COUP ONC sample
is shown as an upper solid histogram on Figure \ref{imf_fig}.
Comparison with the ONC IMF fit based on KLFs obtained by
\citet{Muench02} (dashed-dotted line) indicates that the sample is
complete down to $0.1\, M_{\odot}$ and about 50\% complete (120
versus $\sim 240$ stars) within the low-mass stellar-sub-stellar
range of $(0.03-0.1)\, M_{\odot}$.  This incompleteness at low
masses is an obvious consequence of an interaction between the
observational flux limit and the X-ray/mass correlation
\citep{Preibisch05}.

\begin{figure}
\centering
\includegraphics[angle=0.,width=5.5in]{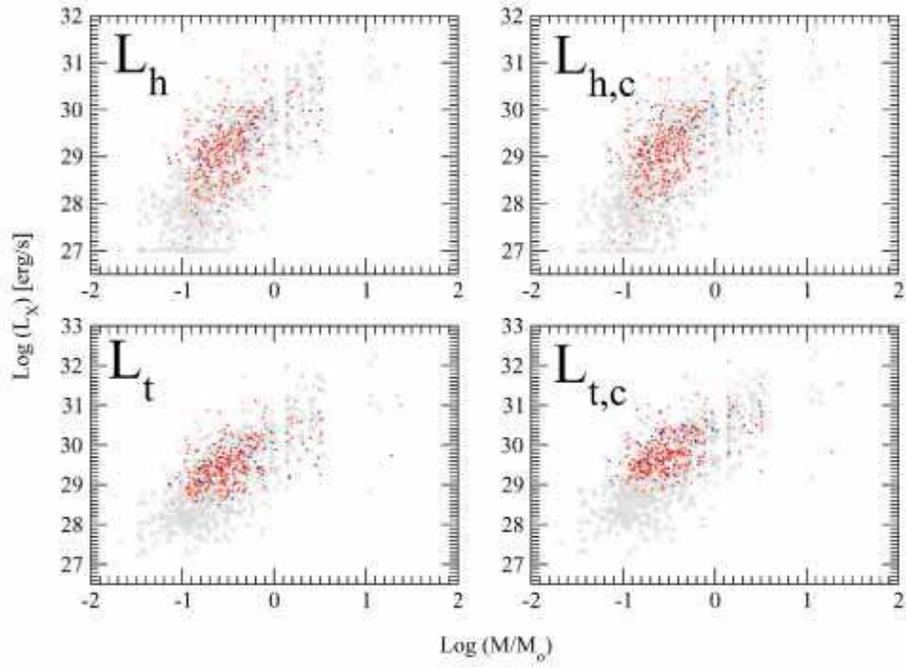}
\caption{X-ray luminosity versus stellar mass.  The grey dots
denote the COUP unobscured cool star sample of 839 stars. To be
consistent with the current work, the plotted stellar masses for
COUP sample have been determined from NIR color-magnitude diagram
similar to stellar masses derived for Cepheus stars. Red and blue
dots represent the Cepheus unabsorbed and embedded populations,
respectively. X-ray luminosities are the same as in Figure
\ref{dlx_vs_nc fig}.\label{lx_vs_mass fig}}
\end{figure}

We now turn to the Cep OB3b IMF derived in a similar manner.  From
321 (64) PMS stars of the $Chandra$ Cepheus unobscured (obscured)
population, 310 (55) have masses estimated from NIR photometry
(Table \ref{tbl_nir}). Figure \ref{lx_vs_mass fig} confirms that
the PMS stars in Cepheus (red - unobscured, blue - obscured)
follow $L_X$ versus mass correlations very similar to that seen in
COUP stars (grey dots).  Even observed luminosity $L_{t}$, prone
to changes as absorption varies, has a similar correlation with
mass, as the median column densities for COUP and Cepheus
unobscured populations are close to each other. Constant and slope
parameters derived from the EM linear regression fit for the COUP
unabsorbed sample are given for all four measures of X-ray
luminosities in Table \ref{tbl_em_regression}. The regression
parameters are slightly different from those derived in
\citet{Preibisch05} for the ONC lightly absorbed optical sample,
but this can arise from the higher sensitivity of the COUP 839
sample than that of the optical sample, and from different methods
used to estimate masses \citep{Preibisch05}. For 11(9) stars of
the Cepheus unobscured (obscured) populations with unknown masses,
we can use their measured X-ray luminosities with the regression
fit parameters from Table \ref{tbl_em_regression} to get rough
estimates on their masses and to add them to the appropriate bins
of the IMF histograms. This is done in the lower histograms of
Figure \ref{imf_fig}.

Though the histogram for the Cep OB3b population within the ACIS-I
field (thick dashed line) is rough and clearly incomplete below
$\sim 0.2$ M$_\odot$, we see a good agreement with the ONC IMF
from 0.5 to 3 M$_\odot$ but an excess of $\sim 20$ or more sources
(depending on the choice of scaling the ONC-Muench IMF) in the
$\log (M/M_{\odot}) \sim -0.5$ (0.3 M$_\odot$) bin. Expressed in
another way, the Cep OB3b IMF in contrast to the ONC does not
exhibit the turnover from the Salpeter IMF around $0.5-0.6$
M$_\odot$.

\subsection{Cep OB3b KLFs \label{KLF_sub.sec}}

We can estimate the total population of the Cepheus stellar
clusters by extrapolating the IMFs below the incompleteness limit
produced by the $Chandra$ flux limit.

We obtain a coarse statistical sample of 2MASS PMS members of the
Cep OB3b population in the ACIS-I field by subtracting a local
background of foreground and background NIR objects from a control
field of the 2MASS catalog. An IMF complete down to $M = 0.08
M_{\odot}$ at a distance of 725 pc corresponds to $J \sim 16$ mag
which is roughly the 2MASS completeness limit.  Figure
\ref{spat_distrib_fig}(c) suggests that $\sim 90\%$ of the Cep
OB3b population are clustered within an area of $\sim 127$
arcmin$^{2}$ lying within the ACIS-I field. This area roughly
matches the ``pentagon'' shown in Figure \ref{extended_smooth_fig}
where X-ray extended emission is seen (\S
\ref{diffuse_emission_section}). We extracted 2MASS sources from
30 control fields with 127 arcmin$^2$ area lying in a $3^{\circ}
\times 3^{\circ}$ neighborhood of the Cep B cloud as shown in
Figure \ref{2mass_bg_fig}. It is not trivial to establish that the
control fields are free from PMS stars, as the entire region is
occupied by the members of the sparse stellar OB3 association and
we trace a gradient in NIR source density as one progresses away
from the cluster core.  Towards the northwest extracted 2MASS
source density clearly reflects the evolution of PMS source
distribution in the Cep OB3 association --- the number of sources
systematically decreases down to 830 objects (at about 45$\arcmin$
($\sim 9.5$ pc) away from the ACIS-I field), but then
systematically rises again. The area with the least number of
objects, 830 in a 127 arcmin$^2$ circle centered on ($\alpha,
\delta$) = (342.92539,+63.26147), perhaps lies at the boundary
between different sub-groups of the Cep OB3 association ("N-W"
background area); see Figure 1 in \citet{Sargent77}. A second
similar area is found toward the west, 814 2MASS sources at
(341.68270,+62.27262) as shown in Figure \ref{2mass_bg_fig} ("W"
background area). Figure \ref{jh_compare_fig} compares cumulative
distributions of $J-H$ color excesses of 2MASS sources seen
towards the two control fields (dashed lines) with that seen
towards the molecular cloud (dotted line). The large population of
$(J-H)<1$ sources in the control fields supports the idea that
they are relatively unobscured areas unaffected by the Cepheus
molecular cloud.

To obtain a coarse IMF trend of the 2MASS unobscured sample of PMS
stars within the ACIS-I field we use the correlation $K_s = 10.86
- 3.32 \times \log(M)$ based on the theoretical synthesized 1 Myr
PMS model, Siess $+$ Baraffe, employed here for $<A_V> \sim 2.5$
mag ($A_K \sim 0.28$ mag). This permits conversion of the KLF to
stellar mass and production of two mass histograms (black solid
lines in Figure \ref{imf2_fig}a) after subtraction of the two
control fields with accounting for 10\% of the unobscured Cep OB3b
members dispersed outside of the ``pentagon'' area.

The result is that our $Chandra$ unobscured sample of 321 stars
(thick dashed in Figure \ref{imf2_fig}a) appears complete down to
$\sim 0.4$ M$_\odot$, and that $400-450$ stars are missing from
the $Chandra$ sample between 0.08 and 0.4 M$_\odot$. Figure
\ref{imf2_fig}a also compares this estimated IMF with that
obtained for the ONC by \citet{Muench02}.

\subsection{Discussion}

Comparison of the IMF based on the 321 X-ray detected PMS stars
with the IMF based on the background-subtracted KLF suggests that
the $Chandra$ image is complete down to $\simeq 0.4$ M$_\odot$
(Figure \ref{imf2_fig}a). The difference between the infrared and
X-ray source counts implies that $400-450$ cluster stars are
missing from the $Chandra$ image, giving a total unobscured
cluster population of $\sim 720-770$ stars in the $\sim 130-170$
arcmin$^2$ area. We estimate the total X-ray luminosity of these
undetected stars by multiplying the number of undetected stars by
a typical X-ray luminosity for the appropriate mass bin using
regression formulae from Table \ref{tbl_em_regression}. Table
\ref{xlf_undetected_tbl} summarizes results of this exercise and
shows that a total of $\log L_t \sim 31.3$ and $\log L_{t,c} \sim
31.5$ ergs s$^{-1}$ is expected from the undetected population.

Assuming that the IMF for the embedded cluster also follows the
mass distribution of the unabsorbed cluster, the embedded cluster
of 64 X-ray-detected members is complete down to $M = 0.8$
M$_{\odot}$.  We then estimate that $300-340$ undetected obscured
stars down to $M \sim 0.08$ M$_{\odot}$ are hidden in the cluster,
giving a total of $\sim 360-400$ obscured stars.

We find that both the XLF and IMF comparisons indicate that the
Cep OB3b population within the ACIS-I field exhibits, compared to
the ONC, more stars around $M \simeq 0.3$ M$_{\odot}$ relative to
$0.5<M<3$ M$_\odot$ stars. This difference cannot be attributed to
incompleteness in the COUP observation, as the addition of 82
stars missing from COUP does not significantly change the shape of
COUP's IMF and hence XLF, and will not remove the discrepancy. It
also cannot be attributed to incompleteness in the Cep OB3b
observation (Figure \ref{completeness_fig}); inclusion of missing
stars below $\log L_t = 29.0$ erg s$^{-1}$ would only increase the
inferred number of lower mass cluster members.  Similarly, the
inclusion of binaries unresolved in Cep OB3b but resolved in the
ONC \citep{Getman05b} would enhance the higher, not the lower,
ends of the Cep OB3b XLF and IMF.

It is possible that our assumption of similar ($\sim 1$ Myr) ages
for all PMS stars of the Cep OB3b population within the ACIS-I
field is incorrect. Recall that \citet{Pozzo03} found the age of
$\sim 1$ Myr for four out of six and $\ga 3$ Myr for another two
optical PMS candidates in our ACIS-I field. For the larger Cep
OB3b association outside of the ACIS-I field, the observed age
spread on the $V$ vs. $(V-I)$ color-magnitude diagram becomes even
more dramatic ranging from $\la 1$ Myr to $10$ Myr, and is too
broad to be accounted for by binarity, photometric uncertainty and
variability \citep{Burningham05}. Subgroups of different ages may
exist, each with different XLF and IMF characteristics. The XLFs
are relatively insensitive to these age effects; between 1 and 10
Myr, only a slight decay (roughly 0.3 in $\log L_t$) is found in
the COUP population \citep{Preibisch05b}.  An older assumed age
will have a bigger effect on the KLF.  Figure \ref{imf2_fig}b
summarizes a KLF analysis similar to that in \S \ref{KLF_sub.sec}
but assuming a cluster age of 3 Myr. In this case, X-ray detected
stars are more massive with the completeness limit shifted from
$\sim 0.4$ M$_\odot$ (for 1 Myr) down to $\sim 0.8$ M$_\odot$ (for
3 Myr). The excess of lower mass stars over the ONC IMF then
mostly disappears.  However, a younger age for Cep OB3b has two
advantages: the unusual inferred IMF agrees with the unusual
observed XLF (\S \ref{XLF_sub.sec}), which is insensitive to age
assumptions; and the shape of the background-subtracted KLF more
closely agrees with the IMF of the X-ray detected population for 1
Myr compared to 3 Myr (Figure \ref{imf2_fig}).  An argument in
favor of the older age can be seen in Figure \ref{imf2_fig}b: the
background-subtracted KLF more closely matches the Galactic field
IMF (grey line) than in the case of 1 Myr.

We thus emerge with a clear determination that the XLFs of Cep
OB3b and the ONC differ, but without a clear interpretation of
this effect.  It may be due to a different IMF with more
lower-mass stars in Cep OB3b, or to differences in age and history
of star formation in the two clusters. Followup optical photometry
and spectroscopy of 321 Cep OB3b X-ray sources would assist in
unravelling these possibilities, as individual (rather than
statistical) masses and ages can be directly estimated on the
Hertzsprung-Russell diagram.

If we assume that the 1 Myr ages are correct, then the inferred
IMF difference may be related to differences between the physical
processes of star formation in the Cep B and Orion molecular
clouds. The average Jeans masses may be lower in Cep B than in
Orion due to different densities or pressures. However if the
Jeans mass is universal around $M_J \sim 0.1$ M$_{\odot}$
\citep{Elmegreen99}, then the IMF difference could arise the
prevalence of different star formation mechanisms in the two
clouds \citep{Elmegreen04}. In this multi-component scenario,
lower-mass stars would have formed similarly in Cep B and Orion
from pre-stellar condensations of $M_J$ mass, but higher-mass
stars would have grown from coalescence of pre-stellar
condensations and gas accretion more efficiently in the more
crowded Orion environment.

\section{The diffuse X-ray emission}\label{diffuse_emission_section}

\begin{figure}
\centering
\includegraphics[angle=0.,width=4.5in]{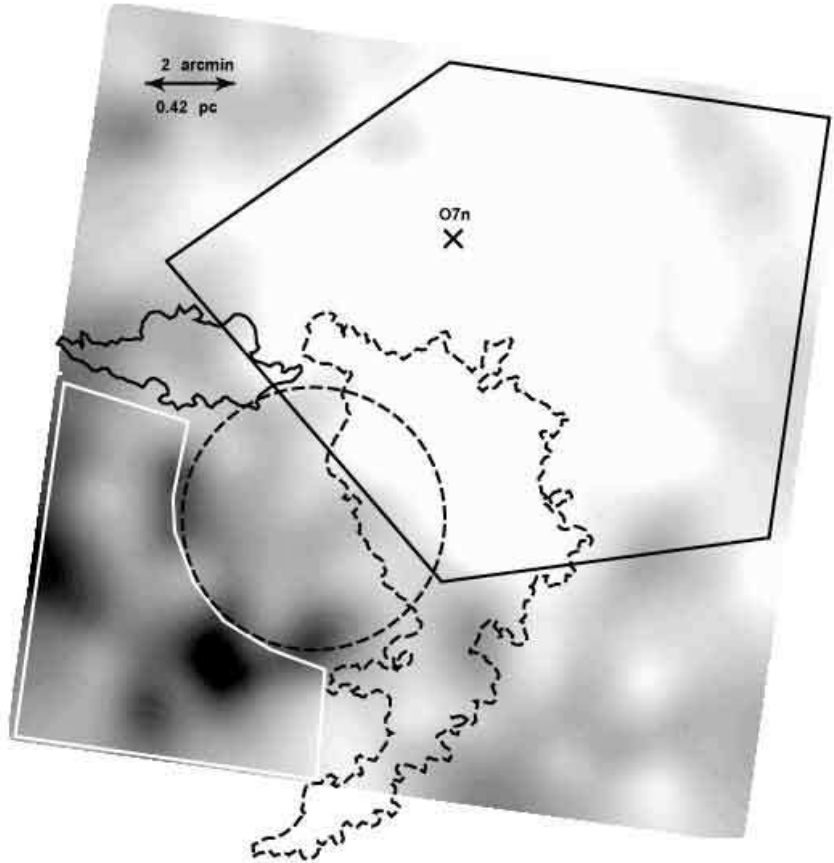}
\caption{The soft-band ($0.5-2$ keV) $Chandra$ image after removal
of point sources with holes smoothed over using an adaptive
smoothing algorithm. The pentagon indicates the extraction region
for the diffuse X-ray emission, and the white line outlines the
area used for background subtraction in spectral analysis. Dashed
lines show the location of the HII region with ionization front
and the Cepheus obscured population, and the $\times$ marks the
location of the HD 217086 O7n star.\label{extended_smooth_fig}}
\end{figure}

\begin{figure}
\centering
\includegraphics[angle=0.,width=6.5in]{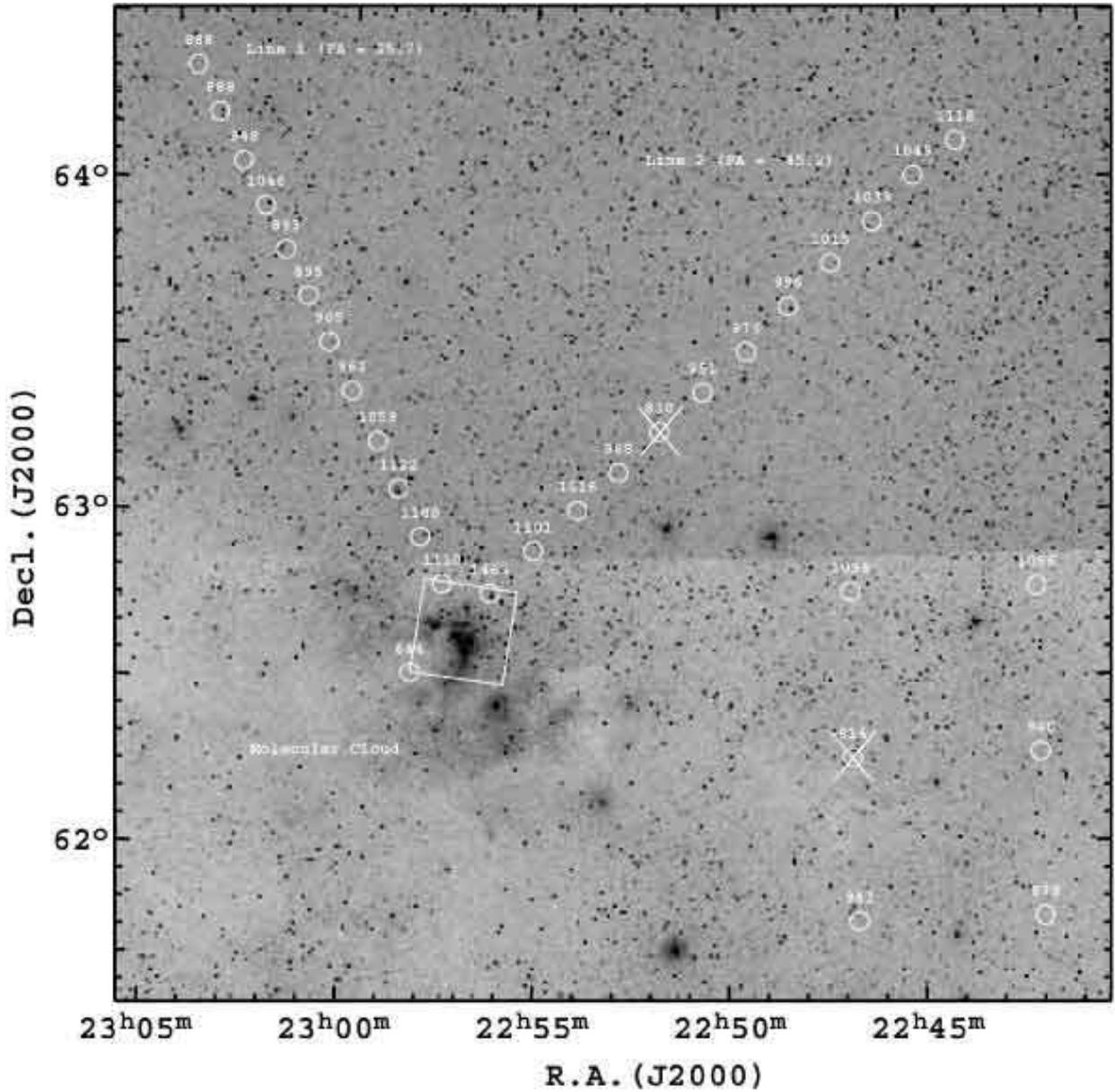}
\caption{Background regions used in estimation of the Cep OB3b
unobscured cluster KLF. This is a $3^{\circ} \times 3^{\circ}$
image from the $R$-band Digital Sky Survey.  The ACIS-I field is
outlined by the white square.  The local number of 2MASS sources
in a 127 arcmin$^2$ region was measured at each of the circled
locations. The two areas which we use for the background
subtraction, with 814 sources in the Western field and 830 sources
in the Northwestern field, are marked by white
crosses.\label{2mass_bg_fig}}
\end{figure}

\begin{figure}
\centering
\includegraphics[angle=0.,width=6.5in]{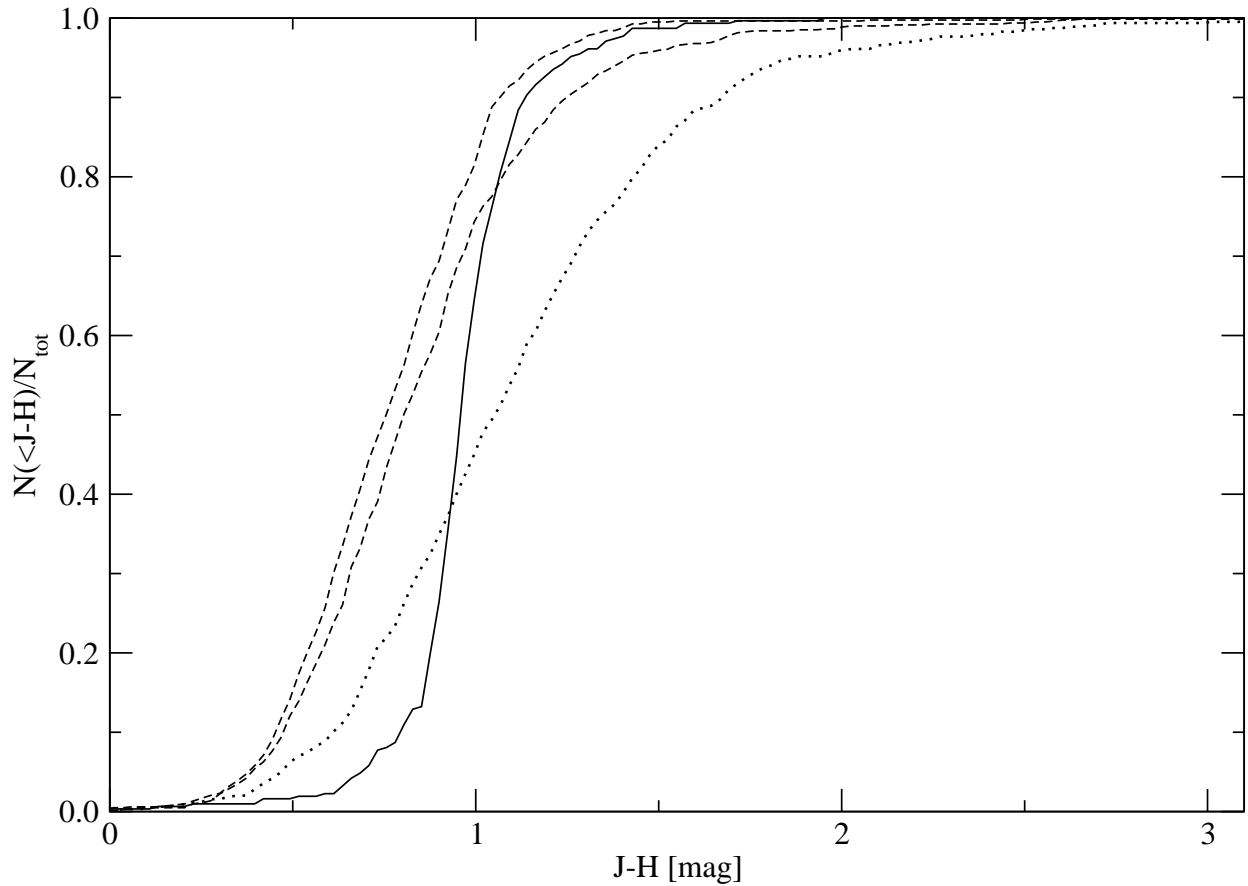}
\caption{Comparison of 2MASS $J-H$ color distributions in the
cluster, molecular cloud, and background fields. The solid line
shows PMS stars of the Cep OB3b unobscured population in the
ACIS-I field. The dotted line shows absorbed 2MASS stars in the
molecular cloud (region around 22$^h$58$^m$, +62$\degr$30$^\prime$
in Figure \ref{2mass_bg_fig}). The upper (lower) dashed lines show
2MASS sources in the Western (Northwestern) background
fields.\label{jh_compare_fig}}
\end{figure}

\begin{figure}
\centering
\includegraphics[angle=0.,width=4.8in]{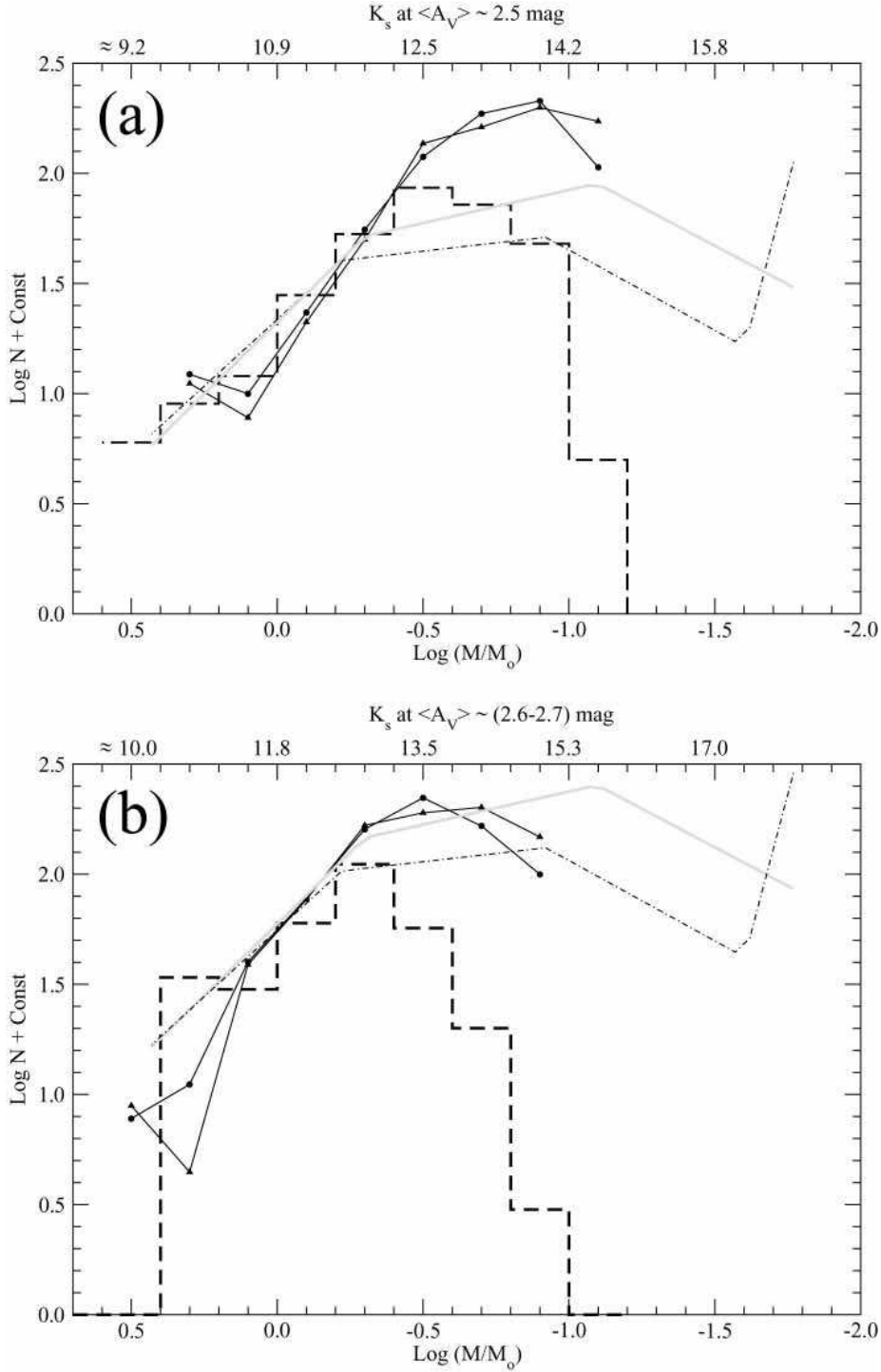}
\caption{Comparison of IMFs assuming an age of 1 Myr (a) and 3 Myr
(b). The dashed histogram shows the IMF of the Cep OB3b unobscured
X-ray population with masses derived from the NIR color-magnitude
diagram (same as in Figure \ref{imf_fig}). The solid lines
represent the IMF trends derived from the 2MASS population within
the $Chandra$ field after subtraction of KLFs from the Western
($\triangle$) and Northwestern ($\circ$) background fields and
conversion them to IMFs (\S \ref{KLF_sub.sec}). The dashed-dotted
broken line and the grey line show scaled versions of the ONC IMF
\citep{Muench02} and Galactic field IMF \citep{Kroupa02}
respectively. \label{imf2_fig}}
\end{figure}

\begin{figure}
\centering
\includegraphics[angle=0.,width=4.5in]{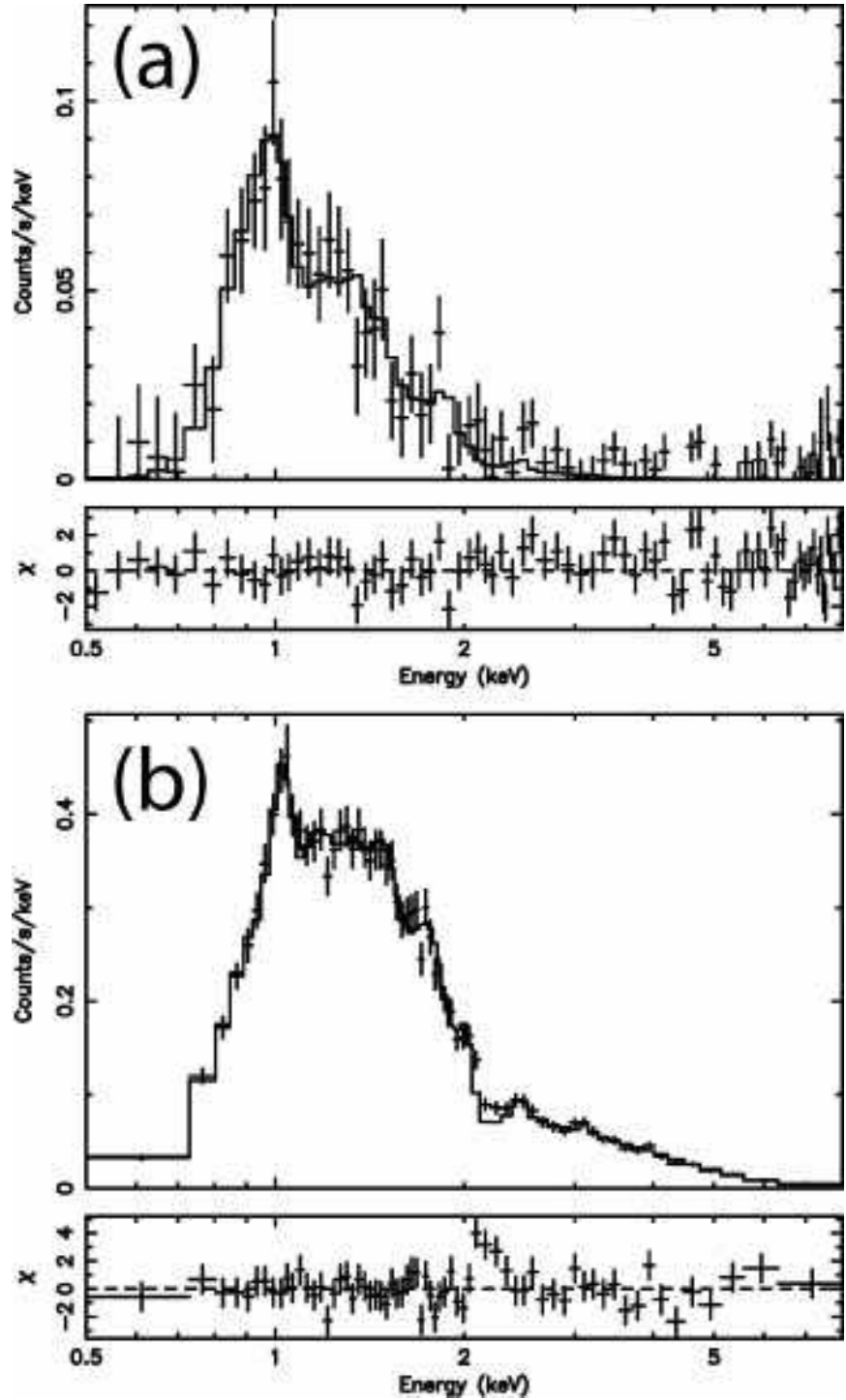}
\caption{Cepheus composite X-ray spectra.  (a) Spectrum of the
diffuse component detected in the 127 arcmin$^2$ region shown in
Figure \ref{extended_smooth_fig} (points with error bars) with
spectral model (histogram). (b) The composite spectrum and
spectral model for 321 PMS stars of the unobscured X-ray
population. The lower plots show residuals between the data and
models in units of $\chi^2$.\label{combined_spectra_fig}}
\end{figure}

Soft X-ray diffuse emission is clearly present in the central part
of the ACIS-I field. Figure \ref{extended_smooth_fig} shows an
adaptively smoothed map of the field with all 431 point sources
removed\footnote{Details on the production of such images, and
extraction source-diffuse and background spectra can be found
on-line in the recipes to {\it AE}
\url{http://www.astro.psu.edu/xray/docs/TARA/ae\_users\_guide/},
\S 7.1.2. They have been previously applied to $Chandra$ images of
high mass star formation regions by \citet{Townsley03}. Appendix C
of Townsley et al. (2003) describes the method.
\label{diffuse_label}}. The diffuse emission occupies the
pentagonal 127 arcmin$^2$ area where 90\% of the clustered X-ray
detected members of the Cep OB3b population reside (\S
\ref{KLF_sub.sec}). We extracted the spectrum of the diffuse
emission from the pentagonal area (see footnote
\ref{diffuse_label}) and subtracted extraneous spectral components
associated with the instrumental, part of cosmic X-ray, and field
star backgrounds measured on the molecular cloud in the
southeastern corner of the field, as shown in Figure
\ref{extended_smooth_fig}. The resulting spectrum of the diffuse
emission is shown in Figure \ref{combined_spectra_fig}(a). It can
be reasonably well fit by a single-temperature optically thin
thermal plasma with $kT \sim 0.8$ keV subject to $\log N_H \sim
21.7$ cm$^{-2}$ ($A_V \sim 2.5$ mag) absorption, which is typical
for the unobscured Cepheus region. The observed
(absorption-corrected) X-ray luminosity of the diffuse emission is
$\log L_{t} \simeq 31.3$ ($\log L_{t,c} \simeq 31.8$) erg
s$^{-1}$.

This is substantially weaker than the $\log L_t \simeq 32-33$ erg
s$^{-1}$ in diffuse X-rays attributed to colliding O star winds in
other H{\sc II} regions such as M 17 \citep{Townsley03}. In theory
a strong wind from a single massive star, interacting with the
surrounding ISM, may fill the surrounding volume with hot shocked
gas and produce an X-ray emission as high as $10^{35-36}$ erg
s$^{-1}$.  The O7n HD 217086 star has a high wind terminal
velocity of $V_{\infty} = 2510$ km/s derived from ultraviolet
spectroscopy by \citet{Prinja90}. Such a fast wind may heat the
gas to the temperatures of $10^{7-8}$ K. On the other hand, HD
217086 has a low mass-loss rate of $\dot{M} \sim 2 \times 10^{-7}$
$M_{\odot}$ yr$^{-1}$ \citep{Howarth89}.  \citet{Stevens03} have
calculated a scaling between expected X-ray emission and the wind
scaling factor  $X_{cl} = \dot{M}^{2}/(R_c V_{\infty})$ at
temperatures higher than $10^{7}$ K.  For a cluster core radius of
$R_c \sim 1$ pc and the HD 217086 wind parameters, the predicted
diffuse X-ray luminosity is $\log L_X < 28$ erg s$^{-1}$ for the
wide range $0.1-1$ of the thermalization efficiency parameter
$\eta$. This is at least 3 orders of magnitude smaller than the
X-ray luminosity derived for the detected diffuse X-ray emission
(Table \ref{combined_spectra_results_tbl}).  This casts
considerable doubt on the idea that the O star wind shocks are
responsible for the emission.

\begin{figure}
\centering
\includegraphics[angle=0.,width=4.5in]{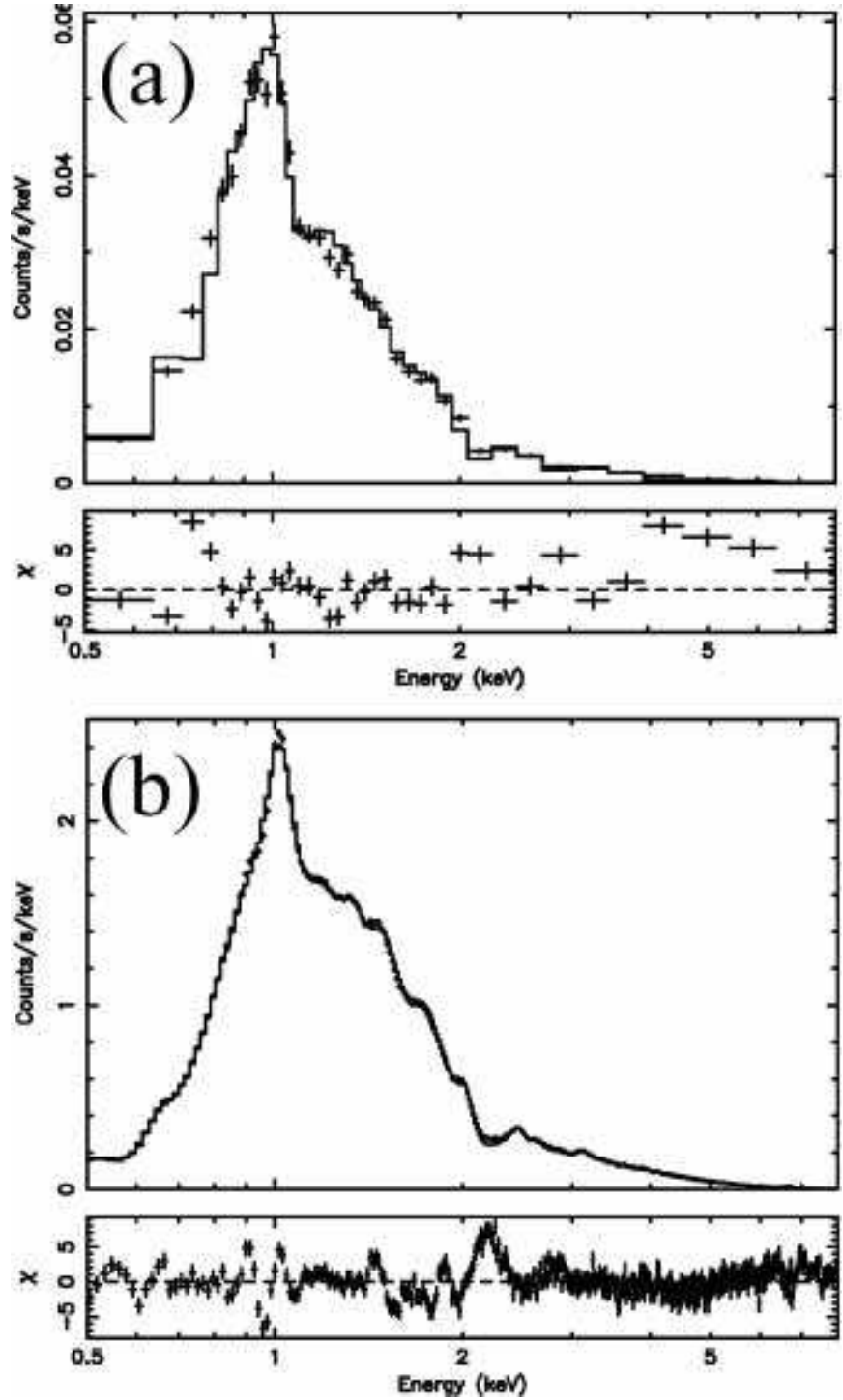}
\caption{Composite X-ray spectra of the cool unobscured ONC
population. (a) Subsample of 326 weak ($\log L_{t,c} < 29$)
sources. (b) Full sample of 820 sources with no
pileup.\label{coup_combined_spectra_fig}} \end{figure}

The alternative and more natural scenario for the origin of the
diffuse emission in Cepheus is X-ray emission from the $400-450$
lower mass PMS stars, which are not individually detected (\S
\ref{KLF_sub.sec}) due to the high sensitivity limits ($\log
L_{t,c,lim} \simeq 29$ for the lightly-obscured population; Figure
\ref{xlf_fig}) of the current $Chandra$ observation. Recall that
the Galactic background contribution to the diffuse emission is
small, $\log L_t \lesssim 30$, as indicated by the simulations
described in \S \ref{galactic_contamination_section}. Several
arguments support the explanation for the diffuse emission from an
unresolved PMS component in Cepheus.
\begin{enumerate}

\item The diffuse emission spatially coincides with the clustered
X-ray detected members of the lightly-obscured Cep OB3b population
as expected if cluster stars are responsible.  Compare the
distribution of detected stars in Figure \ref{spat_distrib_fig}
with the diffuse emission in Figure \ref{extended_smooth_fig}.

\item The spectral shape of the diffuse emission in Cepheus closely
resembles the composite spectrum of the detected X-ray population
as shown in Figure \ref{combined_spectra_fig} for the Cepheus
observation. A similar finding can be seen in the cool lightly
obscured COUP population (839 sources introduced in \S
\ref{XLF_sub.sec}). Figure \ref{coup_combined_spectra_fig}
illustrates the resemblance between (panel a) a COUP sub-sample
containing the weakest ($\log L_{t,c} < 29$)
sources---corresponding to the Cepheus diffuse emission here
attributed to the undetected population---and (panel b) the full
COUP sample (820 sources, excluding those piled-up)---
corresponding to the detected Cepheus population. The plasma
temperature of $kT \sim 0.8$ keV is similar for both the Cepheus
diffuse emission and the COUP weak sources (Table
\ref{combined_spectra_results_tbl}). This is not a surprise, as it
has been established from the COUP project that the temperature of
the hot plasma component for ONC T-Tauri stars having
two-temperature spectral fits decreases as X-ray luminosity drops,
while the cooler plasma component remains around 10 MK ($\sim
0.86$ keV) in nearly all T-Tauri stars \citep{Preibisch05}.

\item The inferred integrated (absorption-corrected) X-ray
luminosity of the missing PMS population of the unobscured Cep
OB3b cluster, estimated in \S \ref{imf_xlf_section} to be $\log
L_{t} \simeq 31.3$ ($\log L_{t,c} \simeq 31.5$) erg s$^{-1}$, is
very close to the observed diffuse X-ray luminosity of $\log L_{t}
\simeq 31.3$ ($\log L_{t,c} \simeq 31.8$) erg
s$^{-1}$.\footnote{The difference of a factor of $\sim 2$ between
absorption-corrected X-ray luminosities of $\log L_{t,c} \simeq
31.5$ expected from faint cluster stars and 31.8 erg s$^{-1}$
observed in diffuse emission can be explained as a statistical
artifact.  The same discrepancy is seen for the integrated
absorption-corrected luminosities of 326 COUP weak sources
considered individually in comparison with the
absorption-corrected luminosity derived from their composite
spectrum. When considered individually, the poor statistics of
these sources leads to successful fits by one-temperature with
median plasma energy $kT \sim 1.6$ keV.  This is noticeably harder
than the $kT \simeq 0.8$ kev energy derived from the combined fit.
This arises because of a bias towards detecting photons at
energies where the $Chandra$-ACIS effective area is greatest
around 1.5 keV; substantially softer or harder photons are not
detected in sufficient quantity in individual weak sources to be
represented in the $XSPEC$ fitting procedure. Due to this bias,
absorption-corrected X-ray luminosities $L_{t,c}$ of weak sources
may often be underestimated due to the unseen soft temperature
component, as discussed by \citet{Getman05, Feigelson05}.}

\end{enumerate}

A strong Ne IX and Ne X line complex is present in the composite
spectrum of the Cep OB3b unobscured PMS members (Figure
\ref{combined_spectra_fig}(b)).  This emission line is prominent
in the composite spectrum of COUP ONC stars (Figure
\ref{coup_combined_spectra_fig}(b)), and \citet{Feigelson05} have
argued that this line should be a useful diagnostic of
magnetically active PMS stars.\footnote{One may note strong
residuals between the spectral model and data around 2.1 keV in
(Figure \ref{combined_spectra_fig}(b)) and
\ref{coup_combined_spectra_fig}(b). These are likely instrumental
features, related to the fact the effective area of the $Chandra$
telescope mirrors is not properly calibrated at the iridium M edge
near 2.1 keV, possibly due to a thin contaminant layer on the
mirrors. For more details see
\url{http://cxc.harvard.edu/ccw/proceedings/04\_proc/presentations/jerius/}.}

\section{Two interesting X-ray stars}

\subsection{Wind shocks in HD 217086 (O7n) \label{OB_section}}

Among the members of the Cep OB3b population within the ACIS-I
field, only one massive star is seen in X-rays: the O7n star HD
217086, $Chandra$ source \#240.  It is thought to be the principal
ionizing source of the S 155 H{\sc II} region.

Our understanding of the production of X-rays in the inner winds
of O stars is somewhat confused.  Some stars show the soft,
constant X-ray emission expected from a myriad small shocks in
unstable radiatively accelerated winds \citep{Lucy80}.  But others
show hard spectral components, rotational modulations, flaring
light curves, and unexpected spectral line shapes that indicate
other processes are operative. \citet{Stelzer05} performed an
in-depth study of the X-ray emission of the early-type stars of
the ONC and found that only two of nine hot massive stars with
spectral class O7-B3 exhibit exclusively the constant
soft-spectrum at $\log(L_X/L_{bol}) \sim -7$ expected from the
standard model involving small-scale shocks from unstable
radiatively accelerated winds. Most of the other massive stars,
including the brightest $\theta^1$ Ori C (O7), show combinations
of soft wind emission and other phenomena which could be explained
by a magnetically confined wind shock model, where the harder and
variable X-rays are produced by magnetically-mediated large-scale
shocks \citep{Babel97, Uddoula02}.

\begin{figure}
\centering
\includegraphics[angle=0.,width=5.5in]{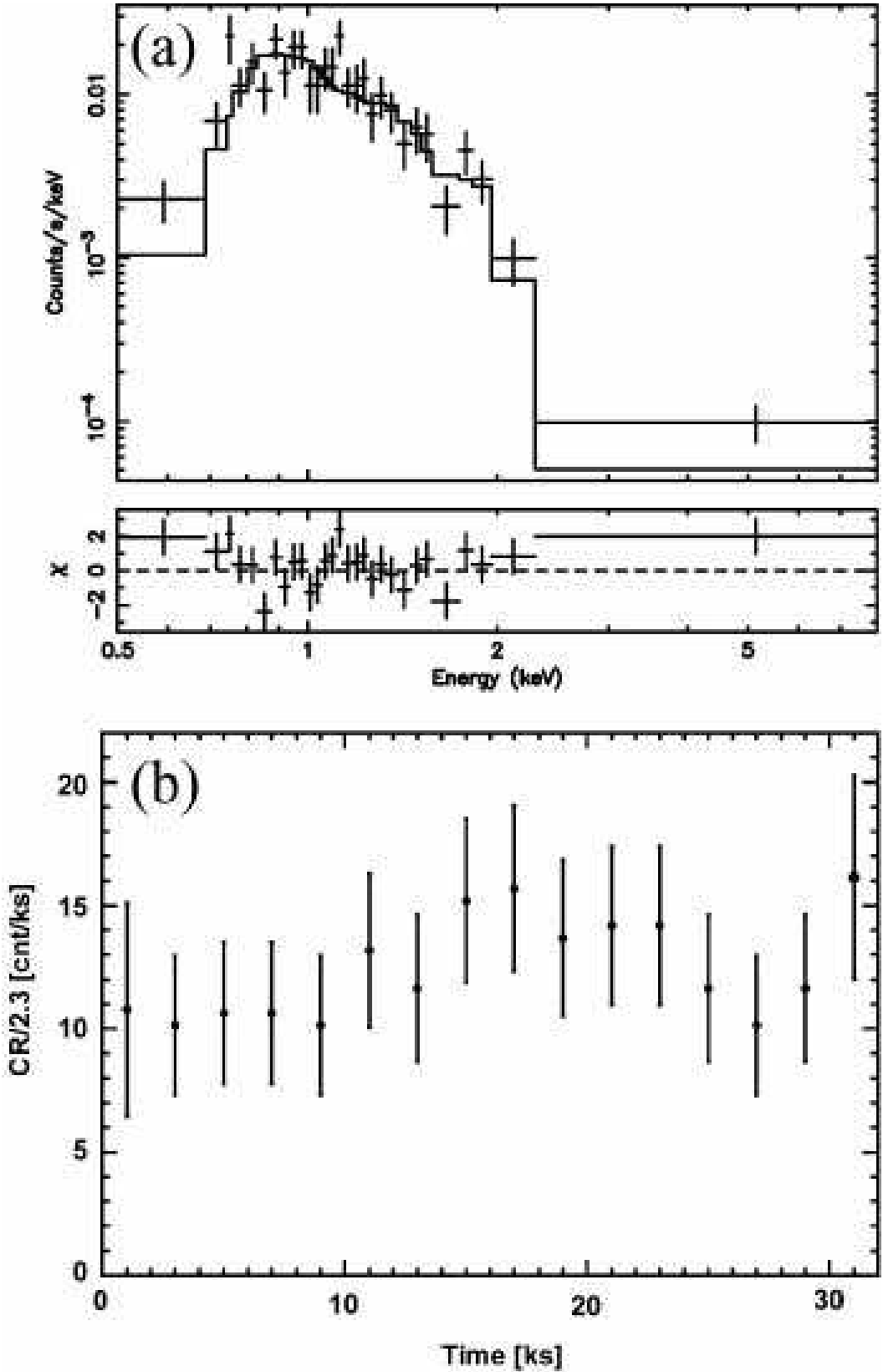}
\caption{X-ray properties of the O7n star HD 217086 ($Chandra$
source \#240): (a) ACIS spectrum, and (b)
lightcurve.\label{o_sp_lc fig}}
\end{figure}

Though the $Chandra$ source \#240 falls by chance on an ACIS CCD
chip gap, this should have little effects on its properties once
we evaluate the exposure map value at source's location. Its
effective exposure is 12.9 ks (Table \ref{tbl_bsp}), about 2.3
times shorter than the 30.1 ks effective exposure for the whole
observation. The source spectrum and lightcurve are shown in
Figure \ref{o_sp_lc fig}. To ensure that the interaction of
satellite dithering with the chip gap does not affect the shape of
the lightcurve, the lightcurve is shown with 2 ks binsize, which
is two times larger than the dithering period. A
Kolmogorov-Smirnov test shows only marginal evidence ($P_{KS} \sim
0.07$) for variability during the 8 hour observation; it was
similarly constant during the earlier $ROSAT$ observation
\citep{Naylor99}. The X-ray spectral analysis gives the count rate
of $\sim 32.3$ cnt ks$^{-1}$, plasma temperature $\sim 0.6$ keV,
and the X-ray luminosity in (0.5-8.0) keV band of $\log L_{t,c}
\sim 31.5$ erg s$^{-1}$ after correction for $\log N_H \simeq
21.7$ cm$^{-2}$ absorption. Comparison with the coarse bolometric
luminosity derived from the 2MASS color-magnitude diagram (Table
\ref{tbl_nir}) gives $\log (L_{t,c}/L_{bol}) = -7$ with a stellar
mass of $\sim 24 M_{\odot}$.

Thus its soft constant X-ray emission at $\log (L_{t,c}/L_{bol}) =
-7$ level is probably attributable to the traditional Lucy-White
model of small-scale wind shocks without invoking magnetic fields.

\subsection{Superflare in Chandra star \#
294}\label{interesting_sources_section}

\begin{figure}
\centering
\includegraphics[angle=0.,width=4.5in]{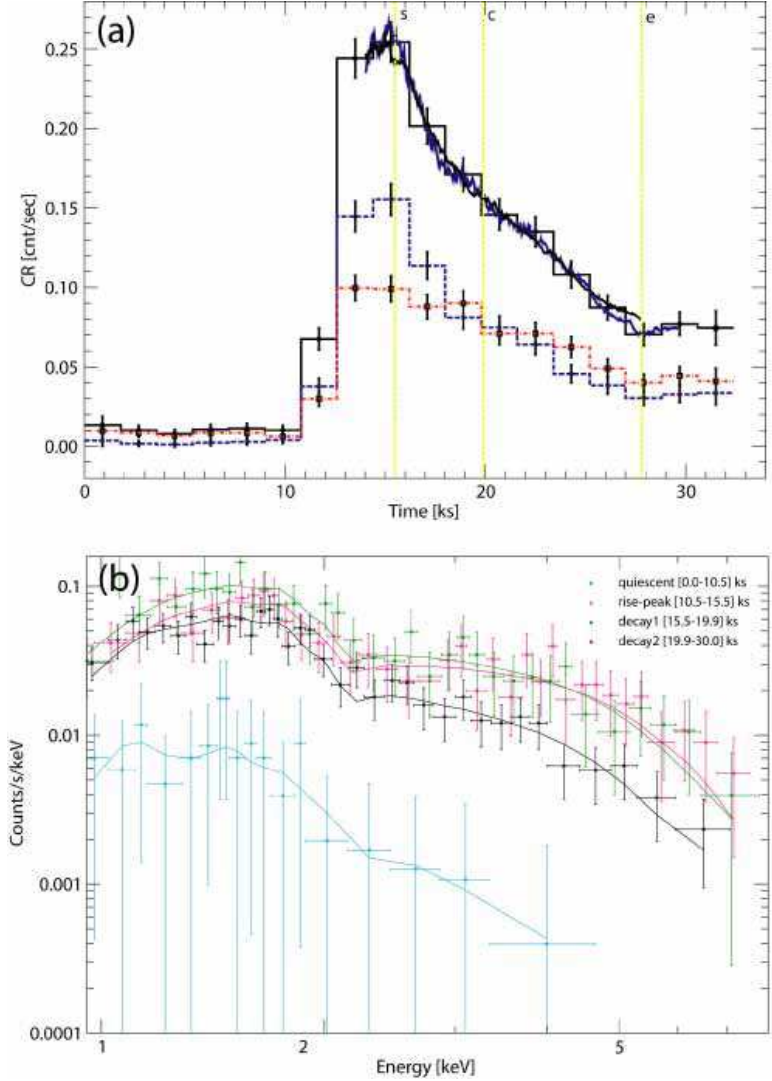}
\caption{X-ray properties of the superflare in Chandra source
\#294. (a) Lightcurve histograms of the source \#294 with a bin
size of a half of an hour for a full energy band $0.5-8.0$ keV
(black histogram), soft band $0.5-2.0$ keV (red histogram), and
hard band $2.0-8.0$ keV (blue histogram). Black and blue curves
represent unbinned but adaptively smoothed versions of the
lightcurve with a window size of at least 256 and 570 counts
respectively. Vertical labelled yellow lines give two
characteristic decay ranges [s,c] ($15.5-19.9$ ks) and [c,e]
($19.9-27.8$ ks), found from our further analysis of the
adaptively smoothed median energies (Figure
\ref{cepb294_me_analysis_fig}). (b) Spectra of four phases of the
lightcurve with 1-T optically thin thermal plasma fits, subject to
$\log N_H \sim 21.8$ cm$^{-2}$ absorption: quiescent level (cyan,
$kT \sim 1.7$ keV), flare rise and peak (magenta, $kT \sim 64$
keV), early decay phase (green, $kT \sim 12$ keV), and late decay
phase (black, $kT \sim 4.5$ keV).\label{cepb294_lc_spectra_fig}}
\end{figure}

The brightest X-ray source in the ACIS field is source \#294 with
2938 counts extracted from the 89\% encircled energy region.
Source \# 294 has an extraordinarily luminous and hard X-ray flare
whose amplitude, power, and temperature are similar to the
brightest flares seen in the Orion population \citep{Grosso04,
Favata05}. Its lightcurve has the typical morphology of stellar
flares with the fast rise and the gradual decay (Figure
\ref{cepb294_lc_spectra_fig}(a)). During the $\sim 1.1$ hour rise
phase the count rate rose by a factor $\sim 20$, $\sim 50$, and
$\sim 10$ in the full, hard, and soft bands, respectively. Due to
the short exposure of our observation, only the first $\sim 4.4$
hours of the decay phase were observed. We divide this phase into
a rapid and a slow cooling phases.

Spectra extracted from the quiescent, rise and peak phases, rapid
cooling, and slow cooling phases are compared in Figure
\ref{cepb294_lc_spectra_fig}(b). They all can be successfully
represented by single-temperature optically thin plasma models
seen through a similar amount of gas absorption of $\log N_H =
21.8$ cm$^{-2}$. During the quiescent phase, the plasma had an
average temperature of $\sim 1.7$ keV ($\sim 20$ MK) which heated
rapidly during the flare rise phase of $\sim 5$ ks up to $<$kT$>$
 $\gg 15$ keV. The nominal $XSPEC$ fit gives 64 keV ($\sim 740$ MK).
Over the next $\sim 4.4$ ks, the plasma rapidly cooled down to
$<$kT$>$ $\sim 12$ keV ($\sim 140$ MK), and cooling continued
during the slow cooling phase to $<$kT$>$ $\sim 4.5$ keV ($\sim
50$ MK) over the last $10$ ks. X-ray luminosities rose from a
quiescent level of $L_{t,c} = 9.7 \times 10^{30}$ ($L_{h,c} = 3.5
\times 10^{30}$) ergs s$^{-1}$ in the full (hard) energy bands to
$L_{t,c} = 2.2 \times 10^{32}$ ($L_{h,c} = 1.8 \times 10^{32}$)
erg s$^{-1}$ during the rise-peak phase. These temperatures and
luminosities are similar to the most luminous X-ray flares ever
recorded in young stellar objects: V773 Tau in the Taurus-Auriga
clouds with $L_{pk} \ga 10^{33}$ erg s$^{-1}$ and $T_{pk} \ga 100$
MK \citep{Tsuboi98}, LkH$\alpha$ 312 in the M 78 reflection nebula
of Orion with $L_{pk} \simeq 10^{32}$ erg s$^{-1}$ and $T_{pk}
\simeq 90$ MK \citep{Grosso04}, and COUP \#1568 in the Orion
Nebula Cluster with $L_{pk} \simeq 7.8 \times 10^{32}$ erg
s$^{-1}$ and $T_{pk} \sim 500$ MK \citep{Favata05}.

With detailed data on the decay of flux and temperature during the
cooling phase, we can infer properties of the radiating plasma
structures within the framework of the time-dependent hydrodynamic
model in a coronal magnetic loop. \citet{Reale97} establish a
formula for estimating loop size with a proper account for the
possibility of prolonged heating during the decay phase. They find
$L \propto \tau_{lc} \sqrt{T_{pk}}/F(\zeta)$, where $L$ is the
length of the loop, $\tau_{lc}$ is the flare decay time, and
$T_{pk}$ is the plasma temperature at the peak.  $F(\zeta)$ is a
correction factor for prolonged heating that is a function of the
slope $\zeta$ of the trajectory in the temperature-density
diagram. $F(\zeta)$ must be calibrated for each X-ray telescope
and detector. The model assumes the plasma is confined within the
semicircular loop of uniform cross-section, whose geometry remains
unaltered during the whole flare, and the motion and energy
transport occur along magnetic field lines of the loop.
\citet{Favata05} applied this approach to infer the size of the
flaring structures for bright X-ray flares in the Orion Nebula
cluster population with $F(\zeta)$ calibrated to $Chandra$ and
ACIS-I.  They found several superflares with inferred loop sizes
long enough to link the stellar photosphere with the inner rim of
the circumstellar disk.

The slope $\zeta$ is usually measured in the $\log T$ vs. $\log
\sqrt(EM)$ plane, where $\sqrt(EM)$ is used as a proxy for the
plasma density. In the case of the source \# 294 flare, we adopt a
slightly different technical approach. Instead of performing
time-sliced spectroscopy with $XSPEC$ over a few characteristic
time blocks of the lightcurve as shown in Figure
\ref{cepb294_lc_spectra_fig}(b), we employ an adaptively smoothed
median energy time series to derive hundreds of time points of
($kT,EM$) along the decay phase of the lightcurve. This is
achieved by calibrating median energies to temperatures using the
spectral simulations described in \S \ref{flux_det_section}.
Median energies are more statistically stable for small-$N$
datasets than $kT$ parameters of nonlinear $XSPEC$ regressions. We
assume $N_H$ is constant and we ignore the weak soft spectral
component that is likely present from the quiescent background.

\begin{figure}
\centering
\includegraphics[angle=0.,width=4.5in]{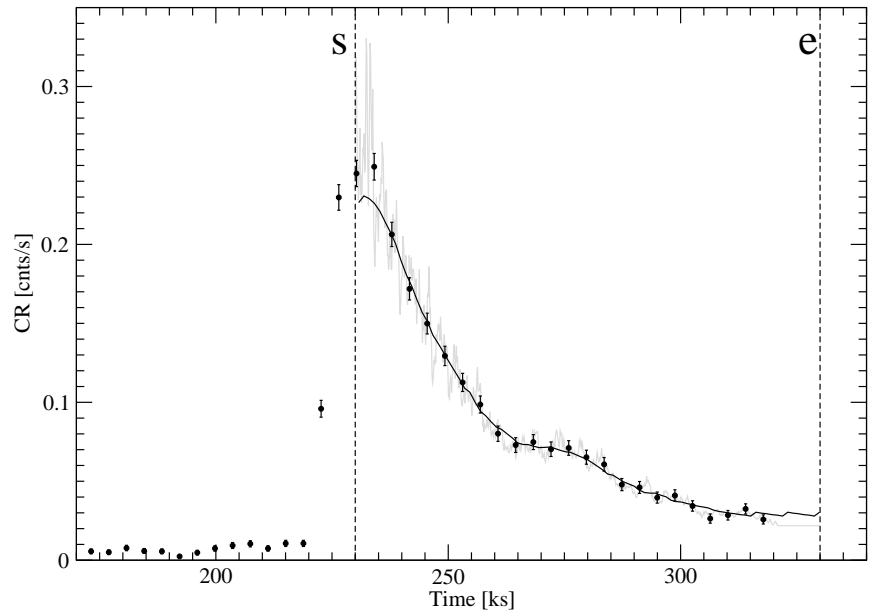}
\caption{The lightcurve histogram (black dots with uncertainties)
with a bin size of 1.06 hours for the COUP source \# 1343, used
here to validate our method of modelling the Cepheus source \#294
flare. Grey and black lines present adaptively smoothed versions
of the lightcurve with a window size of at least 100 and 625
counts respectively. Vertical labelled dashed lines give the time
range [s,e] ($230-330$ ks) over which our analysis of the
adaptively smoothed median energies was performed.
\label{coup1343_lc_fig}}
\end{figure}

\begin{figure}
\centering
\includegraphics[angle=0.,width=6.5in]{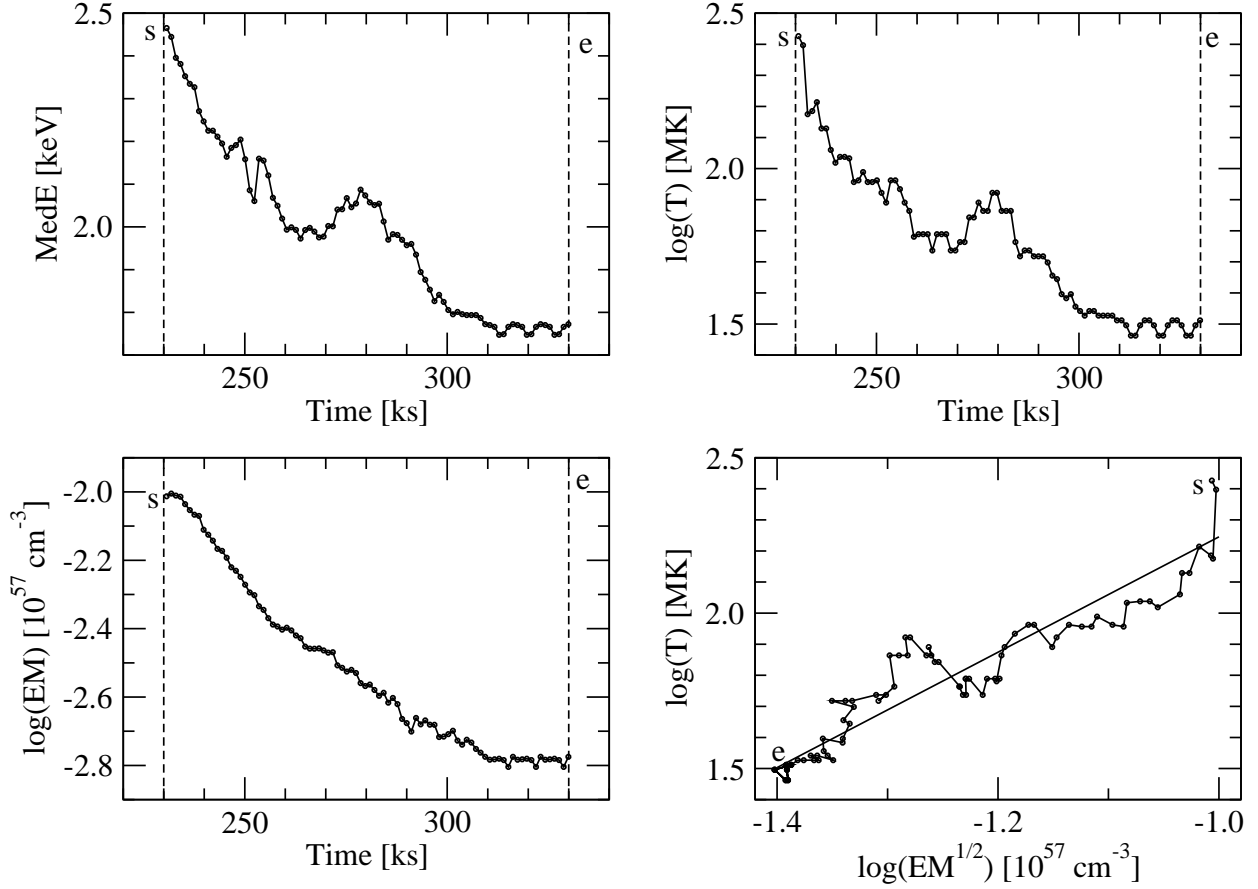}
\caption{X-ray spectral modelling of the superflare in ONC COUP
source \#1343, using the method of the adaptively smoothed median
energy with a window size of at least 625 counts over the time
range [s,e] ($230-330$ ks), including the peak of the flare and
the decay phase. The top left panel shows the evolution of the
observed median energy (88 points), the top right panel -- the
evolution of the plasma temperature, the bottom left panel -- the
evolution of the emission measure. The bottom right panel shows
the evolution of the flare in the $\log T - \log \sqrt(EM)$ plane
with the best fitting decay as a solid line of the slope $\zeta =
1.86 \pm 0.29$. \label{coup1343_me_analysis_fig}}
\end{figure}

We test this method on the decay phase of the flare in the Orion
COUP source \#1343 studied in detail by \citet{Favata05}. The
lightcurve for the COUP source \#1343 is shown in Figure
\ref{coup1343_lc_fig} with black dots representing the lightcurve
histogram, binned with a bin size of $\sim 1$ hour. The peak-decay
phase studied in our analysis ranges from $230$ to $330$ ks from
the beginning of the COUP observations. The black and grey lines
represent smoothed versions of the lightcurve using the adaptively
smoothed algorithm from {\it AE} with different minimum number of
counts in the kernel (625 and 100 count minimum, respectively).
The 100 count kernel is too small as it leaves many noise spikes,
so we use the 625 count kernel although it has oversmoothed the
narrow peak phase of the lightcurve. Results of the analysis
appear in Figure \ref{coup1343_me_analysis_fig}. The top left
panel shows the trend of the median energy with 88 time points
over the 100 ks time range.  The top right and bottom left panels
show the derived decays of temperature and emission measure
assuming a constant absorption of $\log N_H = 21.9$ cm$^{-2}$.
Finally the bottom right panel gives the $\log T - \log \sqrt(EM)$
diagram over the 100 ks peak-decay phase; the flare evolves from
the upper right to the lower left.  The best fit decay slope is
$\zeta = 1.86 \pm 0.29$. These last three panels look very similar
to those shown in Figure 3 of \cite{Favata05} who obtain a similar
slope of $\zeta = 1.95 \pm 0.51$ using coarser time sampling and
$XSPEC$ fits.  This validates our use of adaptively smoothed
median energies as estimates of plasma temperatures in flare loop
models.

\begin{figure}
\centering
\includegraphics[angle=0.,width=6.5in]{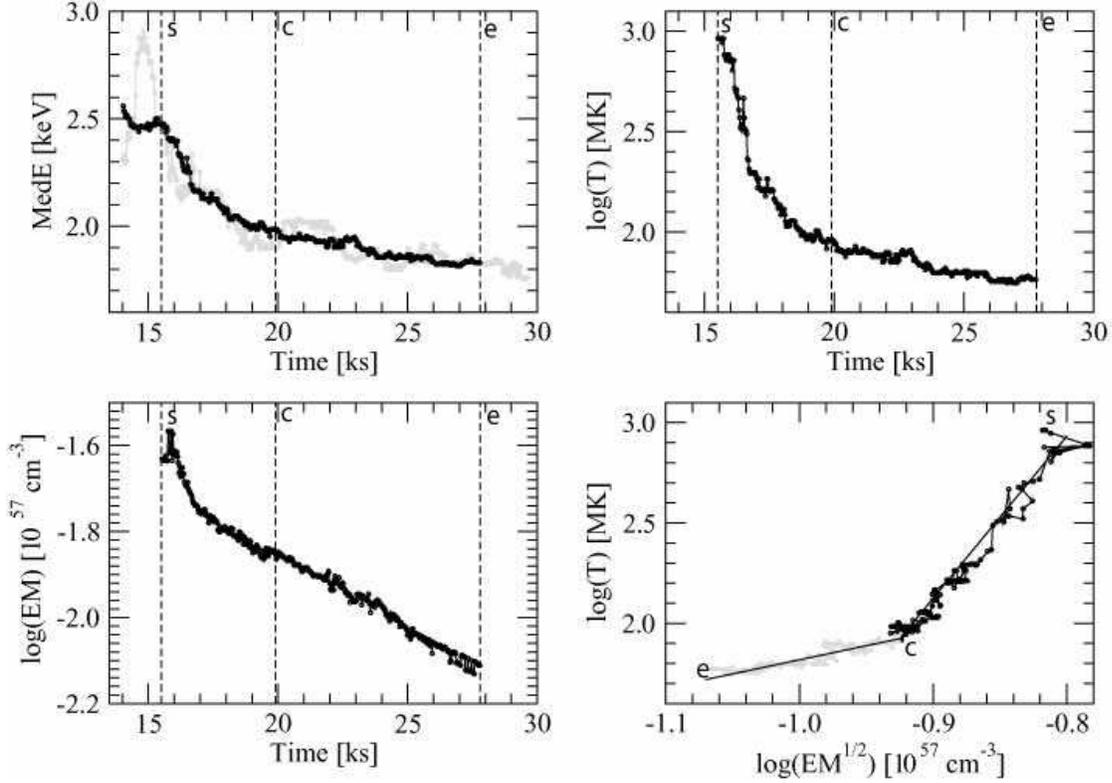}
\caption{X-ray spectral modelling of the superflare in source
\#294, using the adaptively smoothed median energy with a window
size of at least 570 counts over the time range [s,e] ($15.5-27.8$
ks), including the decay phase of the flare (see Figure
\ref{cepb294_lc_spectra_fig}). The top left panel shows the
evolution of the smoothed median energy for the window sizes of
256 (grey) and 570 (black) counts respectively. The top right
panel is the evolution of the temperature (403 points), the bottom
left panel is the evolution of the emission measure. The bottom
right panel shows the evolution of the flare in the $\log T - \log
\sqrt(EM)$ plane with the two best fitting decays: the slope of
$\zeta = 8.08 \pm 0.96$ over the [s,c] ($15.5-19.9$ ks) time
range; and the slope of $\zeta = 1.40 \pm 0.13$ over the [c,e]
($19.9-27.8$ ks) range. \label{cepb294_me_analysis_fig}}
\end{figure}

The method is applied to 12.3 ks ($15.5-27.8$ ks after the start
of the observation) of the ACIS-I source \# 294 lightcurve;
results are shown in Figure \ref{cepb294_me_analysis_fig}. Trends
of the median energy for two different smoothing kernels are alike
within the studied decay phase of the flare.  We report results
for the 570 count kernel, although it does a poor job tracing the
earlier rapidly changing flare peak.  The derived temperatures are
very high, at least $T \sim 300$ MK and conceivably as high as
$800-1000$ MK during the first 1 ks of the decay phase.  But they
rapidly drop to $T \simeq 100$ MK over the next hour and then
decrease more gradually to $T \simeq 60$ MK over the following 2
hours.

The derived $\log T - \log \sqrt(EM)$ correlation (bottom right
panel) suggests two decay phases: rapid cooling over the first
$\sim 4.4$ ks with $\zeta = 8.08 \pm 0.96$, followed by slow
cooling with $\zeta = 1.4 \pm 0.13$. The median temperatures in
the upper right panels are in good agreement with the cruder
time-sliced spectroscopy shown in Figure
\ref{cepb294_lc_spectra_fig}(b). The slow decay value $\zeta =
1.4$ is close to the upper limit of 1.5 for a freely decaying loop
with no prolonged heating input \citep{Favata05,Reale97}.

Since we observe only the first $\sim 4.4$ hours of the decay
phase, we can obtain only a lower limit to the phase duration,
$\tau_{lc} > 15$ ks, and thus only a lower limit to the inferred
loop length. Using formulae (5)--(7) in \citet{Favata05} and
truncating our peak temperature around 300 MK, we derive a loop
length  $L \ga 6.4 \times 10^{11}$ cm.  We roughly estimate the
stellar radius for source \# 294 to be $R_{\star} \sim 2.5
R_{\odot}$ using \citep{Siess00} PMS models for a $\sim 1$ Myr old
star of $\sim 1.4 M_{\odot}$ mass (Table \ref{tbl_nir}). The
magnetic loop is then $\ga 3.7$ times the radius of the star. This
limit is even more conservative if the peak temperature exceeds
300 MK;  the loop would be 2 times larger if a peak temperature
around 800-1000 MK is used, as suggested from the plasma
temperatures derived in Figure \ref{cepb294_me_analysis_fig}.
Source \#294 may thus be another case where magnetic fields
connect the star to the circumstellar disk, as discussed by
\citet{Favata05}. Though there is no an indication for the
presence of $K$-band excess in source \# 294, it is well known
that $JHK$ photometry alone is not sufficient to detect
circumstellar material. Ground-based $JHKLM$ or {\it Spitzer Space
Telescope} photometry would allow much better discrimination
between stars with and without circumstellar disks.

\section{Conclusions}\label{summary_section}

We present a 30 ks (net exposure) $Chandra$ ACIS observation of
the Cepheus B molecular cloud and a portion of the nearby Cep OB3b
OB association. The main conclusions of our study are as follows:

1. We detect 431 X-ray sources with a limiting luminosity for the
region outside of the molecular cloud of $\sim 10^{29}$ ergs/s.
Three hundred eighty four of them are unambiguously identified
with 2MASS NIR sources. Upon visual inspection of 2MASS atlases we
find NIR counterparts for at least 10 (2) more soft (hard) X-ray
sources, which have not been registered in the 2MASS catalog. On
the NIR color-magnitude diagram most of the detected X-ray sources
occupy the locus of low-mass PMS stars with a typical absorption
of $A_V \sim 2.5$ mag (assuming age of $\sim 1$ Myr), but as many
as 60 sources with NIR counterparts, residing close to/in the
molecular cloud, have even much higher absorptions up to $A_V \sim
28$ mag.

2. Twenty four unidentified hard X-ray sources are classified as
possible extragalactic contaminants (Table \ref{tbl_spe}, `a').
Though some of them still may be embedded members of the cloud,
about 70\% percent of them are probably true AGNs, as argued by a
detailed simulation of the extragalactic background population,
including instrumental background and local (Cepheus) plus large
scale Galactic absorptions. Through careful simulations of X-ray
emission from the Galactic foreground and background populations,
employing a stellar population synthesis model of the Galactic
disk, and through the inspection of NIR colors of X-ray detected
sources we expect to have no more than a few Galactic background
stars and we suggest 13 field star candidates (Table
\ref{tbl_spe}, `f').

3. In order to obtain a reliable X-ray luminosity function of the
detected PMS stars we determine source X-ray fluxes based on
knowledge of their NIR absorptions and X-ray median energies. We
then perform careful MARX simulations to derive errors on median
energy, combine them with known errors on count rate and
absorption, and propagate them to the flux errors through
Monte-Carlo simulations.

4. We discover more than 380 PMS stars in the ACIS-I field of the
Cepheus region, which provide the best census to date for the
stellar population of the cloud. Based on $N_H$ and $K$-band
stratified spatial distributions of the detected PMS stars we
separate them into two rich clusters: the lightly obscured cluster
(= ``Cep OB3b within ACIS-I field'') (321 stars, Table
\ref{tbl_spe}, `l') and the embedded cluster (64 stars, Table
\ref{tbl_spe}, `e').  We estimate that another $400-450$
($300-340$) low mass PMS stars are present in the unobscured
(obscured) clusters within the $Chandra$ field but are too
X-ray-faint to be individually identified.

5. We find that the XLF of the lightly-obscured Cep OB3b cluster
region within the ACIS-I field shows an excess of stars around
$\log L_X \sim 29.7$ erg s$^{-1}$ ($0.5-8$ keV, after individual
correction for absorption) compared to the well-studied Orion
Nebula Cluster. This may be caused by a nonstandard IMF in Cep
OB3b with an excess of $\simeq 0.3$ M$_\odot$ stars if the cluster
age is $\sim 1$ Myr. This excess also appears in the independently
derived KLF based on 2MASS source counts (after an uncertain
subtraction of background counts from two control fields) which
show more $K \sim 13-14$ stars than expected from a standard IMF.
However, the discrepancy may also be attributed to different ages
and star formation histories for the Cep OB3b and ONC clusters.

6. We show that the HD 217086 (O7n) star with its known wind
terminal velocity and mass-loss rate is unable to produce the
bright soft X-ray diffuse emission seen in the field. An
alternative explanation --- the X-ray diffuse emission originates
from the unresolved PMS point sources in the region --- is
suggested by (1) the spatial coincidence of the diffuse emission
with the detected point source population, (2) the similarity of
the plasma temperature of the diffuse emission with that of the
combined weak PMS Orion-COUP sources, (3) the equivalence of the
amount of the observed X-ray luminosity for the diffuse emission
with that of the estimated undetected members of the unobscured
population.

7. We establish that the X-ray emission from the HD 217086 (O7n)
star is a constant soft-spectrum emission at the level $\log
(L_{X}/L_{bol}) \sim -7$ that is expected from the standard model
involving many small shocks in an un-magnetized radiatively
accelerated wind.

8. From the evolution of an adaptively smoothed median energy
during the unusually hot and strong flare in source \# 294 we
infer the slope of the trajectory in the temperature-density
diagram, estimate the lower limit on the size of the flaring
structure, and suggest that the source \# 294 may join a number of
similar super-flare sources seen in Orion clouds where long
magnetic structures may connect the protoplanetory disk to the
stellar surface.

We thank the referee, Tim Naylor, for his time and many useful
comments that improved this work. This work was supported by the
ACIS Team (G. Garmire, PI) through NASA contract NAS8-38252 and
from {\it Chandra} contract SV4-74018 issued by the {\it Chandra}
X-ray Center operated by SAO under NASA contract NAS8-03060. M.T.
acknowledges financial support from Japan Society for the
Promotion of Science. This publication makes use of data products
from the Two Micron All Sky Survey, which is a joint project of
the University of Massachusetts and the Infrared Processing and
Analysis Center/California Institute of Technology, funded by the
National Aeronautics and Space Administration and the National
Science Foundation.

\clearpage

\begin{deluxetable}{rcrrrrrrrrrrrrcccc}
\centering \rotate \tabletypesize{\tiny} \tablewidth{0pt}
\tablecolumns{18}

\tablecaption{Basic Source Properties \label{tbl_bsp}}

\tablehead{ \multicolumn{2}{c}{Source} &
  &
\multicolumn{3}{c}{Position} &
  &
\multicolumn{5}{c}{Extracted Counts} &
  &
\multicolumn{5}{c}{Characteristics} \\
\cline{1-2} \cline{4-6} \cline{8-12} \cline{14-18}

\colhead{Seq} & \colhead{CXOCEPB} &
  &
\colhead{$\alpha_{\rm J2000}$} & \colhead{$\delta_{\rm J2000}$} &
\colhead{$\theta$} &
  &
\colhead{Net} & \colhead{$\Delta$Net} & \colhead{Bkgd} &
\colhead{Net} & \colhead{PSF} &
  &
\colhead{Signif} & \colhead{Anom\tablenotemark{a}} & \colhead{Var\tablenotemark{b}} &\colhead{EffExp} & \colhead{Med E}  \\

\colhead{\#} & \colhead{} &
  &
\colhead{(deg)} & \colhead{(deg)} & \colhead{(\arcmin)} &
  &
\colhead{Full} & \colhead{Full} & \colhead{Full} & \colhead{Hard}
& \colhead{Frac} &
  &
\colhead{} & \colhead{} & \colhead{} & \colhead{(ks)} &
\colhead{(keV)}
 \\

\colhead{(1)} & \colhead{(2)} &
  &
\colhead{(3)} & \colhead{(4)} & \colhead{(5)} &
  &
\colhead{(6)} & \colhead{(7)} & \colhead{(8)} & \colhead{(9)} &
\colhead{(10)} &
  &
\colhead{(11)} & \colhead{(12)} & \colhead{(13)} & \colhead{(14)}
& \colhead{(15)} }

\startdata
1 & 225531.25+624451.7 &  &  343.880235 &  62.747708 &  10.22 &  &    10.2 &   4.0 &   1.8 &     2.9 & 0.70 &  &   2.2 & ge.. & \nodata &   11.7 & 1.70 \\
2 & 225533.30+624457.2 &  &  343.888775 &  62.749237 &  10.06 &  &     9.1 &   4.9 &   8.9 &     0.0 & 0.90 &  &   1.7 & .... & a &   25.2 & 1.16 \\
3 & 225535.75+624235.4 &  &  343.898960 &  62.709839 &   8.85 &  &     8.0 &   3.9 &   3.0 &     0.2 & 0.90 &  &   1.8 & g... & \nodata &   21.2 & 1.26 \\
4 & 225537.53+624128.1 &  &  343.906375 &  62.691159 &   8.38 &  &   101.7 &  10.8 &   3.3 &    67.9 & 0.89 &  &   9.0 & .... & b &   26.3 & 2.65 \\
5 & 225538.33+624117.8 &  &  343.909731 &  62.688296 &   8.26 &  &   542.3 &  23.9 &   3.7 &   178.4 & 0.89 &  &  22.2 & .... & c &   26.6 & 1.64 \\
6 & 225540.99+624625.9 &  &  343.920829 &  62.773882 &  10.18 &  &    27.6 &   6.7 &   9.4 &    15.2 & 0.90 &  &   3.8 & .... & a &   25.3 & 2.65 \\
7 & 225541.56+623905.9 &  &  343.923167 &  62.651662 &   7.83 &  &     6.4 &   3.5 &   2.6 &     0.0 & 0.90 &  &   1.5 & .... & a &   27.0 & 1.01 \\
8 & 225541.74+624232.1 &  &  343.923948 &  62.708936 &   8.18 &  &    17.1 &   5.0 &   2.9 &     2.3 & 0.90 &  &   3.1 & .... & a &   24.5 & 1.45 \\
9 & 225542.16+623742.3 &  &  343.925698 &  62.628442 &   8.03 &  &     6.6 &   3.7 &   3.4 &     0.0 & 0.90 &  &   1.5 & .... & a &   24.9 & 1.20 \\
10 & 225544.19+624122.1 &  &  343.934156 &  62.689499 &   7.61 &  &    29.4 &   6.2 &   2.6 &     8.4 & 0.89 &  &   4.4 & .... & a &   26.9 & 1.38 \\
11 & 225544.21+624213.4 &  &  343.934247 &  62.703740 &   7.82 &  &    19.0 &   5.1 &   2.0 &     2.7 & 0.89 &  &   3.3 & .... & a &   26.8 & 1.20 \\
12 & 225545.08+623854.6 &  &  343.937848 &  62.648524 &   7.45 &  &    69.6 &   9.0 &   2.4 &    20.4 & 0.89 &  &   7.3 & .... & c &   27.2 & 1.48 \\
13 & 225547.50+624157.6 &  &  343.947954 &  62.699356 &   7.38 &  &    28.1 &   6.0 &   1.9 &     0.8 & 0.89 &  &   4.3 & .... & a &   27.0 & 1.34 \\
14 & 225547.59+623614.8 &  &  343.948310 &  62.604131 &   7.99 &  &    38.6 &   7.0 &   3.4 &     9.2 & 0.90 &  &   5.1 & g... & \nodata &   26.2 & 1.23 \\
15 & 225549.82+623421.9 &  &  343.957606 &  62.572759 &   8.82 &  &    15.2 &   4.9 &   3.8 &    12.5 & 0.90 &  &   2.8 & .... & a &   26.8 & 2.99 \\
\enddata

\tablecomments{Column 1: X-ray source number. Column 2: IAU
designation. Columns 3-4: Right ascension and declination for
epoch J2000.0 in degrees. Column 5: Off-axis angle. Columns 6-7:
Estimated net counts from extracted area in total energy band
$(0.5-8.0)$ keV and their 1\/$\sigma$ errors. Column 8: Estimated
background counts from extracted area in total energy band. Column
9: Estimated net counts from extracted area in hard energy band
$(2.0-8.0)$ keV. Column 10: Fraction of the PSF at the fiducial
energy of 1.497 keV enclosed within the extracted area. Note that
a reduced PSF fraction (significantly below 90\%) may indicate
that the source is in a crowded region or on the edge of the
field. Column 11: Source significance in sigma. Column 12. Source
anomalies described by a set of flags, see the text
note$^{\rm{a}}$ below for details. Column 13: Variability
characterization based on K-S statistics, see the text
note$^{\rm{b}}$ below for details. Column 14: Source's effective
exposure. Column 15: The background-corrected median photon energy
in the total energy band.}

\tablenotetext{a}{ Source anomalies:  g = fractional time that
source was on a detector (FRACEXPO from {\em mkarf}) is $<0.9$ ; e
= source on field edge; p = source piled up; s = source on readout
streak.} \tablenotetext{b}{ Source variability:  a = no evidence
for variability; b = possibly variable; c = definitely variable.
No test is performed for sources with fewer than 4 total full-band
counts.  No value is reported for sources in chip gaps or on field
edges.}

\tablecomments{The full table of 431 {\it Chandra} sources is
available in the electronic edition of the Journal.}

\end{deluxetable}

\clearpage

\begin{deluxetable}{rccrrrrccrrc}
\centering \rotate \tabletypesize{\tiny} \tablewidth{0pt}
\tablecolumns{12}

\tablecaption{2MASS Counterparts \label{tbl_nir}}

\tablehead{

\colhead{Seq} & \colhead{CXOCEPB} & \colhead{2MASS} &
\colhead{Off} & \colhead{$J$} & \colhead{$H$} & \colhead{$K_s$} &
\colhead{Ph\tablenotemark{a}} &
\colhead{Con\tablenotemark{b}} & \colhead{$A_{V,IR}$} & \colhead{$M_{IR}$} & \colhead{$\log L_{\rm bol,IR}$} \\

\colhead{\#} & \colhead{} & \colhead{} & \colhead{\arcsec} &
\colhead{mag} & \colhead{mag} & \colhead{mag} & \colhead{Flag} &
\colhead{Flag} & \colhead{mag} & \colhead{M$_\odot$} &
\colhead{L$_\odot$} \\

\colhead{(1)} & \colhead{(2)} & \colhead{(3)} & \colhead{(4)} &
\colhead{(5)} & \colhead{(6)} & \colhead{(7)} & \colhead{(8)} &
\colhead{(9)} & \colhead{(10)} & \colhead{(11)} & \colhead{(12)} }

\startdata
1 & 225531.25+624451.7 & 22553069+6244482 & 5.16 & 14.75 & 13.85 & 13.53 & AAA & 000 & 1.84 & 0.15 & -0.86 \\
2 & 225533.30+624457.2 & 22553312+6244596 & 2.72 & 13.93 & 13.00 & 12.69 & AAA & 000 & 1.39 & 0.27 & -0.57 \\
3 & 225535.75+624235.4 & 22553545+6242343 & 2.29 & 12.54 & 11.67 & 11.38 & AEE & 000 & 0.46 & 0.68 & -0.10 \\
4 & 225537.53+624128.1 & 22553745+6241280 & 0.57 & 12.57 & 11.25 & 10.47 & AAA & 000 & 6.15 & 1.81 & 0.74 \\
5 & 225538.33+624117.8 & 22553834+6241177 & 0.14 & 12.99 & 11.96 & 11.44 & AAA & 000 & 1.62 & 0.63 & -0.14 \\
6 & 225540.99+624625.9 & \nodata & \nodata & \nodata & \nodata & \nodata & \nodata & \nodata & \nodata & \nodata & \nodata \\
7 & 225541.56+623905.9 & 22554154+6239030 & 2.92 & 12.05 & 11.72 & 11.56 & AAA & c00 & \nodata & \nodata & \nodata \\
8 & 225541.74+624232.1 & 22554177+6242322 & 0.19 & 14.11 & 13.17 & 12.72 & AAA & c00 & 1.49 & 0.24 & -0.63 \\
9 & 225542.16+623742.3 & 22554205+6237426 & 0.84 & 15.28 & 14.32 & 14.01 & AAA & 000 & 2.47 & 0.12 & -1.01 \\
10 & 225544.19+624122.1 & 22554423+6241220 & 0.29 & 14.35 & 13.45 & 13.15 & AAA & 000 & 1.51 & 0.18 & -0.74 \\
11 & 225544.21+624213.4 & 22554407+6242130 & 1.05 & 13.94 & 13.04 & 12.80 & AAA & 000 & 1.14 & 0.25 & -0.61 \\
12 & 225545.08+623854.6 & 22554507+6238547 & 0.07 & 14.05 & 13.13 & 12.86 & AAA & 000 & 1.45 & 0.24 & -0.62 \\
13 & 225547.50+624157.6 & 22554749+6241576 & 0.12 & 12.96 & 12.05 & 11.78 & AAA & 000 & 0.70 & 0.53 & -0.25 \\
14 & 225547.59+623614.8 & 22554765+6236152 & 0.58 & 13.21 & 12.29 & 11.87 & EAA & 000 & 0.69 & 0.44 & -0.35 \\
15 & 225549.82+623421.9 & \nodata & \nodata & \nodata & \nodata & \nodata & \nodata & \nodata & \nodata & \nodata & \nodata \\
\enddata

\tablecomments{Column 1: X-ray source number. Column 2: IAU
designation. Column 3: 2MASS source. Column 4: $Chandra$-2MASS
positional offset. Columns 5-7: 2MASS $JHK_s$ magnitudes. Columns
8-9: 2MASS photometry quality, and confusion and contamination
flags. See text notes$^{{\rm a,b}}$ below for more details.
Columns 10-12: Visual absorption, stellar mass, and bolometric
luminosity derived from color-magnitude diagram by de-reddening
2MASS photometric colors to the 1 Myr PMS isochrone.}

\tablenotetext{a}{Photometry quality flag with three characters
refer to the $JHK_s$ bands:
     A = very high significance detection ($>10$ SNR);
     B = high significance detection ($>7$ SNR);
     C = moderate significance detection ($>5$ SNR);
     D = low significance detection;
     E = point spread fitting poor;
     F = reliable photometric errors not available;
     U = upper limit on magnitude (source not detected);
     X = source is detected but no valid brightness estimate is
     available.
} \tablenotetext{b}{2MASS confusion and contamination flag with
three characters refer to the $JHK_s$ bands:  0 = source
unaffected by artifacts;
     b = bandmerge confusion (possible multiple source);
     c = photometric confusion from nearby star;
     d = diffraction spike confusion from nearby star;
     p = persistence contamination from nearby star;
     s = electronic stripe from nearby star.}

\tablecomments{The full table of 431 {\it Chandra} sources is
available in the electronic edition of the Journal.}

\end{deluxetable}

\clearpage

\begin{deluxetable}{rcrrccccccccc}
\centering \rotate \tabletypesize{\tiny} \tablewidth{0pt}
\tablecolumns{13}

\tablecaption{X-ray Spectroscopy and Luminosities \label{tbl_spe}}

\tablehead{

\colhead{Seq} & \colhead{CXOCEPB} & \colhead{$\log N_H$} &
\colhead{$kT$} & \colhead{$\log L_h$} & \colhead{$\Delta \log
L_h$} & \colhead{$\log L_{h,c}$} & \colhead{$\Delta \log L_{h,c}$}
& \colhead{$\log L_t$} & \colhead{$\Delta \log L_t$} & \colhead{$\log L_{t,c}$} & \colhead{$\Delta \log L_{t,c}$} & \colhead{MMD} \\

\colhead{\#} & \colhead{} & \colhead{cm$^{-2}$} & \colhead{keV} &
\colhead{erg s$^{-1}$} & \colhead{erg s$^{-1}$} & \colhead{erg
s$^{-1}$} & \colhead{erg s$^{-1}$} & \colhead{erg s$^{-1}$} &
\colhead{erg s$^{-1}$} & \colhead{erg s$^{-1}$} & \colhead{erg s$^{-1}$} & \colhead{Flag} \\

\colhead{(1)} & \colhead{(2)} & \colhead{(3)} & \colhead{(4)} &
\colhead{(5)} & \colhead{(6)} & \colhead{(7)} & \colhead{(8)} &
\colhead{(9)} & \colhead{(10)} & \colhead{(11)} & \colhead{(12)} &
\colhead{(13)}}

\startdata
  1 & 225531.25+624451.7 & 21.57 &  4.70 & 29.75 & 0.34 & 29.77 & 0.33 & 29.88 & 0.18 & 30.01 & 0.25 & wl0 \\
  2 & 225533.30+624457.2 & 21.44 &  1.10 & 28.50 & 0.40 & 28.53 & 0.39 & 29.09 & 0.21 & 29.36 & 0.25 & wl0 \\
  3 & 225535.75+624235.4 & 20.97 &  1.90 & 28.76 & 0.51 & 28.77 & 0.51 & 29.18 & 0.21 & 29.26 & 0.20 & wl0 \\
  4 & 225537.53+624128.1 & 22.14 &  4.65 & 30.52 & 0.07 & 30.58 & 0.15 & 30.57 & 0.06 & 30.83 & 0.38 & bl0 \\
  5 & 225538.33+624117.8 & 21.59 &  2.53 & 30.90 & 0.10 & 30.93 & 0.07 & 31.11 & 0.06 & 31.29 & 0.21 & bl0 \\
  6 & 225540.99+624625.9 & 22.51 &  1.88 & 29.71 & 0.11 & 29.92 & 0.25 & 29.76 & 0.10 & 30.38 & 0.50 & ba0 \\
  7 & 225541.56+623905.9 & 19.00 &  0.81 & 27.89 & 0.64 & 27.89 & 0.64 & 29.05 & 0.22 & 29.06 & 0.21 & vf0 \\
  8 & 225541.74+624232.1 & 21.47 &  2.10 & 29.21 & 0.33 & 29.23 & 0.32 & 29.49 & 0.17 & 29.67 & 0.22 & wl0 \\
  9 & 225542.16+623742.3 & 21.69 &  1.00 & 28.37 & 0.54 & 28.43 & 0.52 & 28.94 & 0.24 & 29.39 & 0.35 & wl0 \\
 10 & 225544.19+624122.1 & 21.77 &  0.59 & 28.45 & 0.21 & 28.51 & 0.19 & 29.39 & 0.13 & 30.07 & 0.31 & bl0 \\
 11 & 225544.21+624213.4 & 21.36 &  1.40 & 28.93 & 0.25 & 28.95 & 0.24 & 29.42 & 0.15 & 29.57 & 0.17 & wl0 \\
 12 & 225545.08+623854.6 & 21.30 &  1.32 & 29.36 & 0.16 & 29.38 & 0.14 & 29.89 & 0.09 & 30.06 & 0.12 & bl0 \\
 13 & 225547.50+624157.6 & 21.24 &  1.10 & 28.69 & 0.21 & 28.70 & 0.20 & 29.38 & 0.13 & 29.54 & 0.13 & bl0 \\
 14 & 225547.59+623614.8 & 21.02 &  1.37 & 29.26 & 0.18 & 29.27 & 0.17 & 29.82 & 0.11 & 29.91 & 0.10 & bl0 \\
 15 & 225549.82+623421.9 & 22.21 & 15.00 & 29.74 & 0.18 & 29.80 & 0.21 & 29.77 & 0.14 & 29.96 & 0.38 & va0 \\
\enddata

\tablecomments{Column 1: X-ray source number. Column 2: IAU
designation. Columns 3-4: Estimated column density and plasma
energy. Columns 5-12: Observed and corrected for absorption X-ray
luminosities, obtained from our spectral analysis (see section
\ref{flux_det_section}), and their errors estimated from
simulations (see section \ref{flux_uncertainties_section}). $h=$
hard band $(2.0-8.0)$ keV, $t=$ total band $(0.5-8.0)$ keV, $c=$
corrected for absorption. Column 13: Three-part flag summarizing
spectral method, source membership, and $K$-disk star information:\\
b..  =  spectral results from spectral fits of 174 relatively
"bright" sources with $> 20$ net counts.\\
w..  =  spectral results from $N_{H,IR} - MedE$ interpolation for
212 "weak" sources with known $N_{H,IR}$.\\
v..  =  spectral results from fits of un-grouped spectra for 45
"weak" sources with unknown $N_{H,IR}$.\\
.e.  =  possible members of the embedded cluster.\\
.l.  =  possible members of the lightly absorbed cluster.\\
.o.  =  other possible members, found in the molecular cloud, not
related to the embedded cluster.\\
.a.  =  AGN candidates.\\
.f.  =  field star candidates.\\
..d  =  possible $K$-band excesses associated circumstellar disks.\\
..0  =  stars with no $K$-band excesses or unknown.}

\tablecomments{The full table of 431 {\it Chandra} sources is
available in the electronic edition of the Journal.}

\end{deluxetable}

\clearpage

\begin{deluxetable}{rcrrc}
\centering \tabletypesize{\tiny} \tablewidth{0pt} \tablecolumns{5}

\tablecaption{Power Law Fits for Extragalactic Candidates
\label{tbl_pow24}}

\tablehead{

\colhead{Seq} & \colhead{CXOCEPB} & \colhead{$\log N_H$} &
\colhead{$\Gamma$} & \colhead{$\log F_h$} \\

\colhead{\#} & \colhead{} & \colhead{cm$^{-2}$} & \colhead{} &
\colhead{erg s$^{-1}$ cm$^{-2}$} \\

\colhead{(1)} & \colhead{(2)} & \colhead{(3)} & \colhead{(4)} &
\colhead{(5)}}

\startdata
  6 & 225540.99+624625.9 & 22.05 &  1.71 & -13.73 \\
 15 & 225549.82+623421.9 & 22.30 &  1.06 & -13.84 \\
 27 & 225555.20+624356.6 & 22.22 &  2.02 & -13.01 \\
 34 & 225557.13+623751.7 & 22.71 & -1.32 & -14.12 \\
 41 & 225601.45+623923.1 & 22.11 &  0.11 & -14.13 \\
 45 & 225603.10+623714.7 & 23.11 &  1.17 & -14.13 \\
 64 & 225608.82+623440.5 & 20.00 & -0.11 & -14.27 \\
 86 & 225614.07+623552.7 & 21.88 &  1.56 & -14.25 \\
109 & 225620.22+624321.8 & 20.00 &  0.68 & -14.36 \\
123 & 225623.11+623557.1 & 22.53 &  2.93 & -14.08 \\
132 & 225624.66+624548.7 & 22.81 &  2.16 & -13.70 \\
144 & 225627.11+623442.8 & 23.14 &  9.50 & -14.32 \\
209 & 225641.33+624259.4 & 22.17 &  1.48 & -14.10 \\
215 & 225642.23+624016.3 & 22.17 &  0.21 & -14.20 \\
218 & 225642.78+623858.8 & 20.00 &  0.82 & -14.61 \\
220 & 225642.80+623211.9 & 21.48 &  1.41 & -14.30 \\
227 & 225643.88+623832.4 & 21.60 &  1.46 & -14.55 \\
244 & 225647.41+623057.9 & 23.22 &  0.95 & -13.41 \\
305 & 225658.56+624434.8 & 22.66 &  3.01 & -14.39 \\
398 & 225726.61+624628.1 & 22.46 &  3.11 & -14.37 \\
401 & 225727.44+624516.5 & 21.67 &  1.07 & -14.29 \\
415 & 225734.11+624251.1 & 22.08 &  0.88 & -13.93 \\
417 & 225738.06+624238.2 & 22.87 &  4.20 & -14.21 \\
429 & 225751.63+624813.0 & 23.38 &  9.46 & -13.86 \\
\enddata

\tablecomments{Column 1: X-ray source number. Column 2: IAU
designation. Columns 3-5: Column density, photon index and
observed flux in hard $(2.0-8.0)$ keV energy band from spline-like
fit of un-grouped spectrum.}

\end{deluxetable}

\clearpage

\begin{deluxetable}{rrrrrr}
\centering \tabletypesize{\tiny} \tablewidth{0pt} \tablecolumns{6}

\tablecaption{Integrated X-ray luminosities of detected PMS
stars\label{tbl_xlf_detected}}

\tablehead{

\colhead{Population} & \colhead{$N_{det}$} & \colhead{$\sum L_t$}
& \colhead{$\sum L_{t,c}$} & \colhead{$\sum L_h$} & \colhead{$\sum
L_{h,c}$} \\

\colhead{} & \colhead{} & \colhead{$10^{32}$ erg s$^{-1}$} & \colhead{$10^{32}$ erg s$^{-1}$} & \colhead{$10^{32}$ erg s$^{-1}$} & \colhead{$10^{32}$ erg s$^{-1}$} \\

\colhead{(1)} & \colhead{(2)} & \colhead{(3)} & \colhead{(4)} &
\colhead{(5)} & \colhead{(6)}}

\startdata
 Cep OB3b unobscured & $321$ & $3.97 \pm 0.10$ & $7.38 \pm 0.89$ & $2.64 \pm 0.11$ & $2.82 \pm 0.14$ \\
 Cep B obscured & $64$ & $0.50 \pm 0.03$ & $2.68 \pm 1.49$ & $0.39 \pm 0.03$ & $0.52 \pm 0.05$ \\
\enddata

\tablecomments{Column 1: Type of Cepheus PMS population. Column
2-6: Number of X-ray detected in this $Chandra$ observation PMS
stars and their total X-ray luminosities.}

\end{deluxetable}

\clearpage

\begin{deluxetable}{crrrr}
\centering \tabletypesize{\tiny} \tablewidth{0pt} \tablecolumns{5}

\tablecaption{EM linear regression fits \label{tbl_em_regression}}

\tablehead{ \multicolumn{5}{c}{$\log L_X = A+B \times \log(M/M_{\odot})$}  \\
\cline{1-5}

\colhead{$L_X$} & \colhead{$A$} & \colhead{$\Delta A$} &
\colhead{$B$} & \colhead{$\Delta B$} \\

\colhead{(1)} & \colhead{(2)} & \colhead{(3)} & \colhead{(4)} &
\colhead{(5)}}

\startdata
 $L_h$ & 29.95 & 0.08 &  2.25 & 0.13 \\
 $L_{h,c}$ & 29.97 & 0.08 &  2.23 & 0.13 \\
 $L_t$ & 30.24 & 0.07 &  1.99 & 0.11 \\
 $L_{t,c}$ & 30.43 & 0.07 & 1.99 & 0.11 \\
\enddata

\tablecomments{Column 1: Type of $L_X$. Column 2-3: Constant
parameter of the linear regression and its standard deviation.
Columns 4-5: Slope of the linear regression and its standard
deviation.}

\end{deluxetable}

\clearpage

\begin{deluxetable}{rrrr}
\centering \tabletypesize{\tiny} \tablewidth{0pt} \tablecolumns{4}

\tablecaption{Integrated X-ray luminosities of undetected
unobscured PMS stars \label{xlf_undetected_tbl}}

\tablehead{ \colhead{Log Mass Bin} & \colhead{$N_{undet}$} &
\colhead{$\sum L_t$} & \colhead{$\sum L_{t,c}$} \\
\colhead{} & \colhead{(NW-W)} & \colhead{(NW-W)} & \colhead{(NW-W)} \\
\colhead{} & \colhead{} & \colhead{$10^{30}$ erg s$^{-1}$} & \colhead{$10^{30}$ erg s$^{-1}$} \\

\colhead{(1)} & \colhead{(2)} & \colhead{(3)} & \colhead{(4)}}

\startdata
 -0.5 & $32-50$ & $5.62-8.79$ & $8.71-13.61$ \\
-0.7 & $115-90$ & $8.08-6.33$ & $12.52-9.80$ \\
-0.9 & $165-151$ & $4.64-4.25$ & $7.18-6.58$ \\
-1.1 & $102-167$ & $1.15-1.88$ & $1.78-2.91$ \\
$\sum$ & $414-458$ & $19.49-21.24$ & $30.20-32.90$ \\
\enddata

\tablecomments{Column 1: Mass bin in log. Column 2: Estimated
number of X-ray undetected stars for both, subtractions of "NW"
and "W" local backgrounds. Columns 3-4: Expected observed and
corrected for absorption X-ray luminosities.}

\end{deluxetable}

\clearpage

\begin{deluxetable}{rrrrrrr}
\centering \tabletypesize{\tiny} \tablewidth{0pt} \tablecolumns{7}

\tablecaption{X-ray properties of the diffuse component and
   PMS populations
\label{combined_spectra_results_tbl}}

\tablehead{ \colhead{Object} & \colhead{CR} & \colhead{$N_H$} &
\colhead{$kT_1$} & \colhead{$kT_2$} & \colhead{$L_t$} & \colhead{$L_{t,c}$} \\
\colhead{} & \colhead{c/ks} & \colhead{$10^{22}$ cm$^{-2}$} & \colhead{keV} & \colhead{keV} & \colhead{$10^{31}$ erg s$^{-1}$} & \colhead{$10^{31}$ erg s$^{-1}$} \\

\colhead{(1)} & \colhead{(2)} & \colhead{(3)} & \colhead{(4)} &
\colhead{(5)} & \colhead{(6)} & \colhead{(7)}}

\startdata
 Cepheus diffuse & 33.5 & $0.49 \pm 0.13$ & $0.78 \pm 0.14$ & \nodata & $2.21 \pm 0.07$ & $6.92 \pm 2.01$ \\
 Cepheus point sources & 570.0 & $0.33 \pm 0.23$ &  $0.61 \pm 0.23$ & $4.51 \pm 0.62$ & $54.34 \pm 4.41$ & $76.74 \pm 19.91$ \\
 COUP 326 weak sources & 43.1 & $0.34 \pm 0.15$ & $0.77 \pm 0.18$ & \nodata &
 $1.93 \pm 0.06$ & $4.84 \pm 2.10$ \\
\enddata

\tablecomments{Column 1: Defines object: either Cepheus X-ray
diffuse emission, or 321 X-ray detected point source members of
the Cep OB3b unobscured population, or 326 weak ($\log L_{t,c} <
29$) members of the COUP unobscured population. Column 2: Net
count rate. Column 3: Column density inferred from $XSPEC$ fits,
using wabs(apec) model for the diffuse emission and
wabs(vapec+vapec) for the point source population. Columns 4-5:
Plasma temperatures inferred from $XSPEC$ fits. Columns 6-7:
Observed and absorption corrected X-ray luminosities in
$(0.5-8.0)$ keV band, inferred from the X-ray spectral fits.}

\end{deluxetable}

\clearpage
\newpage

\clearpage
\newpage

\clearpage
\newpage

\clearpage
\newpage

\clearpage
\newpage

\clearpage
\newpage

\clearpage
\newpage

\clearpage
\newpage

\clearpage
\newpage

\clearpage
\newpage

\clearpage
\newpage

\clearpage
\newpage

\clearpage
\newpage

\clearpage
\newpage

\clearpage
\newpage

\clearpage
\newpage

\clearpage
\newpage

\clearpage
\newpage

\clearpage
\newpage

\clearpage

\clearpage
\newpage

\clearpage
\newpage

\clearpage
\newpage

\clearpage
\newpage

\clearpage
\newpage

\clearpage
\newpage

\clearpage
\newpage

\clearpage
\newpage

\clearpage
\newpage

\clearpage
\newpage

\clearpage
\newpage

\clearpage
\newpage

\end{document}